\documentclass[12pt]{article}
\usepackage{times,latexsym,amssymb,calc,alg,ifthen}
\usepackage{graphicx,color,geometry}
\newcommand{\ignore}[1]{}
\ignore{

latex nmrprhtml
bibtex nmrprhtml
latex nmrprhtml
latex nmrprhtml

latex2html -reuse 1 nmrprhtml

rsync -v -a --delete -e ssh nmrprhtml traal:public_html/qip
ssh traal 'chmod -R a+rX public_html/qip/nmrprhtml'

pdflatex nmrprpdf
bibtex nmrprpdf
pdflatex nmrprpdf
pdflatex nmrprpdf

scp nmrprpdf.pdf traal:public_html/drafts
ssh traal 'chmod -R a+rX public_html/drafts'

latex2html -reuse 1 -dir nmrprhtml1 -split 0 -no_navigation nmrprhtml

rsync -v -a --delete -e ssh nmrprhtml1 traal:public_html/drafts
ssh traal 'chmod -R a+rX public_html/drafts'

rsync -v -a --delete -e ssh nmrprhtml traal:/n/u1/www-external/users/knill/public_html/qip
scp nmrprpdf.pdf traal:/n/u1/www-external/users/knill/public_html/qip/nmrprhtml
ssh traal 'cdwww; chmod -R a+rX qip/nmrprhtml'

ssh traal
cdwww; cd ..
rsync -avH --delete -e ssh public_html www.c3.lanl.gov:/u2/users/knill 

rm -R /tmp/nmrprhtml; mkdir /tmp/nmrprhtml
cp `texfls nmrprhtml.log` /tmp/nmrprhtml
(cd /tmp/nmrprhtml; tar czvf nmrprhtml.tar.gz *)

}
\definecolor{snow}{rgb}{0.99609375,0.9765625,0.9765625}
\definecolor{ghost}{rgb}{0.96875,0.96875,0.99609375}
\definecolor{GhostWhite}{rgb}{0.96875,0.96875,0.99609375}
\definecolor{WhiteSmoke}{rgb}{0.95703125,0.95703125,0.95703125}
\definecolor{gainsboro}{rgb}{0.859375,0.859375,0.859375}
\definecolor{floral}{rgb}{0.99609375,0.9765625,0.9375}
\definecolor{FloralWhite}{rgb}{0.99609375,0.9765625,0.9375}
\definecolor{old}{rgb}{0.98828125,0.95703125,0.8984375}
\definecolor{OldLace}{rgb}{0.98828125,0.95703125,0.8984375}
\definecolor{linen}{rgb}{0.9765625,0.9375,0.8984375}
\definecolor{antique}{rgb}{0.9765625,0.91796875,0.83984375}
\definecolor{AntiqueWhite}{rgb}{0.9765625,0.91796875,0.83984375}
\definecolor{papaya}{rgb}{0.99609375,0.93359375,0.83203125}
\definecolor{PapayaWhip}{rgb}{0.99609375,0.93359375,0.83203125}
\definecolor{blanched}{rgb}{0.99609375,0.91796875,0.80078125}
\definecolor{BlanchedAlmond}{rgb}{0.99609375,0.91796875,0.80078125}
\definecolor{bisque}{rgb}{0.99609375,0.890625,0.765625}
\definecolor{peach}{rgb}{0.99609375,0.8515625,0.72265625}
\definecolor{PeachPuff}{rgb}{0.99609375,0.8515625,0.72265625}
\definecolor{navajo}{rgb}{0.99609375,0.8671875,0.67578125}
\definecolor{NavajoWhite}{rgb}{0.99609375,0.8671875,0.67578125}
\definecolor{moccasin}{rgb}{0.99609375,0.890625,0.70703125}
\definecolor{cornsilk}{rgb}{0.99609375,0.96875,0.859375}
\definecolor{ivory}{rgb}{0.99609375,0.99609375,0.9375}
\definecolor{lemon}{rgb}{0.99609375,0.9765625,0.80078125}
\definecolor{LemonChiffon}{rgb}{0.99609375,0.9765625,0.80078125}
\definecolor{seashell}{rgb}{0.99609375,0.95703125,0.9296875}
\definecolor{honeydew}{rgb}{0.9375,0.99609375,0.9375}
\definecolor{mint}{rgb}{0.95703125,0.99609375,0.9765625}
\definecolor{MintCream}{rgb}{0.95703125,0.99609375,0.9765625}
\definecolor{azure}{rgb}{0.9375,0.99609375,0.99609375}
\definecolor{alice}{rgb}{0.9375,0.96875,0.99609375}
\definecolor{AliceBlue}{rgb}{0.9375,0.96875,0.99609375}
\definecolor{lavender}{rgb}{0.99609375,0.9375,0.95703125}
\definecolor{LavenderBlush}{rgb}{0.99609375,0.9375,0.95703125}
\definecolor{misty}{rgb}{0.99609375,0.890625,0.87890625}
\definecolor{MistyRose}{rgb}{0.99609375,0.890625,0.87890625}
\definecolor{DarkSlateGray}{rgb}{0.18359375,0.30859375,0.30859375}
\definecolor{dim}{rgb}{0.41015625,0.41015625,0.41015625}
\definecolor{DimGray}{rgb}{0.41015625,0.41015625,0.41015625}
\definecolor{dim}{rgb}{0.41015625,0.41015625,0.41015625}
\definecolor{DimGrey}{rgb}{0.41015625,0.41015625,0.41015625}
\definecolor{SlateGray}{rgb}{0.4375,0.5,0.5625}
\definecolor{SlateGrey}{rgb}{0.4375,0.5,0.5625}
\definecolor{LightSlateGray}{rgb}{0.46484375,0.53125,0.59765625}
\definecolor{LightSlateGrey}{rgb}{0.46484375,0.53125,0.59765625}
\definecolor{gray}{rgb}{0.7421875,0.7421875,0.7421875}
\definecolor{grey}{rgb}{0.7421875,0.7421875,0.7421875}
\definecolor{LightGrey}{rgb}{0.82421875,0.82421875,0.82421875}
\definecolor{LightGray}{rgb}{0.82421875,0.82421875,0.82421875}
\definecolor{midnight}{rgb}{0.09765625,0.09765625,0.4375}
\definecolor{MidnightBlue}{rgb}{0.09765625,0.09765625,0.4375}
\definecolor{NavyBlue}{rgb}{0,0,0.5}
\definecolor{cornflower}{rgb}{0.390625,0.58203125,0.92578125}
\definecolor{CornflowerBlue}{rgb}{0.390625,0.58203125,0.92578125}
\definecolor{DarkSlateBlue}{rgb}{0.28125,0.23828125,0.54296875}
\definecolor{SlateBlue}{rgb}{0.4140625,0.3515625,0.80078125}
\definecolor{MediumSlateBlue}{rgb}{0.48046875,0.40625,0.9296875}
\definecolor{light}{rgb}{0.515625,0.4375,0.99609375}
\definecolor{LightSlateBlue}{rgb}{0.515625,0.4375,0.99609375}
\definecolor{MediumBlue}{rgb}{0,0,0.80078125}
\definecolor{royal}{rgb}{0.25390625,0.41015625,0.87890625}
\definecolor{RoyalBlue}{rgb}{0.25390625,0.41015625,0.87890625}
\definecolor{dodger}{rgb}{0.1171875,0.5625,0.99609375}
\definecolor{DodgerBlue}{rgb}{0.1171875,0.5625,0.99609375}
\definecolor{deep}{rgb}{0,0.74609375,0.99609375}
\definecolor{DeepSkyBlue}{rgb}{0,0.74609375,0.99609375}
\definecolor{sky}{rgb}{0.52734375,0.8046875,0.91796875}
\definecolor{SkyBlue}{rgb}{0.52734375,0.8046875,0.91796875}
\definecolor{LightSkyBlue}{rgb}{0.52734375,0.8046875,0.9765625}
\definecolor{steel}{rgb}{0.2734375,0.5078125,0.703125}
\definecolor{SteelBlue}{rgb}{0.2734375,0.5078125,0.703125}
\definecolor{LightSteelBlue}{rgb}{0.6875,0.765625,0.8671875}
\definecolor{LightBlue}{rgb}{0.67578125,0.84375,0.8984375}
\definecolor{powder}{rgb}{0.6875,0.875,0.8984375}
\definecolor{PowderBlue}{rgb}{0.6875,0.875,0.8984375}
\definecolor{PaleTurquoise}{rgb}{0.68359375,0.9296875,0.9296875}
\definecolor{DarkTurquoise}{rgb}{0,0.8046875,0.81640625}
\definecolor{MediumTurquoise}{rgb}{0.28125,0.81640625,0.796875}
\definecolor{turquoise}{rgb}{0.25,0.875,0.8125}
\definecolor{LightCyan}{rgb}{0.875,0.99609375,0.99609375}
\definecolor{cadet}{rgb}{0.37109375,0.6171875,0.625}
\definecolor{CadetBlue}{rgb}{0.37109375,0.6171875,0.625}
\definecolor{MediumAquamarine}{rgb}{0.3984375,0.80078125,0.6640625}
\definecolor{aquamarine}{rgb}{0.49609375,0.99609375,0.828125}
\definecolor{DarkGreen}{rgb}{0,0.390625,0}
\definecolor{DarkOliveGreen}{rgb}{0.33203125,0.41796875,0.18359375}
\definecolor{DarkSeaGreen}{rgb}{0.55859375,0.734375,0.55859375}
\definecolor{sea}{rgb}{0.1796875,0.54296875,0.33984375}
\definecolor{SeaGreen}{rgb}{0.1796875,0.54296875,0.33984375}
\definecolor{MediumSeaGreen}{rgb}{0.234375,0.69921875,0.44140625}
\definecolor{LightSeaGreen}{rgb}{0.125,0.6953125,0.6640625}
\definecolor{PaleGreen}{rgb}{0.59375,0.98046875,0.59375}
\definecolor{spring}{rgb}{0,0.99609375,0.49609375}
\definecolor{SpringGreen}{rgb}{0,0.99609375,0.49609375}
\definecolor{lawn}{rgb}{0.484375,0.984375,0}
\definecolor{LawnGreen}{rgb}{0.484375,0.984375,0}
\definecolor{chartreuse}{rgb}{0.49609375,0.99609375,0}
\definecolor{MediumSpringGreen}{rgb}{0,0.9765625,0.6015625}
\definecolor{GreenYellow}{rgb}{0.67578125,0.99609375,0.18359375}
\definecolor{lime}{rgb}{0.1953125,0.80078125,0.1953125}
\definecolor{LimeGreen}{rgb}{0.1953125,0.80078125,0.1953125}
\definecolor{YellowGreen}{rgb}{0.6015625,0.80078125,0.1953125}
\definecolor{forest}{rgb}{0.1328125,0.54296875,0.1328125}
\definecolor{ForestGreen}{rgb}{0.1328125,0.54296875,0.1328125}
\definecolor{olive}{rgb}{0.41796875,0.5546875,0.13671875}
\definecolor{OliveDrab}{rgb}{0.41796875,0.5546875,0.13671875}
\definecolor{DarkKhaki}{rgb}{0.73828125,0.71484375,0.41796875}
\definecolor{khaki}{rgb}{0.9375,0.8984375,0.546875}
\definecolor{PaleGoldenrod}{rgb}{0.9296875,0.90625,0.6640625}
\definecolor{LightGoldenrodYellow}{rgb}{0.9765625,0.9765625,0.8203125}
\definecolor{LightYellow}{rgb}{0.99609375,0.99609375,0.875}
\definecolor{gold}{rgb}{0.99609375,0.83984375,0}
\definecolor{LightGoldenrod}{rgb}{0.9296875,0.86328125,0.5078125}
\definecolor{goldenrod}{rgb}{0.8515625,0.64453125,0.125}
\definecolor{DarkGoldenrod}{rgb}{0.71875,0.5234375,0.04296875}
\definecolor{rosy}{rgb}{0.734375,0.55859375,0.55859375}
\definecolor{RosyBrown}{rgb}{0.734375,0.55859375,0.55859375}
\definecolor{indian}{rgb}{0.80078125,0.359375,0.359375}
\definecolor{IndianRed}{rgb}{0.80078125,0.359375,0.359375}
\definecolor{saddle}{rgb}{0.54296875,0.26953125,0.07421875}
\definecolor{SaddleBrown}{rgb}{0.54296875,0.26953125,0.07421875}
\definecolor{sienna}{rgb}{0.625,0.3203125,0.17578125}
\definecolor{peru}{rgb}{0.80078125,0.51953125,0.24609375}
\definecolor{burlywood}{rgb}{0.8671875,0.71875,0.52734375}
\definecolor{beige}{rgb}{0.95703125,0.95703125,0.859375}
\definecolor{wheat}{rgb}{0.95703125,0.8671875,0.69921875}
\definecolor{sandy}{rgb}{0.953125,0.640625,0.375}
\definecolor{SandyBrown}{rgb}{0.953125,0.640625,0.375}
\definecolor{tan}{rgb}{0.8203125,0.703125,0.546875}
\definecolor{chocolate}{rgb}{0.8203125,0.41015625,0.1171875}
\definecolor{firebrick}{rgb}{0.6953125,0.1328125,0.1328125}
\definecolor{brown}{rgb}{0.64453125,0.1640625,0.1640625}
\definecolor{DarkSalmon}{rgb}{0.91015625,0.5859375,0.4765625}
\definecolor{salmon}{rgb}{0.9765625,0.5,0.4453125}
\definecolor{LightSalmon}{rgb}{0.99609375,0.625,0.4765625}
\definecolor{orange}{rgb}{0.99609375,0.64453125,0}
\definecolor{DarkOrange}{rgb}{0.99609375,0.546875,0}
\definecolor{coral}{rgb}{0.99609375,0.49609375,0.3125}
\definecolor{LightCoral}{rgb}{0.9375,0.5,0.5}
\definecolor{tomato}{rgb}{0.99609375,0.38671875,0.27734375}
\definecolor{OrangeRed}{rgb}{0.99609375,0.26953125,0}
\definecolor{HotPink}{rgb}{0.99609375,0.41015625,0.703125}
\definecolor{DeepPink}{rgb}{0.99609375,0.078125,0.57421875}
\definecolor{pink}{rgb}{0.99609375,0.75,0.79296875}
\definecolor{LightPink}{rgb}{0.99609375,0.7109375,0.75390625}
\definecolor{PaleVioletRed}{rgb}{0.85546875,0.4375,0.57421875}
\definecolor{maroon}{rgb}{0.6875,0.1875,0.375}
\definecolor{MediumVioletRed}{rgb}{0.77734375,0.08203125,0.51953125}
\definecolor{violet}{rgb}{0.8125,0.125,0.5625}
\definecolor{VioletRed}{rgb}{0.8125,0.125,0.5625}
\definecolor{plum}{rgb}{0.86328125,0.625,0.86328125}
\definecolor{orchid}{rgb}{0.8515625,0.4375,0.8359375}
\definecolor{MediumOrchid}{rgb}{0.7265625,0.33203125,0.82421875}
\definecolor{DarkOrchid}{rgb}{0.59765625,0.1953125,0.796875}
\definecolor{DarkViolet}{rgb}{0.578125,0,0.82421875}
\definecolor{blue}{rgb}{0.5390625,0.16796875,0.8828125}
\definecolor{BlueViolet}{rgb}{0.5390625,0.16796875,0.8828125}
\definecolor{purple}{rgb}{0.625,0.125,0.9375}
\definecolor{MediumPurple}{rgb}{0.57421875,0.4375,0.85546875}
\definecolor{thistle}{rgb}{0.84375,0.74609375,0.84375}
\definecolor{snow1}{rgb}{0.99609375,0.9765625,0.9765625}
\definecolor{snow2}{rgb}{0.9296875,0.91015625,0.91015625}
\definecolor{snow3}{rgb}{0.80078125,0.78515625,0.78515625}
\definecolor{snow4}{rgb}{0.54296875,0.53515625,0.53515625}
\definecolor{seashell1}{rgb}{0.99609375,0.95703125,0.9296875}
\definecolor{seashell2}{rgb}{0.9296875,0.89453125,0.8671875}
\definecolor{seashell3}{rgb}{0.80078125,0.76953125,0.74609375}
\definecolor{seashell4}{rgb}{0.54296875,0.5234375,0.5078125}
\definecolor{AntiqueWhite1}{rgb}{0.99609375,0.93359375,0.85546875}
\definecolor{AntiqueWhite2}{rgb}{0.9296875,0.87109375,0.796875}
\definecolor{AntiqueWhite3}{rgb}{0.80078125,0.75,0.6875}
\definecolor{AntiqueWhite4}{rgb}{0.54296875,0.51171875,0.46875}
\definecolor{bisque1}{rgb}{0.99609375,0.890625,0.765625}
\definecolor{bisque2}{rgb}{0.9296875,0.83203125,0.71484375}
\definecolor{bisque3}{rgb}{0.80078125,0.71484375,0.6171875}
\definecolor{bisque4}{rgb}{0.54296875,0.48828125,0.41796875}
\definecolor{PeachPuff1}{rgb}{0.99609375,0.8515625,0.72265625}
\definecolor{PeachPuff2}{rgb}{0.9296875,0.79296875,0.67578125}
\definecolor{PeachPuff3}{rgb}{0.80078125,0.68359375,0.58203125}
\definecolor{PeachPuff4}{rgb}{0.54296875,0.46484375,0.39453125}
\definecolor{NavajoWhite1}{rgb}{0.99609375,0.8671875,0.67578125}
\definecolor{NavajoWhite2}{rgb}{0.9296875,0.80859375,0.62890625}
\definecolor{NavajoWhite3}{rgb}{0.80078125,0.69921875,0.54296875}
\definecolor{NavajoWhite4}{rgb}{0.54296875,0.47265625,0.3671875}
\definecolor{LemonChiffon1}{rgb}{0.99609375,0.9765625,0.80078125}
\definecolor{LemonChiffon2}{rgb}{0.9296875,0.91015625,0.74609375}
\definecolor{LemonChiffon3}{rgb}{0.80078125,0.78515625,0.64453125}
\definecolor{LemonChiffon4}{rgb}{0.54296875,0.53515625,0.4375}
\definecolor{cornsilk1}{rgb}{0.99609375,0.96875,0.859375}
\definecolor{cornsilk2}{rgb}{0.9296875,0.90625,0.80078125}
\definecolor{cornsilk3}{rgb}{0.80078125,0.78125,0.69140625}
\definecolor{cornsilk4}{rgb}{0.54296875,0.53125,0.46875}
\definecolor{ivory1}{rgb}{0.99609375,0.99609375,0.9375}
\definecolor{ivory2}{rgb}{0.9296875,0.9296875,0.875}
\definecolor{ivory3}{rgb}{0.80078125,0.80078125,0.75390625}
\definecolor{ivory4}{rgb}{0.54296875,0.54296875,0.51171875}
\definecolor{honeydew1}{rgb}{0.9375,0.99609375,0.9375}
\definecolor{honeydew2}{rgb}{0.875,0.9296875,0.875}
\definecolor{honeydew3}{rgb}{0.75390625,0.80078125,0.75390625}
\definecolor{honeydew4}{rgb}{0.51171875,0.54296875,0.51171875}
\definecolor{LavenderBlush1}{rgb}{0.99609375,0.9375,0.95703125}
\definecolor{LavenderBlush2}{rgb}{0.9296875,0.875,0.89453125}
\definecolor{LavenderBlush3}{rgb}{0.80078125,0.75390625,0.76953125}
\definecolor{LavenderBlush4}{rgb}{0.54296875,0.51171875,0.5234375}
\definecolor{MistyRose1}{rgb}{0.99609375,0.890625,0.87890625}
\definecolor{MistyRose2}{rgb}{0.9296875,0.83203125,0.8203125}
\definecolor{MistyRose3}{rgb}{0.80078125,0.71484375,0.70703125}
\definecolor{MistyRose4}{rgb}{0.54296875,0.48828125,0.48046875}
\definecolor{azure1}{rgb}{0.9375,0.99609375,0.99609375}
\definecolor{azure2}{rgb}{0.875,0.9296875,0.9296875}
\definecolor{azure3}{rgb}{0.75390625,0.80078125,0.80078125}
\definecolor{azure4}{rgb}{0.51171875,0.54296875,0.54296875}
\definecolor{SlateBlue1}{rgb}{0.51171875,0.43359375,0.99609375}
\definecolor{SlateBlue2}{rgb}{0.4765625,0.40234375,0.9296875}
\definecolor{SlateBlue3}{rgb}{0.41015625,0.34765625,0.80078125}
\definecolor{SlateBlue4}{rgb}{0.27734375,0.234375,0.54296875}
\definecolor{RoyalBlue1}{rgb}{0.28125,0.4609375,0.99609375}
\definecolor{RoyalBlue2}{rgb}{0.26171875,0.4296875,0.9296875}
\definecolor{RoyalBlue3}{rgb}{0.2265625,0.37109375,0.80078125}
\definecolor{RoyalBlue4}{rgb}{0.15234375,0.25,0.54296875}
\definecolor{blue1}{rgb}{0,0,0.99609375}
\definecolor{blue2}{rgb}{0,0,0.9296875}
\definecolor{blue3}{rgb}{0,0,0.80078125}
\definecolor{blue4}{rgb}{0,0,0.54296875}
\definecolor{DodgerBlue1}{rgb}{0.1171875,0.5625,0.99609375}
\definecolor{DodgerBlue2}{rgb}{0.109375,0.5234375,0.9296875}
\definecolor{DodgerBlue3}{rgb}{0.09375,0.453125,0.80078125}
\definecolor{DodgerBlue4}{rgb}{0.0625,0.3046875,0.54296875}
\definecolor{SteelBlue1}{rgb}{0.38671875,0.71875,0.99609375}
\definecolor{SteelBlue2}{rgb}{0.359375,0.671875,0.9296875}
\definecolor{SteelBlue3}{rgb}{0.30859375,0.578125,0.80078125}
\definecolor{SteelBlue4}{rgb}{0.2109375,0.390625,0.54296875}
\definecolor{DeepSkyBlue1}{rgb}{0,0.74609375,0.99609375}
\definecolor{DeepSkyBlue2}{rgb}{0,0.6953125,0.9296875}
\definecolor{DeepSkyBlue3}{rgb}{0,0.6015625,0.80078125}
\definecolor{DeepSkyBlue4}{rgb}{0,0.40625,0.54296875}
\definecolor{SkyBlue1}{rgb}{0.52734375,0.8046875,0.99609375}
\definecolor{SkyBlue2}{rgb}{0.4921875,0.75,0.9296875}
\definecolor{SkyBlue3}{rgb}{0.421875,0.6484375,0.80078125}
\definecolor{SkyBlue4}{rgb}{0.2890625,0.4375,0.54296875}
\definecolor{LightSkyBlue1}{rgb}{0.6875,0.8828125,0.99609375}
\definecolor{LightSkyBlue2}{rgb}{0.640625,0.82421875,0.9296875}
\definecolor{LightSkyBlue3}{rgb}{0.55078125,0.7109375,0.80078125}
\definecolor{LightSkyBlue4}{rgb}{0.375,0.48046875,0.54296875}
\definecolor{SlateGray1}{rgb}{0.7734375,0.8828125,0.99609375}
\definecolor{SlateGray2}{rgb}{0.72265625,0.82421875,0.9296875}
\definecolor{SlateGray3}{rgb}{0.62109375,0.7109375,0.80078125}
\definecolor{SlateGray4}{rgb}{0.421875,0.48046875,0.54296875}
\definecolor{LightSteelBlue1}{rgb}{0.7890625,0.87890625,0.99609375}
\definecolor{LightSteelBlue2}{rgb}{0.734375,0.8203125,0.9296875}
\definecolor{LightSteelBlue3}{rgb}{0.6328125,0.70703125,0.80078125}
\definecolor{LightSteelBlue4}{rgb}{0.4296875,0.48046875,0.54296875}
\definecolor{LightBlue1}{rgb}{0.74609375,0.93359375,0.99609375}
\definecolor{LightBlue2}{rgb}{0.6953125,0.87109375,0.9296875}
\definecolor{LightBlue3}{rgb}{0.6015625,0.75,0.80078125}
\definecolor{LightBlue4}{rgb}{0.40625,0.51171875,0.54296875}
\definecolor{LightCyan1}{rgb}{0.875,0.99609375,0.99609375}
\definecolor{LightCyan2}{rgb}{0.81640625,0.9296875,0.9296875}
\definecolor{LightCyan3}{rgb}{0.703125,0.80078125,0.80078125}
\definecolor{LightCyan4}{rgb}{0.4765625,0.54296875,0.54296875}
\definecolor{PaleTurquoise1}{rgb}{0.73046875,0.99609375,0.99609375}
\definecolor{PaleTurquoise2}{rgb}{0.6796875,0.9296875,0.9296875}
\definecolor{PaleTurquoise3}{rgb}{0.5859375,0.80078125,0.80078125}
\definecolor{PaleTurquoise4}{rgb}{0.3984375,0.54296875,0.54296875}
\definecolor{CadetBlue1}{rgb}{0.59375,0.95703125,0.99609375}
\definecolor{CadetBlue2}{rgb}{0.5546875,0.89453125,0.9296875}
\definecolor{CadetBlue3}{rgb}{0.4765625,0.76953125,0.80078125}
\definecolor{CadetBlue4}{rgb}{0.32421875,0.5234375,0.54296875}
\definecolor{turquoise1}{rgb}{0,0.95703125,0.99609375}
\definecolor{turquoise2}{rgb}{0,0.89453125,0.9296875}
\definecolor{turquoise3}{rgb}{0,0.76953125,0.80078125}
\definecolor{turquoise4}{rgb}{0,0.5234375,0.54296875}
\definecolor{cyan1}{rgb}{0,0.99609375,0.99609375}
\definecolor{cyan2}{rgb}{0,0.9296875,0.9296875}
\definecolor{cyan3}{rgb}{0,0.80078125,0.80078125}
\definecolor{cyan4}{rgb}{0,0.54296875,0.54296875}
\definecolor{DarkSlateGray1}{rgb}{0.58984375,0.99609375,0.99609375}
\definecolor{DarkSlateGray2}{rgb}{0.55078125,0.9296875,0.9296875}
\definecolor{DarkSlateGray3}{rgb}{0.47265625,0.80078125,0.80078125}
\definecolor{DarkSlateGray4}{rgb}{0.3203125,0.54296875,0.54296875}
\definecolor{aquamarine1}{rgb}{0.49609375,0.99609375,0.828125}
\definecolor{aquamarine2}{rgb}{0.4609375,0.9296875,0.7734375}
\definecolor{aquamarine3}{rgb}{0.3984375,0.80078125,0.6640625}
\definecolor{aquamarine4}{rgb}{0.26953125,0.54296875,0.453125}
\definecolor{DarkSeaGreen1}{rgb}{0.75390625,0.99609375,0.75390625}
\definecolor{DarkSeaGreen2}{rgb}{0.703125,0.9296875,0.703125}
\definecolor{DarkSeaGreen3}{rgb}{0.60546875,0.80078125,0.60546875}
\definecolor{DarkSeaGreen4}{rgb}{0.41015625,0.54296875,0.41015625}
\definecolor{SeaGreen1}{rgb}{0.328125,0.99609375,0.62109375}
\definecolor{SeaGreen2}{rgb}{0.3046875,0.9296875,0.578125}
\definecolor{SeaGreen3}{rgb}{0.26171875,0.80078125,0.5}
\definecolor{SeaGreen4}{rgb}{0.1796875,0.54296875,0.33984375}
\definecolor{PaleGreen1}{rgb}{0.6015625,0.99609375,0.6015625}
\definecolor{PaleGreen2}{rgb}{0.5625,0.9296875,0.5625}
\definecolor{PaleGreen3}{rgb}{0.484375,0.80078125,0.484375}
\definecolor{PaleGreen4}{rgb}{0.328125,0.54296875,0.328125}
\definecolor{SpringGreen1}{rgb}{0,0.99609375,0.49609375}
\definecolor{SpringGreen2}{rgb}{0,0.9296875,0.4609375}
\definecolor{SpringGreen3}{rgb}{0,0.80078125,0.3984375}
\definecolor{SpringGreen4}{rgb}{0,0.54296875,0.26953125}
\definecolor{green1}{rgb}{0,0.99609375,0}
\definecolor{green2}{rgb}{0,0.9296875,0}
\definecolor{green3}{rgb}{0,0.80078125,0}
\definecolor{green4}{rgb}{0,0.54296875,0}
\definecolor{chartreuse1}{rgb}{0.49609375,0.99609375,0}
\definecolor{chartreuse2}{rgb}{0.4609375,0.9296875,0}
\definecolor{chartreuse3}{rgb}{0.3984375,0.80078125,0}
\definecolor{chartreuse4}{rgb}{0.26953125,0.54296875,0}
\definecolor{OliveDrab1}{rgb}{0.75,0.99609375,0.2421875}
\definecolor{OliveDrab2}{rgb}{0.69921875,0.9296875,0.2265625}
\definecolor{OliveDrab3}{rgb}{0.6015625,0.80078125,0.1953125}
\definecolor{OliveDrab4}{rgb}{0.41015625,0.54296875,0.1328125}
\definecolor{DarkOliveGreen1}{rgb}{0.7890625,0.99609375,0.4375}
\definecolor{DarkOliveGreen2}{rgb}{0.734375,0.9296875,0.40625}
\definecolor{DarkOliveGreen3}{rgb}{0.6328125,0.80078125,0.3515625}
\definecolor{DarkOliveGreen4}{rgb}{0.4296875,0.54296875,0.23828125}
\definecolor{khaki1}{rgb}{0.99609375,0.9609375,0.55859375}
\definecolor{khaki2}{rgb}{0.9296875,0.8984375,0.51953125}
\definecolor{khaki3}{rgb}{0.80078125,0.7734375,0.44921875}
\definecolor{khaki4}{rgb}{0.54296875,0.5234375,0.3046875}
\definecolor{LightGoldenrod1}{rgb}{0.99609375,0.921875,0.54296875}
\definecolor{LightGoldenrod2}{rgb}{0.9296875,0.859375,0.5078125}
\definecolor{LightGoldenrod3}{rgb}{0.80078125,0.7421875,0.4375}
\definecolor{LightGoldenrod4}{rgb}{0.54296875,0.50390625,0.296875}
\definecolor{LightYellow1}{rgb}{0.99609375,0.99609375,0.875}
\definecolor{LightYellow2}{rgb}{0.9296875,0.9296875,0.81640625}
\definecolor{LightYellow3}{rgb}{0.80078125,0.80078125,0.703125}
\definecolor{LightYellow4}{rgb}{0.54296875,0.54296875,0.4765625}
\definecolor{yellow1}{rgb}{0.99609375,0.99609375,0}
\definecolor{yellow2}{rgb}{0.9296875,0.9296875,0}
\definecolor{yellow3}{rgb}{0.80078125,0.80078125,0}
\definecolor{yellow4}{rgb}{0.54296875,0.54296875,0}
\definecolor{gold1}{rgb}{0.99609375,0.83984375,0}
\definecolor{gold2}{rgb}{0.9296875,0.78515625,0}
\definecolor{gold3}{rgb}{0.80078125,0.67578125,0}
\definecolor{gold4}{rgb}{0.54296875,0.45703125,0}
\definecolor{goldenrod1}{rgb}{0.99609375,0.75390625,0.14453125}
\definecolor{goldenrod2}{rgb}{0.9296875,0.703125,0.1328125}
\definecolor{goldenrod3}{rgb}{0.80078125,0.60546875,0.11328125}
\definecolor{goldenrod4}{rgb}{0.54296875,0.41015625,0.078125}
\definecolor{DarkGoldenrod1}{rgb}{0.99609375,0.72265625,0.05859375}
\definecolor{DarkGoldenrod2}{rgb}{0.9296875,0.67578125,0.0546875}
\definecolor{DarkGoldenrod3}{rgb}{0.80078125,0.58203125,0.046875}
\definecolor{DarkGoldenrod4}{rgb}{0.54296875,0.39453125,0.03125}
\definecolor{RosyBrown1}{rgb}{0.99609375,0.75390625,0.75390625}
\definecolor{RosyBrown2}{rgb}{0.9296875,0.703125,0.703125}
\definecolor{RosyBrown3}{rgb}{0.80078125,0.60546875,0.60546875}
\definecolor{RosyBrown4}{rgb}{0.54296875,0.41015625,0.41015625}
\definecolor{IndianRed1}{rgb}{0.99609375,0.4140625,0.4140625}
\definecolor{IndianRed2}{rgb}{0.9296875,0.38671875,0.38671875}
\definecolor{IndianRed3}{rgb}{0.80078125,0.33203125,0.33203125}
\definecolor{IndianRed4}{rgb}{0.54296875,0.2265625,0.2265625}
\definecolor{sienna1}{rgb}{0.99609375,0.5078125,0.27734375}
\definecolor{sienna2}{rgb}{0.9296875,0.47265625,0.2578125}
\definecolor{sienna3}{rgb}{0.80078125,0.40625,0.22265625}
\definecolor{sienna4}{rgb}{0.54296875,0.27734375,0.1484375}
\definecolor{burlywood1}{rgb}{0.99609375,0.82421875,0.60546875}
\definecolor{burlywood2}{rgb}{0.9296875,0.76953125,0.56640625}
\definecolor{burlywood3}{rgb}{0.80078125,0.6640625,0.48828125}
\definecolor{burlywood4}{rgb}{0.54296875,0.44921875,0.33203125}
\definecolor{wheat1}{rgb}{0.99609375,0.90234375,0.7265625}
\definecolor{wheat2}{rgb}{0.9296875,0.84375,0.6796875}
\definecolor{wheat3}{rgb}{0.80078125,0.7265625,0.5859375}
\definecolor{wheat4}{rgb}{0.54296875,0.4921875,0.3984375}
\definecolor{tan1}{rgb}{0.99609375,0.64453125,0.30859375}
\definecolor{tan2}{rgb}{0.9296875,0.6015625,0.28515625}
\definecolor{tan3}{rgb}{0.80078125,0.51953125,0.24609375}
\definecolor{tan4}{rgb}{0.54296875,0.3515625,0.16796875}
\definecolor{chocolate1}{rgb}{0.99609375,0.49609375,0.140625}
\definecolor{chocolate2}{rgb}{0.9296875,0.4609375,0.12890625}
\definecolor{chocolate3}{rgb}{0.80078125,0.3984375,0.11328125}
\definecolor{chocolate4}{rgb}{0.54296875,0.26953125,0.07421875}
\definecolor{firebrick1}{rgb}{0.99609375,0.1875,0.1875}
\definecolor{firebrick2}{rgb}{0.9296875,0.171875,0.171875}
\definecolor{firebrick3}{rgb}{0.80078125,0.1484375,0.1484375}
\definecolor{firebrick4}{rgb}{0.54296875,0.1015625,0.1015625}
\definecolor{brown1}{rgb}{0.99609375,0.25,0.25}
\definecolor{brown2}{rgb}{0.9296875,0.23046875,0.23046875}
\definecolor{brown3}{rgb}{0.80078125,0.19921875,0.19921875}
\definecolor{brown4}{rgb}{0.54296875,0.13671875,0.13671875}
\definecolor{salmon1}{rgb}{0.99609375,0.546875,0.41015625}
\definecolor{salmon2}{rgb}{0.9296875,0.5078125,0.3828125}
\definecolor{salmon3}{rgb}{0.80078125,0.4375,0.328125}
\definecolor{salmon4}{rgb}{0.54296875,0.296875,0.22265625}
\definecolor{LightSalmon1}{rgb}{0.99609375,0.625,0.4765625}
\definecolor{LightSalmon2}{rgb}{0.9296875,0.58203125,0.4453125}
\definecolor{LightSalmon3}{rgb}{0.80078125,0.50390625,0.3828125}
\definecolor{LightSalmon4}{rgb}{0.54296875,0.33984375,0.2578125}
\definecolor{orange1}{rgb}{0.99609375,0.64453125,0}
\definecolor{orange2}{rgb}{0.9296875,0.6015625,0}
\definecolor{orange3}{rgb}{0.80078125,0.51953125,0}
\definecolor{orange4}{rgb}{0.54296875,0.3515625,0}
\definecolor{DarkOrange1}{rgb}{0.99609375,0.49609375,0}
\definecolor{DarkOrange2}{rgb}{0.9296875,0.4609375,0}
\definecolor{DarkOrange3}{rgb}{0.80078125,0.3984375,0}
\definecolor{DarkOrange4}{rgb}{0.54296875,0.26953125,0}
\definecolor{coral1}{rgb}{0.99609375,0.4453125,0.3359375}
\definecolor{coral2}{rgb}{0.9296875,0.4140625,0.3125}
\definecolor{coral3}{rgb}{0.80078125,0.35546875,0.26953125}
\definecolor{coral4}{rgb}{0.54296875,0.2421875,0.18359375}
\definecolor{tomato1}{rgb}{0.99609375,0.38671875,0.27734375}
\definecolor{tomato2}{rgb}{0.9296875,0.359375,0.2578125}
\definecolor{tomato3}{rgb}{0.80078125,0.30859375,0.22265625}
\definecolor{tomato4}{rgb}{0.54296875,0.2109375,0.1484375}
\definecolor{OrangeRed1}{rgb}{0.99609375,0.26953125,0}
\definecolor{OrangeRed2}{rgb}{0.9296875,0.25,0}
\definecolor{OrangeRed3}{rgb}{0.80078125,0.21484375,0}
\definecolor{OrangeRed4}{rgb}{0.54296875,0.14453125,0}
\definecolor{red1}{rgb}{0.99609375,0,0}
\definecolor{red2}{rgb}{0.9296875,0,0}
\definecolor{red3}{rgb}{0.80078125,0,0}
\definecolor{red4}{rgb}{0.54296875,0,0}
\definecolor{DeepPink1}{rgb}{0.99609375,0.078125,0.57421875}
\definecolor{DeepPink2}{rgb}{0.9296875,0.0703125,0.53515625}
\definecolor{DeepPink3}{rgb}{0.80078125,0.0625,0.4609375}
\definecolor{DeepPink4}{rgb}{0.54296875,0.0390625,0.3125}
\definecolor{HotPink1}{rgb}{0.99609375,0.4296875,0.703125}
\definecolor{HotPink2}{rgb}{0.9296875,0.4140625,0.65234375}
\definecolor{HotPink3}{rgb}{0.80078125,0.375,0.5625}
\definecolor{HotPink4}{rgb}{0.54296875,0.2265625,0.3828125}
\definecolor{pink1}{rgb}{0.99609375,0.70703125,0.76953125}
\definecolor{pink2}{rgb}{0.9296875,0.66015625,0.71875}
\definecolor{pink3}{rgb}{0.80078125,0.56640625,0.6171875}
\definecolor{pink4}{rgb}{0.54296875,0.38671875,0.421875}
\definecolor{LightPink1}{rgb}{0.99609375,0.6796875,0.72265625}
\definecolor{LightPink2}{rgb}{0.9296875,0.6328125,0.67578125}
\definecolor{LightPink3}{rgb}{0.80078125,0.546875,0.58203125}
\definecolor{LightPink4}{rgb}{0.54296875,0.37109375,0.39453125}
\definecolor{PaleVioletRed1}{rgb}{0.99609375,0.5078125,0.66796875}
\definecolor{PaleVioletRed2}{rgb}{0.9296875,0.47265625,0.62109375}
\definecolor{PaleVioletRed3}{rgb}{0.80078125,0.40625,0.53515625}
\definecolor{PaleVioletRed4}{rgb}{0.54296875,0.27734375,0.36328125}
\definecolor{maroon1}{rgb}{0.99609375,0.203125,0.69921875}
\definecolor{maroon2}{rgb}{0.9296875,0.1875,0.65234375}
\definecolor{maroon3}{rgb}{0.80078125,0.16015625,0.5625}
\definecolor{maroon4}{rgb}{0.54296875,0.109375,0.3828125}
\definecolor{VioletRed1}{rgb}{0.99609375,0.2421875,0.5859375}
\definecolor{VioletRed2}{rgb}{0.9296875,0.2265625,0.546875}
\definecolor{VioletRed3}{rgb}{0.80078125,0.1953125,0.46875}
\definecolor{VioletRed4}{rgb}{0.54296875,0.1328125,0.3203125}
\definecolor{magenta1}{rgb}{0.99609375,0,0.99609375}
\definecolor{magenta2}{rgb}{0.9296875,0,0.9296875}
\definecolor{magenta3}{rgb}{0.80078125,0,0.80078125}
\definecolor{magenta4}{rgb}{0.54296875,0,0.54296875}
\definecolor{orchid1}{rgb}{0.99609375,0.51171875,0.9765625}
\definecolor{orchid2}{rgb}{0.9296875,0.4765625,0.91015625}
\definecolor{orchid3}{rgb}{0.80078125,0.41015625,0.78515625}
\definecolor{orchid4}{rgb}{0.54296875,0.27734375,0.53515625}
\definecolor{plum1}{rgb}{0.99609375,0.73046875,0.99609375}
\definecolor{plum2}{rgb}{0.9296875,0.6796875,0.9296875}
\definecolor{plum3}{rgb}{0.80078125,0.5859375,0.80078125}
\definecolor{plum4}{rgb}{0.54296875,0.3984375,0.54296875}
\definecolor{MediumOrchid1}{rgb}{0.875,0.3984375,0.99609375}
\definecolor{MediumOrchid2}{rgb}{0.81640625,0.37109375,0.9296875}
\definecolor{MediumOrchid3}{rgb}{0.703125,0.3203125,0.80078125}
\definecolor{MediumOrchid4}{rgb}{0.4765625,0.21484375,0.54296875}
\definecolor{DarkOrchid1}{rgb}{0.74609375,0.2421875,0.99609375}
\definecolor{DarkOrchid2}{rgb}{0.6953125,0.2265625,0.9296875}
\definecolor{DarkOrchid3}{rgb}{0.6015625,0.1953125,0.80078125}
\definecolor{DarkOrchid4}{rgb}{0.40625,0.1328125,0.54296875}
\definecolor{purple1}{rgb}{0.60546875,0.1875,0.99609375}
\definecolor{purple2}{rgb}{0.56640625,0.171875,0.9296875}
\definecolor{purple3}{rgb}{0.48828125,0.1484375,0.80078125}
\definecolor{purple4}{rgb}{0.33203125,0.1015625,0.54296875}
\definecolor{MediumPurple1}{rgb}{0.66796875,0.5078125,0.99609375}
\definecolor{MediumPurple2}{rgb}{0.62109375,0.47265625,0.9296875}
\definecolor{MediumPurple3}{rgb}{0.53515625,0.40625,0.80078125}
\definecolor{MediumPurple4}{rgb}{0.36328125,0.27734375,0.54296875}
\definecolor{thistle1}{rgb}{0.99609375,0.87890625,0.99609375}
\definecolor{thistle2}{rgb}{0.9296875,0.8203125,0.9296875}
\definecolor{thistle3}{rgb}{0.80078125,0.70703125,0.80078125}
\definecolor{thistle4}{rgb}{0.54296875,0.48046875,0.54296875}
\definecolor{gray0}{rgb}{0,0,0}
\definecolor{grey0}{rgb}{0,0,0}
\definecolor{gray1}{rgb}{0.01171875,0.01171875,0.01171875}
\definecolor{grey1}{rgb}{0.01171875,0.01171875,0.01171875}
\definecolor{gray2}{rgb}{0.01953125,0.01953125,0.01953125}
\definecolor{grey2}{rgb}{0.01953125,0.01953125,0.01953125}
\definecolor{gray3}{rgb}{0.03125,0.03125,0.03125}
\definecolor{grey3}{rgb}{0.03125,0.03125,0.03125}
\definecolor{gray4}{rgb}{0.0390625,0.0390625,0.0390625}
\definecolor{grey4}{rgb}{0.0390625,0.0390625,0.0390625}
\definecolor{gray5}{rgb}{0.05078125,0.05078125,0.05078125}
\definecolor{grey5}{rgb}{0.05078125,0.05078125,0.05078125}
\definecolor{gray6}{rgb}{0.05859375,0.05859375,0.05859375}
\definecolor{grey6}{rgb}{0.05859375,0.05859375,0.05859375}
\definecolor{gray7}{rgb}{0.0703125,0.0703125,0.0703125}
\definecolor{grey7}{rgb}{0.0703125,0.0703125,0.0703125}
\definecolor{gray8}{rgb}{0.078125,0.078125,0.078125}
\definecolor{grey8}{rgb}{0.078125,0.078125,0.078125}
\definecolor{gray9}{rgb}{0.08984375,0.08984375,0.08984375}
\definecolor{grey9}{rgb}{0.08984375,0.08984375,0.08984375}
\definecolor{gray10}{rgb}{0.1015625,0.1015625,0.1015625}
\definecolor{grey10}{rgb}{0.1015625,0.1015625,0.1015625}
\definecolor{gray11}{rgb}{0.109375,0.109375,0.109375}
\definecolor{grey11}{rgb}{0.109375,0.109375,0.109375}
\definecolor{gray12}{rgb}{0.12109375,0.12109375,0.12109375}
\definecolor{grey12}{rgb}{0.12109375,0.12109375,0.12109375}
\definecolor{gray13}{rgb}{0.12890625,0.12890625,0.12890625}
\definecolor{grey13}{rgb}{0.12890625,0.12890625,0.12890625}
\definecolor{gray14}{rgb}{0.140625,0.140625,0.140625}
\definecolor{grey14}{rgb}{0.140625,0.140625,0.140625}
\definecolor{gray15}{rgb}{0.1484375,0.1484375,0.1484375}
\definecolor{grey15}{rgb}{0.1484375,0.1484375,0.1484375}
\definecolor{gray16}{rgb}{0.16015625,0.16015625,0.16015625}
\definecolor{grey16}{rgb}{0.16015625,0.16015625,0.16015625}
\definecolor{gray17}{rgb}{0.16796875,0.16796875,0.16796875}
\definecolor{grey17}{rgb}{0.16796875,0.16796875,0.16796875}
\definecolor{gray18}{rgb}{0.1796875,0.1796875,0.1796875}
\definecolor{grey18}{rgb}{0.1796875,0.1796875,0.1796875}
\definecolor{gray19}{rgb}{0.1875,0.1875,0.1875}
\definecolor{grey19}{rgb}{0.1875,0.1875,0.1875}
\definecolor{gray20}{rgb}{0.19921875,0.19921875,0.19921875}
\definecolor{grey20}{rgb}{0.19921875,0.19921875,0.19921875}
\definecolor{gray21}{rgb}{0.2109375,0.2109375,0.2109375}
\definecolor{grey21}{rgb}{0.2109375,0.2109375,0.2109375}
\definecolor{gray22}{rgb}{0.21875,0.21875,0.21875}
\definecolor{grey22}{rgb}{0.21875,0.21875,0.21875}
\definecolor{gray23}{rgb}{0.23046875,0.23046875,0.23046875}
\definecolor{grey23}{rgb}{0.23046875,0.23046875,0.23046875}
\definecolor{gray24}{rgb}{0.23828125,0.23828125,0.23828125}
\definecolor{grey24}{rgb}{0.23828125,0.23828125,0.23828125}
\definecolor{gray25}{rgb}{0.25,0.25,0.25}
\definecolor{grey25}{rgb}{0.25,0.25,0.25}
\definecolor{gray26}{rgb}{0.2578125,0.2578125,0.2578125}
\definecolor{grey26}{rgb}{0.2578125,0.2578125,0.2578125}
\definecolor{gray27}{rgb}{0.26953125,0.26953125,0.26953125}
\definecolor{grey27}{rgb}{0.26953125,0.26953125,0.26953125}
\definecolor{gray28}{rgb}{0.27734375,0.27734375,0.27734375}
\definecolor{grey28}{rgb}{0.27734375,0.27734375,0.27734375}
\definecolor{gray29}{rgb}{0.2890625,0.2890625,0.2890625}
\definecolor{grey29}{rgb}{0.2890625,0.2890625,0.2890625}
\definecolor{gray30}{rgb}{0.30078125,0.30078125,0.30078125}
\definecolor{grey30}{rgb}{0.30078125,0.30078125,0.30078125}
\definecolor{gray31}{rgb}{0.30859375,0.30859375,0.30859375}
\definecolor{grey31}{rgb}{0.30859375,0.30859375,0.30859375}
\definecolor{gray32}{rgb}{0.3203125,0.3203125,0.3203125}
\definecolor{grey32}{rgb}{0.3203125,0.3203125,0.3203125}
\definecolor{gray33}{rgb}{0.328125,0.328125,0.328125}
\definecolor{grey33}{rgb}{0.328125,0.328125,0.328125}
\definecolor{gray34}{rgb}{0.33984375,0.33984375,0.33984375}
\definecolor{grey34}{rgb}{0.33984375,0.33984375,0.33984375}
\definecolor{gray35}{rgb}{0.34765625,0.34765625,0.34765625}
\definecolor{grey35}{rgb}{0.34765625,0.34765625,0.34765625}
\definecolor{gray36}{rgb}{0.359375,0.359375,0.359375}
\definecolor{grey36}{rgb}{0.359375,0.359375,0.359375}
\definecolor{gray37}{rgb}{0.3671875,0.3671875,0.3671875}
\definecolor{grey37}{rgb}{0.3671875,0.3671875,0.3671875}
\definecolor{gray38}{rgb}{0.37890625,0.37890625,0.37890625}
\definecolor{grey38}{rgb}{0.37890625,0.37890625,0.37890625}
\definecolor{gray39}{rgb}{0.38671875,0.38671875,0.38671875}
\definecolor{grey39}{rgb}{0.38671875,0.38671875,0.38671875}
\definecolor{gray40}{rgb}{0.3984375,0.3984375,0.3984375}
\definecolor{grey40}{rgb}{0.3984375,0.3984375,0.3984375}
\definecolor{gray41}{rgb}{0.41015625,0.41015625,0.41015625}
\definecolor{grey41}{rgb}{0.41015625,0.41015625,0.41015625}
\definecolor{gray42}{rgb}{0.41796875,0.41796875,0.41796875}
\definecolor{grey42}{rgb}{0.41796875,0.41796875,0.41796875}
\definecolor{gray43}{rgb}{0.4296875,0.4296875,0.4296875}
\definecolor{grey43}{rgb}{0.4296875,0.4296875,0.4296875}
\definecolor{gray44}{rgb}{0.4375,0.4375,0.4375}
\definecolor{grey44}{rgb}{0.4375,0.4375,0.4375}
\definecolor{gray45}{rgb}{0.44921875,0.44921875,0.44921875}
\definecolor{grey45}{rgb}{0.44921875,0.44921875,0.44921875}
\definecolor{gray46}{rgb}{0.45703125,0.45703125,0.45703125}
\definecolor{grey46}{rgb}{0.45703125,0.45703125,0.45703125}
\definecolor{gray47}{rgb}{0.46875,0.46875,0.46875}
\definecolor{grey47}{rgb}{0.46875,0.46875,0.46875}
\definecolor{gray48}{rgb}{0.4765625,0.4765625,0.4765625}
\definecolor{grey48}{rgb}{0.4765625,0.4765625,0.4765625}
\definecolor{gray49}{rgb}{0.48828125,0.48828125,0.48828125}
\definecolor{grey49}{rgb}{0.48828125,0.48828125,0.48828125}
\definecolor{gray50}{rgb}{0.49609375,0.49609375,0.49609375}
\definecolor{grey50}{rgb}{0.49609375,0.49609375,0.49609375}
\definecolor{gray51}{rgb}{0.5078125,0.5078125,0.5078125}
\definecolor{grey51}{rgb}{0.5078125,0.5078125,0.5078125}
\definecolor{gray52}{rgb}{0.51953125,0.51953125,0.51953125}
\definecolor{grey52}{rgb}{0.51953125,0.51953125,0.51953125}
\definecolor{gray53}{rgb}{0.52734375,0.52734375,0.52734375}
\definecolor{grey53}{rgb}{0.52734375,0.52734375,0.52734375}
\definecolor{gray54}{rgb}{0.5390625,0.5390625,0.5390625}
\definecolor{grey54}{rgb}{0.5390625,0.5390625,0.5390625}
\definecolor{gray55}{rgb}{0.546875,0.546875,0.546875}
\definecolor{grey55}{rgb}{0.546875,0.546875,0.546875}
\definecolor{gray56}{rgb}{0.55859375,0.55859375,0.55859375}
\definecolor{grey56}{rgb}{0.55859375,0.55859375,0.55859375}
\definecolor{gray57}{rgb}{0.56640625,0.56640625,0.56640625}
\definecolor{grey57}{rgb}{0.56640625,0.56640625,0.56640625}
\definecolor{gray58}{rgb}{0.578125,0.578125,0.578125}
\definecolor{grey58}{rgb}{0.578125,0.578125,0.578125}
\definecolor{gray59}{rgb}{0.5859375,0.5859375,0.5859375}
\definecolor{grey59}{rgb}{0.5859375,0.5859375,0.5859375}
\definecolor{gray60}{rgb}{0.59765625,0.59765625,0.59765625}
\definecolor{grey60}{rgb}{0.59765625,0.59765625,0.59765625}
\definecolor{gray61}{rgb}{0.609375,0.609375,0.609375}
\definecolor{grey61}{rgb}{0.609375,0.609375,0.609375}
\definecolor{gray62}{rgb}{0.6171875,0.6171875,0.6171875}
\definecolor{grey62}{rgb}{0.6171875,0.6171875,0.6171875}
\definecolor{gray63}{rgb}{0.62890625,0.62890625,0.62890625}
\definecolor{grey63}{rgb}{0.62890625,0.62890625,0.62890625}
\definecolor{gray64}{rgb}{0.63671875,0.63671875,0.63671875}
\definecolor{grey64}{rgb}{0.63671875,0.63671875,0.63671875}
\definecolor{gray65}{rgb}{0.6484375,0.6484375,0.6484375}
\definecolor{grey65}{rgb}{0.6484375,0.6484375,0.6484375}
\definecolor{gray66}{rgb}{0.65625,0.65625,0.65625}
\definecolor{grey66}{rgb}{0.65625,0.65625,0.65625}
\definecolor{gray67}{rgb}{0.66796875,0.66796875,0.66796875}
\definecolor{grey67}{rgb}{0.66796875,0.66796875,0.66796875}
\definecolor{gray68}{rgb}{0.67578125,0.67578125,0.67578125}
\definecolor{grey68}{rgb}{0.67578125,0.67578125,0.67578125}
\definecolor{gray69}{rgb}{0.6875,0.6875,0.6875}
\definecolor{grey69}{rgb}{0.6875,0.6875,0.6875}
\definecolor{gray70}{rgb}{0.69921875,0.69921875,0.69921875}
\definecolor{grey70}{rgb}{0.69921875,0.69921875,0.69921875}
\definecolor{gray71}{rgb}{0.70703125,0.70703125,0.70703125}
\definecolor{grey71}{rgb}{0.70703125,0.70703125,0.70703125}
\definecolor{gray72}{rgb}{0.71875,0.71875,0.71875}
\definecolor{grey72}{rgb}{0.71875,0.71875,0.71875}
\definecolor{gray73}{rgb}{0.7265625,0.7265625,0.7265625}
\definecolor{grey73}{rgb}{0.7265625,0.7265625,0.7265625}
\definecolor{gray74}{rgb}{0.73828125,0.73828125,0.73828125}
\definecolor{grey74}{rgb}{0.73828125,0.73828125,0.73828125}
\definecolor{gray75}{rgb}{0.74609375,0.74609375,0.74609375}
\definecolor{grey75}{rgb}{0.74609375,0.74609375,0.74609375}
\definecolor{gray76}{rgb}{0.7578125,0.7578125,0.7578125}
\definecolor{grey76}{rgb}{0.7578125,0.7578125,0.7578125}
\definecolor{gray77}{rgb}{0.765625,0.765625,0.765625}
\definecolor{grey77}{rgb}{0.765625,0.765625,0.765625}
\definecolor{gray78}{rgb}{0.77734375,0.77734375,0.77734375}
\definecolor{grey78}{rgb}{0.77734375,0.77734375,0.77734375}
\definecolor{gray79}{rgb}{0.78515625,0.78515625,0.78515625}
\definecolor{grey79}{rgb}{0.78515625,0.78515625,0.78515625}
\definecolor{gray80}{rgb}{0.796875,0.796875,0.796875}
\definecolor{grey80}{rgb}{0.796875,0.796875,0.796875}
\definecolor{gray81}{rgb}{0.80859375,0.80859375,0.80859375}
\definecolor{grey81}{rgb}{0.80859375,0.80859375,0.80859375}
\definecolor{gray82}{rgb}{0.81640625,0.81640625,0.81640625}
\definecolor{grey82}{rgb}{0.81640625,0.81640625,0.81640625}
\definecolor{gray83}{rgb}{0.828125,0.828125,0.828125}
\definecolor{grey83}{rgb}{0.828125,0.828125,0.828125}
\definecolor{gray84}{rgb}{0.8359375,0.8359375,0.8359375}
\definecolor{grey84}{rgb}{0.8359375,0.8359375,0.8359375}
\definecolor{gray85}{rgb}{0.84765625,0.84765625,0.84765625}
\definecolor{grey85}{rgb}{0.84765625,0.84765625,0.84765625}
\definecolor{gray86}{rgb}{0.85546875,0.85546875,0.85546875}
\definecolor{grey86}{rgb}{0.85546875,0.85546875,0.85546875}
\definecolor{gray87}{rgb}{0.8671875,0.8671875,0.8671875}
\definecolor{grey87}{rgb}{0.8671875,0.8671875,0.8671875}
\definecolor{gray88}{rgb}{0.875,0.875,0.875}
\definecolor{grey88}{rgb}{0.875,0.875,0.875}
\definecolor{gray89}{rgb}{0.88671875,0.88671875,0.88671875}
\definecolor{grey89}{rgb}{0.88671875,0.88671875,0.88671875}
\definecolor{gray90}{rgb}{0.89453125,0.89453125,0.89453125}
\definecolor{grey90}{rgb}{0.89453125,0.89453125,0.89453125}
\definecolor{gray91}{rgb}{0.90625,0.90625,0.90625}
\definecolor{grey91}{rgb}{0.90625,0.90625,0.90625}
\definecolor{gray92}{rgb}{0.91796875,0.91796875,0.91796875}
\definecolor{grey92}{rgb}{0.91796875,0.91796875,0.91796875}
\definecolor{gray93}{rgb}{0.92578125,0.92578125,0.92578125}
\definecolor{grey93}{rgb}{0.92578125,0.92578125,0.92578125}
\definecolor{gray94}{rgb}{0.9375,0.9375,0.9375}
\definecolor{grey94}{rgb}{0.9375,0.9375,0.9375}
\definecolor{gray95}{rgb}{0.9453125,0.9453125,0.9453125}
\definecolor{grey95}{rgb}{0.9453125,0.9453125,0.9453125}
\definecolor{gray96}{rgb}{0.95703125,0.95703125,0.95703125}
\definecolor{grey96}{rgb}{0.95703125,0.95703125,0.95703125}
\definecolor{gray97}{rgb}{0.96484375,0.96484375,0.96484375}
\definecolor{grey97}{rgb}{0.96484375,0.96484375,0.96484375}
\definecolor{gray98}{rgb}{0.9765625,0.9765625,0.9765625}
\definecolor{grey98}{rgb}{0.9765625,0.9765625,0.9765625}
\definecolor{gray99}{rgb}{0.984375,0.984375,0.984375}
\definecolor{grey99}{rgb}{0.984375,0.984375,0.984375}
\definecolor{gray100}{rgb}{0.99609375,0.99609375,0.99609375}
\definecolor{grey100}{rgb}{0.99609375,0.99609375,0.99609375}
\definecolor{DarkGrey}{rgb}{0.66015625,0.66015625,0.66015625}
\definecolor{DarkGray}{rgb}{0.66015625,0.66015625,0.66015625}
\definecolor{DarkBlue}{rgb}{0,0,0.54296875}
\definecolor{DarkCyan}{rgb}{0,0.54296875,0.54296875}
\definecolor{DarkMagenta}{rgb}{0.54296875,0,0.54296875}
\definecolor{DarkRed}{rgb}{0.54296875,0,0}
\definecolor{LightGreen}{rgb}{0.5625,0.9296875,0.5625}

\usepackage{html}

\newcommand{\makecnts}{\tableofcontents\newpage}

\newcommand{\makeaddress}{
\noindent\textbf{Addresses:}
\begin{tabular}[t]{lcl}
R. Laflamme:&
University of Waterloo and Perimeter Institute
&\htmladdnormallink{\texttt{laflamme@iqc.ca}}{mailto:laflamme@iqc.ca}
\\
E. Knill:&Los Alamos National Laboratory&\htmladdnormallink{\texttt{knill@lanl.gov}}{mailto:knill@lanl.gov}
\\
D. Cory:&MIT&\htmladdnormallink{\texttt{dcory@mit.edu}}{mailto:dcory@mit.edu}
\\
E. M. Fortunato:&''&\htmladdnormallink{\texttt{evanmf@mit.edu}}{mailto:evanmf@mit.edu}
\\
T. Havel:&''&\htmladdnormallink{\texttt{tfhavel@mit.edu}}{mailto:tfhavel@mit.edu}
\\
C. Miquel:&FCEN, Univ. Buenos Aires&\htmladdnormallink{\texttt{miquel@df.uba.ar}}{mailto:miquel@df.uba.ar}
\\
R. Martinez:&Los Alamos National Laboratory&\htmladdnormallink{\texttt{rudy@lanl.gov}}{mailto:rudy@lanl.gov}
\\
C. Negrevergne:&''&\htmladdnormallink{\texttt{cjn@lanl.gov}}{mailto:cjn@lanl.gov}
\\
G. Ortiz: &''&\htmladdnormallink{\texttt{g\string_ortiz@lanl.gov}}{mailto:g_ortiz@lanl.gov}
\\
M. A. Pravia: &MIT&\htmladdnormallink{\texttt{praviam@mit.edu}}{mailto:praviam@mit.edu}
\\
Y. Sharf: &''&\htmladdnormallink{\texttt{ysharf@mit.edu}}{mailto:ysharf@mit.edu}
\\
S. Sinha: &''&\htmladdnormallink{\texttt{suddha@mit.edu}}{mailto:suddha@mit.edu}
\\
R. Somma:&Los Alamos National Laboratory&\htmladdnormallink{\texttt{somma@lanl.gov}}{mailto:somma@lanl.gov}
\\
L. Viola: &''&\htmladdnormallink{\texttt{lviola@lanl.gov}}{mailto:lviola@lanl.gov}
\end{tabular}
}

\graphicspath{{./}{fig.pdf/}{graphics/}}

\newcommand{\kComment}[1]{}
\newcommand{\lComment}[1]{}
\newcommand{\cComment}[1]{}
\newcommand{\nComment}[1]{}
\newcommand{\mComment}[1]{}
\renewcommand{\kComment}[1]{\textcolor{blue}{Manny: #1}}
\renewcommand{\lComment}[1]{\textcolor{red}{Ray: #1}}
\renewcommand{\cComment}[1]{\textcolor{green}{David: #1}}
\renewcommand{\nComment}[1]{\textcolor{magenta}{Camille: #1}}
\renewcommand{\mComment}[1]{\textcolor{purple}{Rudy: #1}}

\def\one{{\mathchoice {\rm 1\mskip-4mu l} {\rm 1\mskip-4mu l} {\rm
1\mskip-4.5mu l} {\rm 1\mskip-5mu l}}}

\newcommand{\qvbar}{\mbox{$|\hspace*{-3pt}|\hspace*{-3pt}|$}}
\newcommand{\qrangle}{\mbox{$\rangle\hspace*{-4.3pt}\rangle\hspace*{-4.3pt}\rangle$}}
\newcommand{\qlangle}{\mbox{$\langle\hspace*{-4.3pt}\langle\hspace*{-4.3pt}\langle$}}

\newcommand{\sysfnt}{\mathsf}

\newcommand{\ket}[1]{\qvbar{#1}\qrangle}
\newcommand{\bra}[1]{\qlangle{#1}\qvbar}

\newcommand{\ketbra}[2]{\ket{#1}\bra{#2}}
\newcommand{\kets}[2]{\qvbar{#1}\qrangle_{{}_{\!\!\scriptstyle{\sysfnt{#2}}}}}
\newcommand{\bras}[2]{{}^{\scriptstyle\sysfnt{ #2}}\!\qlangle{#1}\qvbar}

\newcommand{\ketbras}[3]{\kets{#1}{#3}\!\!\bras{#2}{#3}}
\newcommand{\slb}[2]{{#1}^{({\sysfnt{#2}})}}

\newcommand{\qaop}[4]{\left(\begin{array}{cc}#1&#2\\ #3&#4\end{array}\right)}

\newcommand{\nputbox}[3]{\put(#1){\makebox(0,0)[#2]{#3}}}
\newcommand{\nputgr}[4]{\put(#1){\makebox(0,0)[#2]{\includegraphics[#3]{#4}}}}

\newlength{\elimdepthdim}
\newlength{\elimheightdim}
\newlength{\elimwidthdim}

\newlength{\strutdepthdim}
\newlength{\strutheightdim}
\newlength{\strutwidthdim}

\newcommand{\trace}{\mbox{tr}}

\def\id{{\mathchoice {\rm 1\mskip-4mu l} {\rm 1\mskip-4mu l} {\rm
1\mskip-4.5mu l} {\rm 1\mskip-5mu l}}}

\newcommand{\mb}[1]{\mathbf{#1}}

\def\uts#1{{\mathchoice{\mbox{#1}}{\mbox{#1}}{\mbox{\small #1}}{\mbox{\tiny #1}}}\,}

\newcommand{\bitzero}{\mathfrak{0}}
\newcommand{\bitone}{\mathfrak{1}}

\newcommand{\idop}{\id}

\newcounter{herefignum}
\newenvironment{herefig}{\begin{center}\refstepcounter{herefignum}}{\end{center}}
\newcommand{\herefigcap}[1]{\\\begin{minipage}{\textwidth}{FIG.~\theherefignum: #1}\end{minipage}}

\begin{document}

\title{Introduction to NMR Quantum Information Processing}
\author{R. Laflamme, E. Knill, D. G. Cory, E. M. Fortunato, T. Havel, \\
C. Miquel, R. Martinez, C. Negrevergne,  G. Ortiz, M. A. Pravia, Y. Sharf, \\
S. Sinha, R. Somma and L. Viola}
\date{\today}

\maketitle

\begin{latexonly}
\makecnts
\end{latexonly}

\ignore{
After a general introduction to nuclear magnetic resonance (NMR), we
give the basics of implementing quantum algorithms.  We describe how
qubits are realized and controlled with RF pulses, their internal
interactions, and gradient fields.  A peculiarity of NMR is that the
internal interactions (given by the internal Hamiltonian) are always
on.  We discuss how they can be effectively turned off with the help
of a standard NMR method called ``refocusing''.  Liquid state NMR
experiments are done at room temperature, leading to an extremely
mixed (that is, nearly random) initial state.  Despite this high
degree of randomness, it is possible to investigate QIP because the
relaxation time (the time scale over which useful signal from a
computation is lost) is sufficiently long.  We explain how this
feature leads to the crucial ability of simulating a pure (non-random)
state by using ``pseudopure'' states. We discuss how the ``answer''
provided by a computation is obtained by measurement and how this
measurement differs from the ideal, projective measurement of QIP.  We
then give implementations of some simple quantum algorithms with a
typical experimental result. We conclude with a discussion of what we
have learned from NMR QIP so far and what the prospects for future NMR
QIP experiments are.  
}

Using quantum physics to represent and manipulate information makes
possible surprising improvements in the efficiency with which some
problems can be solved.  But can these improvements be realized
experimentally?  If we consider the history of implementing
theoretical ideas about classical information and computation, we find
that initially, small numbers of simple devices were used to explore
the advantages and the difficulties of information processing.  For
example, in 1933 Atanasoff and his colleagues at the Iowa State
College were able to implement digital calculations using about 300
vacuum tubes (see~\cite{zalta:qc2002a}, the entry for ``computing,
modern history of'').  Although the device was never practical because
its error rate was too large, it was probably the first instance of a
programmable computer using vacuum tubes and it opened the way for
more stable and reliable devices. Progress toward implementing quantum
information processors is also initially confined to limited capacity
and error-prone devices.

There are numerous proposals for implementing quantum information
processing (QIP) prototypes. To date (2002), only three of them have
been used to successfully manipulate more than one qubit: cavity
quantum electrodynamics (cavity QED), ion traps and nuclear magnetic
resonance (NMR) with molecules in a liquid (liquid state NMR). 
The difficulty of realizing QIP devices can be attributed to an
intrinsic conflict between two of the most important requirements: On the
one hand, it is necessary for the device to be well isolated from, and
therefore interact only weakly with, its environment; otherwise, the
crucial quantum correlations on which the advantages of QIP are based
are destroyed. On the other hand, it is necessary for the different
parts of the device to interact strongly with each other and for some
of them to be coupled strongly with the measuring device, which is
needed to read out ``answers''. That few physical systems have these
properties naturally is apparent from the absence of obvious quantum
effects in the macroscopic world.

One system whose properties constitute a reasonable compromise between
the two requirements consists of the nuclear spins in a
molecule in the liquid state.  The spins, particularly those with
spin ${1\over 2}$, provide a natural representation of quantum bits. They
interact weakly but reliably with each other and the effects of the
environment are often small enough. The spins can be controlled with
radio-frequency (RF) pulses and observed with measurements of the
magnetic fields that they generate. Liquid state NMR has so far
been used to demonstrate control of up to seven physical qubits. 

It is important to remember that the idea of QIP is less than two
decades old, and, with the notable exception of quantum cryptography,
experimental proposals and efforts aimed at realizing modern QIP began
only in the last five years of the 20'th century.  Increasingly
advanced experiments are being implemented. But from an information
processing point of view, we are a long way from using quantum
technology to solve an independently posed problem not solvable on a
standard personal computer---a typical ``classical'' computer. In
order to get to the point where such problems can be solved by QIP,
current experimental efforts are devoted to understanding the behavior
of and the methods for controlling various quantum systems, as well as
ways of overcoming their limitations. The work on NMR QIP has focused
on the control of quantum systems by algorithmically implementing
quantum transformations as precisely as possible. Within the
limitations of the device, this approach has been surprisingly successful,
thanks to the many scientists and engineers who have perfected NMR
spectrometers over the past 50 years.

After a general introduction to NMR, we give the basics of
implementing quantum algorithms.  We describe how qubits are realized
and controlled with RF pulses, their internal interactions, and gradient
fields.  A peculiarity of NMR is that the internal interactions (given
by the internal Hamiltonian) are always on.  We discuss how they can be
effectively turned off with the help of a standard NMR method called
``refocusing''.  Liquid state NMR experiments are done at room
temperature, leading to an extremely mixed (that is, nearly random)
initial state.  Despite this high degree of randomness, it is possible
to investigate QIP because the relaxation time (the time scale over
which useful signal from a computation is lost) is sufficiently long.
We explain how this feature leads to the crucial ability of simulating a pure
(non-random) state by using ``pseudopure'' states. We discuss how the
``answer'' provided by a computation is obtained by measurement and
how this measurement differs from the ideal, projective measurement of
QIP.  We then give implementations of some simple quantum algorithms
with a typical experimental result. We conclude with a discussion of
what we have learned from NMR QIP so far and what the prospects for
future NMR QIP experiments are.  For an elementary, device-independent
introduction to quantum information and definitions of the states and
operators used here, see~\cite{knill:qc2001c}.

\section {Liquid-State NMR}

\subsection{NMR Basics}

Many atomic nuclei have a magnetic moment, which means that, like
small bar magnets, they respond to and can be detected by their
magnetic fields.  Although single nuclei are impossible to detect
directly by these means with currently available technology, if
sufficiently many are available so that their contributions to the
magnetic field add, they can be observed as an ensemble. In
liquid-state NMR, the nuclei belong to atoms forming a molecule, a
very large number of which are dissolved in a liquid.  An example is
${}^{13}C$-labeled trichloroethylene (TCE) (Fig.~\ref{fig:tce}).  The hydrogen
nucleus (that is the proton) of each TCE molecule has a relatively
strong magnetic moment. When the sample is placed in a powerful
external magnetic field, each proton's spin prefers to align itself
with the field.  It is possible to induce the spin direction to
``tip'' off-axis by means of RF pulses, at which point the effect of
the static field is to induce a rapid precession of the proton spins.
In this introduction, precession refers to a rotation of a spin
direction around the main axis, here the $z$-axis as determined by the
external magnetic field.  The precession frequency $\omega$ is often
called the Larmor frequency and is linearly related to the strength
$B$ of the external field: $\omega =
\mu B$, where $\mu$ is the magnetic moment. For the proton, the
magnetic moment is $42.7\uts{Mhz}/\uts{T}$. ($\uts{Mhz}$ stands for
``megahertz'', which is a frequency unit equal to $10^6$ rotations per
second.  $\uts{T}$ stands for ``Tesla'', a magnetic field unit.)  At a
typical field of $B=11.7\uts{T}$, the proton's precession frequency is
$500\uts{Mhz}$.  The magnetic field produced by the precessing protons
induces oscillating currents in a coil judiciously placed around the
sample and ``tuned'' to the precession frequency, allowing
observation of the entire ensemble of protons by ``magnetic
induction''. This is the fundamental idea of NMR.  The device that
applies the static magnetic field and RF control pulses and
that detects the magnetic induction is called an NMR spectrometer
(Fig.~\ref{fig:spec}).

\begin{herefig}
\begin{picture}(4,2.8)(-2,-2.7)\large
\nputgr{0,0}{t}{height=2.5in}{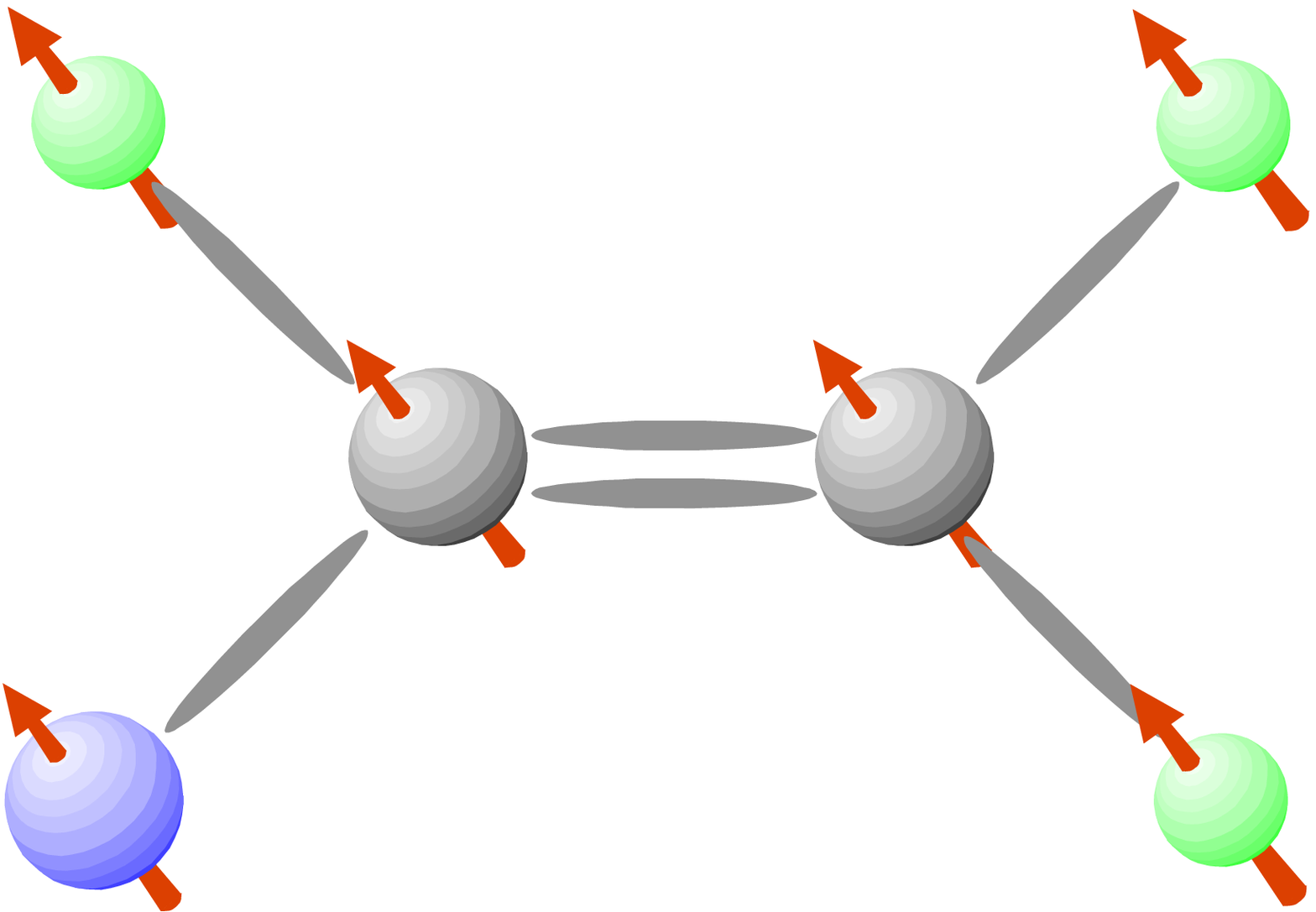}
\nputbox{-1.9,-.2}{t}{Cl}
\nputbox{+1.9,-.2}{t}{Cl}
\nputbox{+1.9,-2.1}{t}{Cl}
\nputbox{-1.9,-2.1}{t}{H}
\nputbox{-.6,-.9}{b}{${}^{13}C$}
\nputbox{+.6,-.9}{b}{${}^{13}C$}
\end{picture}
\label{fig:tce}
\herefigcap{Schematic of trichloroethylene, a typical molecule used
for QIP. There are three useful
nuclei for realizing qubits. They are the proton (H),
and the two carbons (${}^{13}$C). The molecule is ``labeled'',
which means that the nuclei are carefully chosen isotopes.
In this case, the normally predominant isotope of carbon,
${}^{12}$C (a spin-zero nucleus), is replaced by ${}^{13}$C,
which has spin ${1\over 2}$.
}
\end{herefig}

\pagebreak

\begin{herefig}
\begin{picture}(7.5,5.5)(-3.75,-6)
\nputgr{0,0}{t}{height=6in}{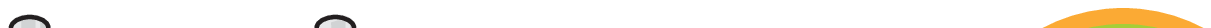}
\end{picture}
\label{fig:spec}
\herefigcap{Schematic of a typical NMR spectrometer (not to scale).
The main components of a spectrometer are the magnet, which is
superconducting, and the console , which has the electronics needed to
control the spectrometer.  The sample containing a liquid solution of
the molecule used for QIP is inserted into the central core of the
magnet, where it is surrounded by the ``probe''. The probe (shown
enlarged in the insert to the right) contains coils for applying the
radio frequency (RF) pulses and magnetic field gradients.}
\end{herefig}

\pagebreak

Magnetic induction by nuclear spins was observed for the first time in
1946 by the groups of E.~Purcell~\cite{purcell:qc1945a} and
F.~Bloch~\cite{bloch:qc1946a}.  This achievement opened a new field of
research, leading to many important applications, such as molecular
structure determination, dynamics studies both in the liquid and solid
state~\cite{ernst:qc1994a}, and magnetic resonance
imaging~\cite{mansfield:qc1982a}. The application of NMR to QIP is
related to methods for molecular structure determination by NMR. Many
of the same techniques are used in QIP, but instead of using
uncharacterized molecules, specific ones with well-defined nuclear
spins are synthesized. In this setting, one can manipulate the nuclear
spins as quantum information so that it becomes possible to
experimentally demonstrate the fundamental ideas of QIP.

Perhaps the clearest example of early connections of NMR to
information theory is the spin echo phenomenon~\cite{hahn:qc1950a}.
When the static magnetic field is not ``homogeneous'' (that is, it is
not constant across the sample), the spins precess at different
frequencies depending on their location in the sample. As a result,
the magnetic induction signal rapidly vanishes because the magnetic
fields produced by the spins are no longer aligned and therefore do
not add. The spin echo is used to ``refocus'' this effect by inverting
the spins, an action that effectively reverses their precession until they are
all aligned again.  Based on spin echoes, the idea of using nuclear
spins for (classical) information storage was suggested and patented
by A.~Anderson and E.~Hahn as early
as~1955~\cite{anderson:qc1955a,hahn:qc1955a}.

NMR spectroscopy would not be possible if it were not for relatively long
``relaxation'' times. Relaxation is the process that tends to
re-align the nuclear spins with the field and randomize their phases, an effect
that leads to complete loss of the information represented in such a
spin. In liquid state, relaxation times of the order
of seconds are common and attributed to the weakness
of nuclear interactions and a fast averaging effect associated
with the rapid, tumbling motions of molecules in the liquid state.

Currently, ``off-the-shelf'' NMR spectrometers are robust and
straightforward to use. The requisite control is to a large extent
computerized, so most NMR experiments involve few custom 
adjustments after the sample has been obtained.  Given that the
underlying nature of the nuclear spins is intrinsically quantum
mechanical, it is not surprising that, soon after P.~Shor's discovery of the
quantum factoring algorithm, NMR was studied as a potentially
useful device for QIP.

\subsection {A Brief Survey of NMR QIP}

Concrete and workable proposals for using liquid-state NMR for quantum
information were first given in 1996/7 by D.~Cory, A.~Fahmy and
T.~Havel~\cite{cory:qc1997a} and by N.~Gershenfeld and
I.~Chuang~\cite{chuang:qc1997a}. Three difficulties had to be overcome
for NMR QIP to become possible. The first was that the standard
definitions of quantum information and computation require that
quantum information be stored in a single physical system.  In NMR, an
obvious such system consists of some of the nuclear spins in a single
molecule. But it is not possible to detect single molecules with
available NMR technology. The solution that makes NMR QIP possible can
be applied to other QIP technologies: Consider the large collection of
available molecules as an ensemble of identical systems.  As long as
they all perform the same task, the desired answers can be read out
collectively. The second difficulty was that the standard definitions
require that read-out take place by a projective quantum measurements
of the qubits. From such a measurement, one learns whether a qubit is
in the state $\ket{\bitzero}$ or $\ket{\bitone}$. The two measurement
outcomes have probabilities determined by the initial state of the
qubits being used, and after the measurement the state ``collapses''
to a state consistent with the outcome. The measurement in NMR is much
too weak to determine the outcome and cause the state's collapse for
each molecule. But because of the additive effects of the ensemble,
one can observe a (noisy) signal that represents the average, over all
the molecules of the probability that $\ket{\bitone}$ would be the
outcome of a projective measurement. It turns out that this so-called
``weak measurement'' suffices for realizing most quantum algorithms,
in particular those whose ultimate answer is deterministic. Shor's
factoring and Grover's search algorithm can be modified to satisfy
this property.  The final and most severe difficulty was that, even
though in equilibrium there is a tendency for the spins to align with
the magnetic field, the energy associated with this tendency is very
small compared to room temperature. Therefore, the equilibrium states
of the molecules' nuclear spins are nearly random, with only a small
fraction pointing in the right direction. This difficulty was overcome
by methods for singling out the small
fraction of the observable signal that represents the desired initial
state. These methods were anticipated in 1977~\cite{stoll:qc1977a}.

Soon after these difficulties were shown to be overcome or
circumventable, two groups were able to experimentally implement short
quantum algorithms using NMR with small
molecules~\cite{chuang:qc1998a,jones:qc1998b}.  At present it is
considered unlikely that liquid-state NMR algorithms will solve
problems not easily solvable with available classical computing
resources. Nevertheless, experiments in liquid-state NMR QIP are
remarkable for demonstrating that one can control the unitary
evolution of physical qubits sufficiently well to implement simple
QIP tasks.  The control methods borrowed
from NMR and developed for the more complex experiments in NMR QIP are
applicable to other device technologies, enabling better control in
general.

\section{Principles of Liquid-State NMR QIP}
\label{sec:princs}

In order to physically realize quantum information, it is necessary to
find ways of representing, manipulating, and coupling qubits so as to
implement non-trivial quantum gates, prepare a useful initial state
and read out the answer. The next sections show how to accomplish
these tasks in liquid-state NMR.

\subsection {Realizing Qubits}

The first step for implementing QIP is to have a physical system that
can carry quantum information. The preferred system for realizing
qubits in liquid-state NMR consists of spin-${1\over 2}$ nuclei, which
are naturally equivalent to qubits. The nuclear-spin degree of freedom
of a spin-${1\over 2}$ nucleus defines a quantum mechanical two-state
system.  Once the direction along the strong external magnetic field
is fixed, its state space consists of the superpositions of ``up'' and
``down'' states. That is, we can imagine that the nucleus behaves
somewhat like a small magnet, with a definite axis, which can point
either ``up'' (logical state $\ket{\bitzero}$) or ``down'' (logical
state $\ket{\bitone}$). By the superposition principle, every quantum
state of the form
$\ket{\psi_0}=\alpha\ket{\bitzero}+\beta\ket{\bitone}$ with
$|\alpha|^2+|\beta|^2=1$ is a possible (pure) state for the nuclear
spin.  In the external magnetic field, the two logical states have
different energies. The energy difference results in a time evolution
of $\ket{\psi_0}$ given by
\begin{eqnarray}
\ket{\psi_t}
 &=&  e^{-i\omega t/2}\alpha\ket{\bitzero}+e^{i\omega t/2}\beta\ket{\bitone}.
\end{eqnarray}
The constant $\omega$ is the precession frequency of the nuclear spin
in the external magnetic field in units of radians per second if $t$
is in seconds.  The frequency is proportional to the energy difference
$\epsilon$ between the ``up'' and ``down'' states: $\omega =
2\pi\epsilon/h$, where $h$ is Planck's constant.

Although a spin-$1\over 2$ nucleus' state space is the same as that of
a qubit, the precession implies that the state is not constant.
We would like the realization of a qubit to retain its state
over time when we are not intentionally modifying it. For this reason,
in the next section, the qubit state realized by the nuclear spin
will be defined so as to compensate for the precession.

Precession frequencies for nuclear spins can vary substantially
depending on the nuclei's magnetic moments.  For example, at
$11.7\uts{T}$, the precession frequency for protons is $500\uts{Mhz}$
and for ${}^{13}$C it is $125\uts{Mhz}$. These frequency differences
are exploited in measurement and control to distinguish between the
types of nuclei.  The effective magnetic field seen by nuclear spins
also depends on their chemical environment. This dependence causes small
variations in the spins' precession frequencies that can be used to
distinguish, for example, the two ${}^{13}$C nuclei in TCE: The
frequency difference (called the ``chemical shift'') is
$600$--$900\uts{Hz}$ at $11.7\uts{T}$, depending on the solvent, the
temperature and the TCE concentration.

Using the Pauli matrix $\sigma_z=\qaop{1}{0}{0}{-1}$, the time
evolution can be expressed as $\ket{\psi_t}=e^{i\omega\sigma_z
t/2}\ket{\psi_0}$. The operator $\omega\sigma_z/2$ is the internal
Hamiltonian (that is, the energy observable, in units for which
$h/(2\pi)=1$) of the nuclear spin.  The direction of the external
magnetic field determines the $z$-axis.  Given a choice of axes, the
idea that a single nuclear spin-${1\over 2}$ has a spin direction (as
would be expected for a tiny magnet) can be made explicit by means of
the Bloch sphere representation of a nuclear spin's state
(Fig.~\ref{figbloch}).  The Pauli matrix $\sigma_z$ can be thought of
as the observable that measures the nuclear spin along the $z$-axis.
Observables for spin along the $x$- and $y$-axis are given by the
other two Pauli matrices $\sigma_x=\qaop{0}{1}{1}{0}$ and
$\sigma_y=\qaop{0}{-i}{i}{0}$.  Given a state
$\ket{\psi}=\alpha\ket{\bitzero}+\beta\ket{\bitone}$ of the nuclear
spin, one can form the density matrix $\ketbra{\psi}{\psi}$ and
express it in the form
\begin{equation}
\ketbra{\psi}{\psi}=
{1\over 2}(\idop+\alpha_x\sigma_x+\alpha_y\sigma_y+\alpha_z\sigma_z).
\end{equation}
The vector $\vec v = (\alpha_x,\alpha_y,\alpha_z)$ then is a point on
the unit sphere in three-dimensional space.  Conversely, every point
on the unit sphere corresponds to a pure state of the nuclear spin.
The representation also works for ``mixed'' states, which correspond
to points in the interior of the sphere.  As a representation of spin
states, the unit sphere is called the ``Bloch sphere''. Because
quantum evolutions of a spin correspond to rotations of the Bloch
sphere, this sphere is a useful tool for thinking about one- and
sometimes about two-qubit processes.

\begin{herefig}
\begin{picture}(7,3.5)(-3.5,-3.5)
\nputgr{0,0}{t}{height=3.5in}{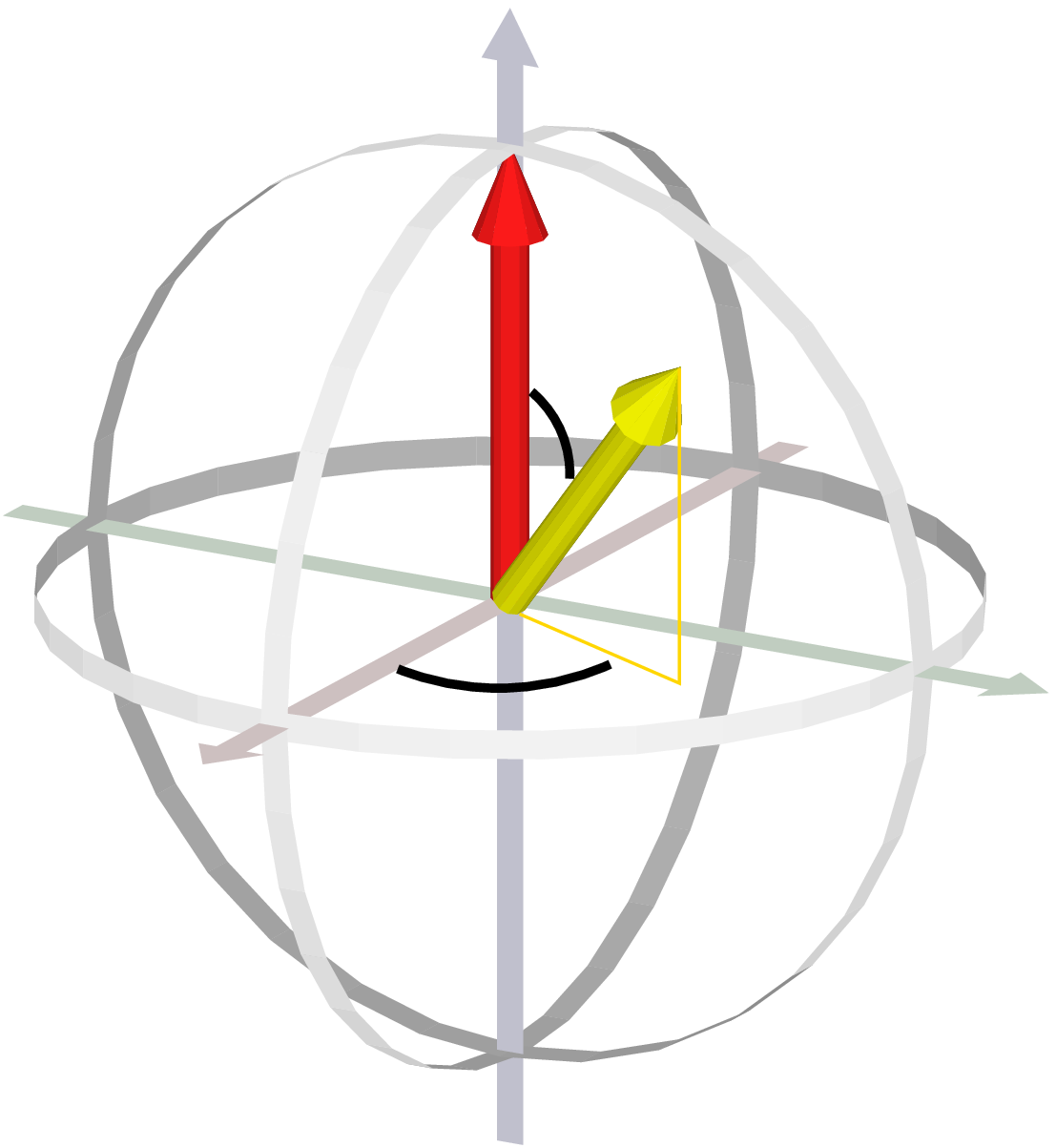}
\Large
\nputbox{.1,-.1}{c}{$z$}
\nputbox{1.7,-2.1}{c}{$y$}
\nputbox{-.95,-2.45}{c}{$x$}
\nputbox{.2,-1.1}{c}{$\theta$}
\nputbox{.1,-2.2}{c}{$\phi$}
\end{picture}
\label{figbloch}
\herefigcap{Bloch sphere representation of a qubit state.
The yellow arrow represents a pure state $\ket{\psi}$ for the qubit or
nuclear spin-${1\over 2}$. The Euler angles are indicated and
determine the state according to the formula
$\ket{\psi}=\cos(\theta/2)\ket{\bitzero} +e^{i\phi}\sin(\theta/2)
\ket{\bitone}$. The red arrow along the $z$-axis indicates the
orientation of the magnetic field and the vector for $\ket{\bitzero}$.
If we write the state as a density matrix $\rho$ and expand it in
terms of Pauli matrices, 
\begin{eqnarray}
\rho=\ketbra{\psi}{\psi}&=&(\idop+x\sigma_x+y\sigma_y+z\sigma_z)/2 
\nonumber\\ &=&
{1\over
2}\left(\one +
\sin(\theta)\cos(\phi) \sigma_{x} + \sin(\theta)\sin(\phi) \sigma_{y}
+ \cos(\theta)\sigma_{z}\right),
\end{eqnarray}
then the coefficients
$(x,y,z)=\left(\sin(\theta)\cos(\phi),\sin(\theta)\sin(\phi),\cos(\theta)\right)$
of the Pauli matrices form the vector for the state.  For a pure state
this vector is on the surface of the unit sphere, and for a mixed
state, it is inside the unit sphere.  The Pauli matrices are
associated with spin observables in the laboratory frame, so that all
axes of the representation are meaningful with respect to real space.
}
\end{herefig}

\subsection{One Qubit Gates}

The second step for realizing QIP is to give a means for controlling
the qubits so that quantum algorithms can be implemented.  The qubits
are controlled with carefully modulated external fields to realize
specific unitary evolutions called ``gates''.  Each such evolution can
be described by a unitary operator applied to one or more qubits. The
simplest method for demonstrating that sufficient control is available
is to show how to realize a set of one- and two-qubit gates that is
``universal'' in the sense that in principle, every unitary operator
can be implemented as a composition of
gates~\cite{barenco:qc1995a,divincenzo:qc1995a,lloyd:qc1995b}.

One-qubit gates can be thought of as rotations of the Bloch sphere and
can be implemented in NMR with electromagnetic pulses.  In
general, the effect of a magnetic field on a nuclear spin is to cause
a rotation around the direction of the field. In terms of the quantum
state of the spin, the effect is described by an internal Hamiltonian
of the form
$H=(\omega_x\sigma_x+\omega_y\sigma_y+\omega_z\sigma_z)/2$. The
coefficients of the Pauli matrices depend on the magnetic field
according to $\vec{\omega} = (\omega_x,\omega_y,\omega_z)= -\mu
\mathbf{B}$, where $\mu$ is the nuclear magnetic moment and
$\mathbf{B}$ is the magnetic field vector.  In terms of the
Hamiltonian, the evolution of the spin's quantum state in the presence
of the magnetic field $\mathbf{B}$ is therefore given by $\ket{\psi_t}
= e^{-iHt}\ket{\psi_0}$, so that the spin direction in the Bloch
sphere rotates around $\vec{\omega}$ with angular frequency
$\omega=|\vec{\omega}|$.

In the case of liquid-state NMR, there is an external, strong magnetic
field along the $z$-axis and the applied electromagnetic pulses add to
this field. One can think of these pulses as contributing a relatively
weak magnetic field (typically less than $.001$ of the external field)
whose orientation is in the $xy$-plane. One use of such a pulse is to
tip the nuclear spin from the $z$-axis to the $xy$-plane.  To see how
that can be done, assume that the spin starts in the state
$\ket{\bitzero}$, which points up along the $z$-axis in the Bloch
sphere representation. Because this state is aligned with the external
field, it does not precess.  To tip the spin, one can start by
applying a pulse field along the $x$-axis. Because the pulse field is
weak compared to the external field, the net field is still almost
along the $z$-axis. The spin now rotates around the net field. Because
it started along $z$, it moves only in a small circle near the
$z$-axis.  To force the spin to tip further, one changes the
orientation of the pulse field at the same frequency as the precession
caused by the external field.  This is called a ``resonant'' pulse.
Because typical precession frequencies are hundreds of $\uts{Mhz}$, such
a pulse consists of radio-frequency (RF) electromagnetic fields.

To better understand how resonant pulses work, it is convenient to use
the ``rotating frame''. In this frame, we imagine that our apparatus
rotates at the precession frequency of the nuclear spin. In this way,
the effect of the external field is removed. In particular, in the
rotating frame the nuclear spin does not precess, and a resonant
pulse's magnetic field looks like a constant magnetic field applied,
for example, along the $(-x)$-axis of the rotating frame.  The nuclear
spin responds to the pulse by rotating around the $x$-axis as
expected: If the spin starts along the $z$-axis, it tips toward the
$(-y)$-axis, then goes to tthe $(-z)$-, the $y$-, and finally back to
the $z$-axis, all in the rotating frame. See Fig.~\ref{figblochrot}.

\begin{herefig}
\begin{picture}(7,3.9)(-3.5,-3.75)
\nputgr{0,0}{t}{height=3.5in}{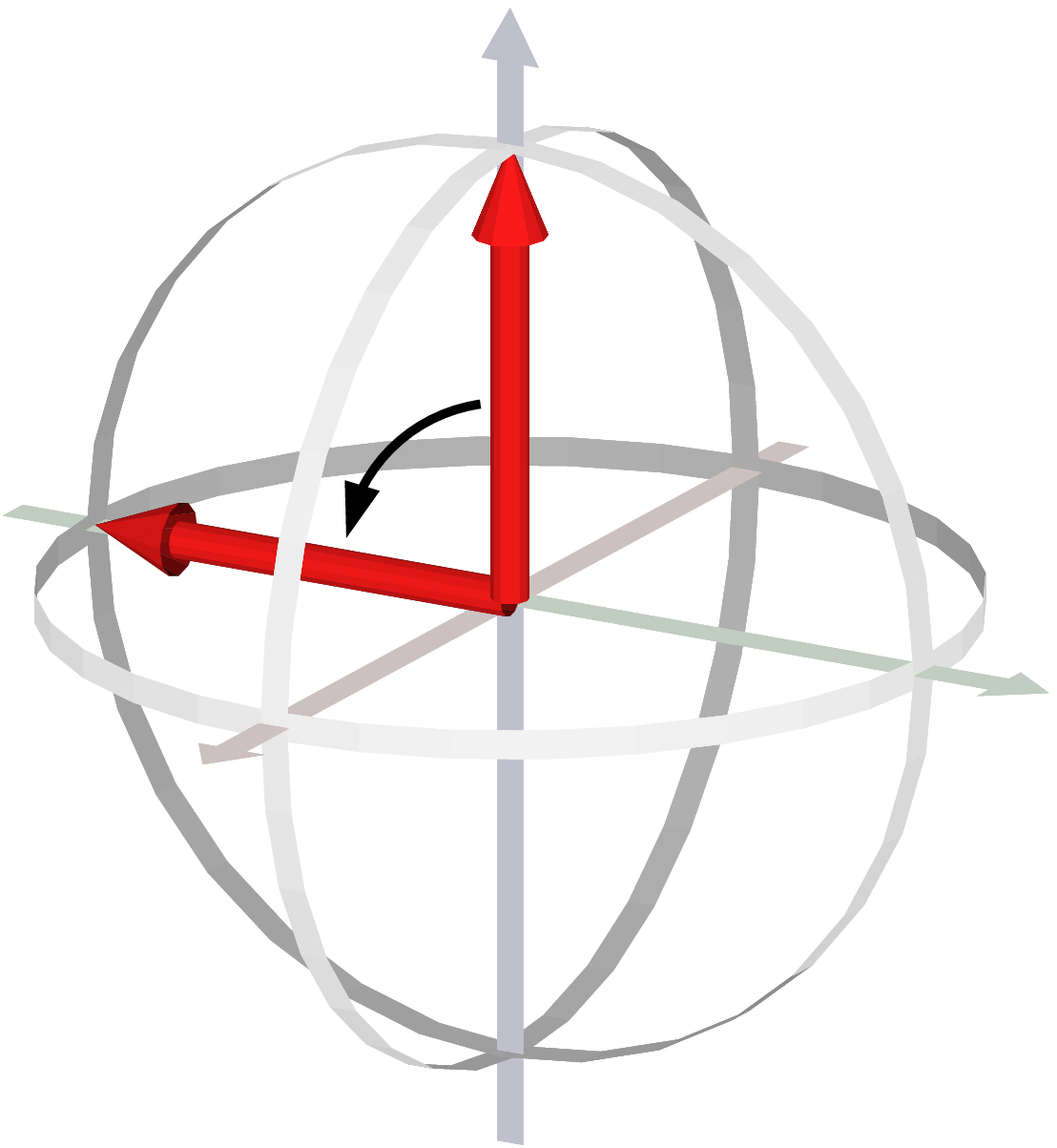}
\Large
\nputbox{.1,-.1}{c}{$z$}
\nputbox{1.7,-2.1}{c}{$y$}
\nputbox{-.95,-2.45}{c}{$x$}
\end{picture}
\herefigcap{Single bit rotation around the $x$-axis in the rotating frame.
An applied magnetic field along the rotating frame's $(-x)$-axis due to a
resonant RF pulse moves the nuclear spin direction from the $z$-axis
toward the $(-y)$-axis. The initial and final states for the nuclear
spin are shown for a $90^\circ$ rotation.  If the strength of the
applied magnetic field is such that the spin evolves according to
the Hamiltonian $\omega_x\sigma_x/2$, then it has to be turned on for
a time $t=\pi/(2\omega_x)$ to cause the rotation shown.}
\label{figblochrot}
\end{herefig}

The rotating frame makes it possible to define the state of the qubit
realized by a nuclear spin as the state with respect to this
frame. As a result, the qubit's state does not change unless RF
pulses are applied.  In the context of the qubit realized by a nuclear
spin, the rotating frame is called the ``logical frame''.  In the
following, references to the Bloch sphere axes and associated
observables are understood to be with respect to an appropriate,
usually rotating, frame. Different frames can be chosen for each
nuclear spin of interest, so we often use multiple independently
rotating frames and refer each spin's state to the appropriate frame.

Use of the rotating frame together with RF pulses makes it possible to
implement all one-qubit gates on a qubit realized by a spin-${1\over
2}$ nucleus.  To apply a rotation around the $x$-axis, a resonant RF
pulse with effective field along the rotating frame's $(-x)$-axis is
applied.  This is called an ``$x$-pulse'', and $x$ is the ``axis''
of the pulse. While the RF pulse is on,
the qubit's state evolves as $e^{-i\omega_x\sigma_x t/2}$.  The
strength (or ``power'') of the pulse is characterized by $\omega_x$,
the ``nutation'' frequency. To implement a rotation by an angle of
$\phi$, the pulse is turned on for a period $t=\phi/\omega_x$.
Rotations around any axis in the plane can be implemented similarly.
The angle of the pulse field with respect to the $(-x)$-axis is called
the ``phase'' of the pulse.  It is a fact that all rotations of the
Bloch sphere can be decomposed into rotations around axes in the
plane. For rotations around the $z$-axis, an easier technique is
possible. The current absolute phase $\theta$ of the rotating frame's
$x$-axis is given by $\theta_0+\omega t$, where $\omega$ is the
precession frequency of the nuclear spin.  Changing the angle
$\theta_0$ by $-\phi$ is equivalent to rotating the qubit's state by
$\phi$ around the $z$-axis. In this sense, $z$-pulses can be
implemented exactly.  In practice, this change of the rotating frame's
phase means that the absolute phases of future pulses must be shifted
accordingly.  This implementation of rotations around the $z$-axis is
possible because phase control in modern equipment is extremely
reliable so that errors in the phase of applied pulses are negligible
compared to other sources of errors.

So far, we have considered just one nuclear spin in a molecule. But
the RF fields are experienced by the other nuclear spins as well.
This side-effect is a problem if only one ``target'' nuclear spin's
state is to be rotated. There are two cases to consider depending on
the precession frequencies of the other, ``non-target'' spins.  Spins
of nuclei of different isotopes, such as those of other species of
atoms, usually have precession frequencies that differ from the
target's by many $\uts{Mhz}$ at $11.7\uts{T}$. A pulse resonant for
the target has little effect on such spins. This is because in the
rotating frames of the non-target spins, the pulse's magnetic field is
not constant but rotates rapidly. The power of a typical pulse is such
that the effect during one rotation of the pulse's field direction is
insignificant and averages to zero over many rotations. This is not
the case for non-target spins of the same isotope. Although the
variations in their chemical environments result in frequency
differences, these differences are much smaller, often only a few
$\uts{kHz}$. The period of a $1\uts{kHz}$ rotation is $1\uts{ms}$,
whereas so-called ``hard'' RF pulses require only $10$'s of
$\mu\uts{s}$ ($.001\uts{ms}$) to complete the typical $90^\circ$ or
$180^\circ$ rotations.  Consequently, in the rotating frame of a
non-target spin with a small frequency difference, a hard RF pulse's
magnetic field is nearly constant for the duration of the pulse.  As a
result, such a spin experiences a rotation similar to the one intended
for the target. To rotate a specific nuclear spin or spins within a
narrow range of precession frequencies, one can use weaker,
longer-lasting ``soft'' pulses instead.  This approach leads to the
following strategies for applying pulses: To rotate all the nuclear
spins of a given species (such as the two ${}^{13}$C of TCE) by a
desired angle, apply a hard RF pulse for as short a time as possible.
To rotate just one spin having a distinct precession frequency, apply
a soft RF pulse of sufficient duration to have little effect on other
spins. The power of soft pulses is usually modulated in time
(``shaped'') to reduce the time needed for a rotation while minimizing
``crosstalk'', a term that describes unintended effects on other
nuclear spins.

\subsection{Two Qubit Gates}

Two nuclear spins in a molecule interact with each other, as one would
expect of two magnets. But the details of the spins' interaction are
more complicated because they are mediated by the electrons. In liquid
state, the interaction is also modulated by the rapid motions of the
molecule.  The resulting effective interaction is called the
$J$-coupling.  When the difference of the precession frequencies
between the coupled nuclear spins is large compared to the strength of
the coupling, it is a good approximation to write the coupling
Hamiltonian as a product of the $z$-Pauli operators for each spin:
$H_J=C\slb{\sigma_{z}}{1}\slb{\sigma_{z}}{2}$. This is the ``weak
coupling'' regime. With this Hamiltonian, an initial state
$\ket{\psi_0}$ of two nuclear-spin qubits evolves as
$\ket{\psi_t}=e^{-iC\slb{\sigma_{z}}{1}\slb{\sigma_{z}}{2}t}\ket{\psi_0}$,
where a different rotating frame is used for each nuclear spin to
eliminate the spin's internal evolution. (The use of rotating frames
is compatible with the coupling Hamiltonian because the Hamlitonian is
invariant under frame rotations.) Because the Hamiltonian is
diagonal in the logical basis, the effect of the coupling can be
understood as an increase of the (signed) precession frequency of the
second spin if the first one is up and a decrease if the first one is
down (Fig.~\ref{figtwobitgate}). The changes in precession frequency
for adjacent nuclear spins in organic molecules are typically in the
range of $20$--$200\uts{Hz}$.  They are normally much smaller for
non-adjacent nuclear spins.  The strength of the coupling is called
the ``coupling constant'' and is given as the change in the precession
frequency.  In terms of the constant $C$ used above, the coupling
constant is given by $J=2C/\pi$ in $\uts{Hz}$.  For example, the
coupling constants in TCE are close to $100\uts{Hz}$ between the two
carbons, $200\uts{Hz}$ between the proton and the adjacent carbon, and
$9\uts{Hz}$ between the proton and the far carbon.

\begin{herefig}
\begin{picture}(7,3.75)(-3.3,-3.75)
\nputgr{-2,0}{t}{height=3.5in}{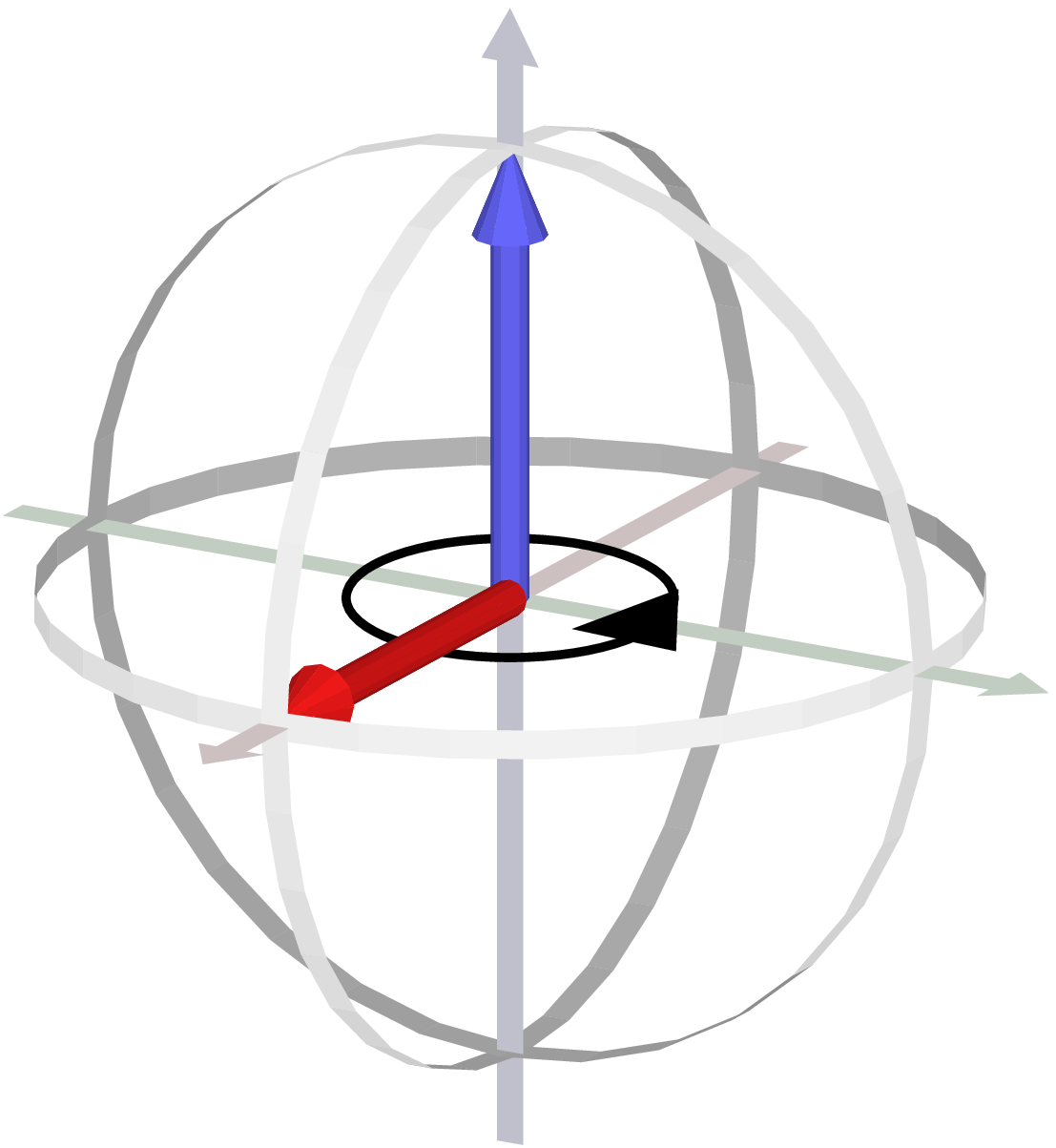}
\nputgr{+2,0}{t}{height=3.5in}{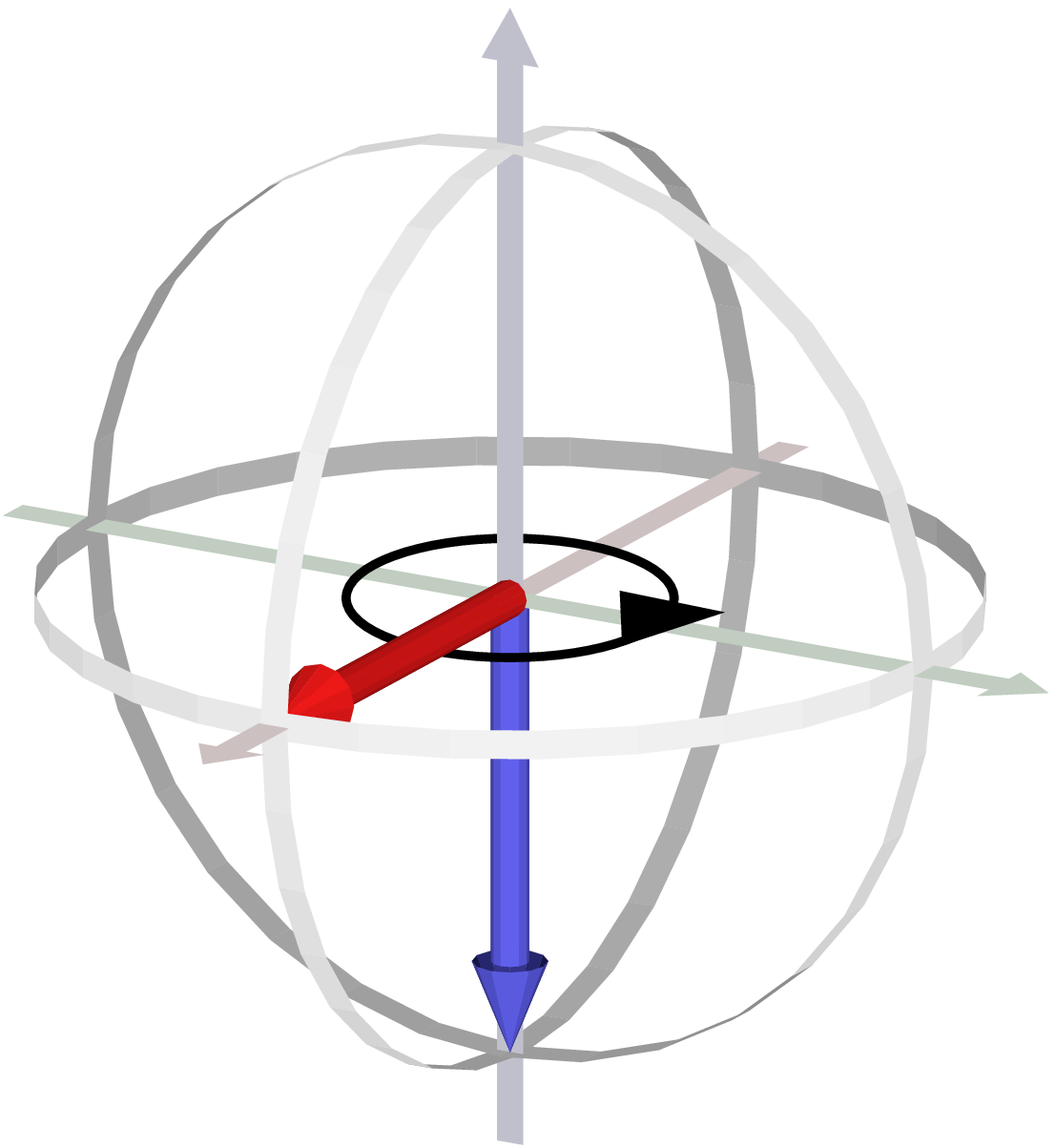}
\Large
\nputbox{-1.9,-.1}{c}{$z$}
\nputbox{-.3,-2.1}{c}{$y$}
\nputbox{-2.95,-2.45}{c}{$x$}
\nputbox{-1.8,-.65}{c}{\huge$\sysfnt{1}$}
\nputbox{-2.3,-2.2}{c}{\huge$\sysfnt{2}$}
\nputbox{2.1,-.1}{c}{$z$}
\nputbox{3.7,-2.1}{c}{$y$}
\nputbox{1.05,-2.45}{c}{$x$}
\nputbox{2.2,-2.8}{c}{\huge$\sysfnt{1}$}
\nputbox{1.7,-2.2}{c}{\huge$\sysfnt{2}$}
\end{picture}
\herefigcap{Effect of the $J$-coupling. 
In the weak-coupling regime with a positive coupling constant, the
coupling between two spins can be interpreted as an increase in
precession frequency of the spin $\sysfnt{2}$ when the spin
$\sysfnt{1}$ is ``up'' and a decrease when spin $\sysfnt{1}$ is
``down''.  The two diagrams depict the situation in which spin
$\sysfnt{2}$ is in the plane. The diagram on the left has spin
$\sysfnt{1}$ pointing up along the $z$ axis. In the rotating frame of
spin $\sysfnt{2}$, it precesses from the $x$-axis to the $y$-axis. The
diagram on the right has spin $\sysfnt{1}$ pointing down, causing a
precession in the opposite direction of spin $\sysfnt{2}$. Note that
neither the coupling nor the external field change the orientation of
a spin pointing up or down along the $z$-axis.  }
\label{figtwobitgate}
\end{herefig}

The $J$-coupling and the one-qubit pulses suffice for realizing the
controlled-not operation usually taken as one of the fundamental gates
of QIP.  A pulse sequence for implementing the controlled-not in terms
of the $J$-coupling constitutes the first quantum algorithm of
Sect.~\ref{sec:examples}.  A problem with the $J$-coupling in
liquid-state NMR is that it cannot be turned off when it is not needed
for implementing a gate.

\subsection{Turning off the $J$-Coupling}

The coupling between the nuclear spins in a molecule cannot be
physically turned off. But for QIP, we need to be able to maintain a
state in memory and to couple qubits selectively.  Fortunately, NMR
spectroscopists solved this problem well before the development of
modern quantum information concepts.  The idea is to use the control
of single spins to cancel the interaction's effect over a given
period.  This technique is called refocusing and requires applying a
$180^\circ$ pulse to one of two coupled spins at the midpoint of the
desired period. To understand how refocusing works, consider again the
visualization of Fig.~\ref{figtwobitgate}.  A general state is in a
superposition of the four logical states of the two spins. By
linearity, it suffices to consider the evolution with spin
$\sysfnt{1}$ being in one of its two logical states, up or down, along
the $z$-axis.  Suppose we wish to remove the effects of the coupling
over a period of $2\uts{ms}$.  To do so, wait $1\uts{ms}$. In a
sequence of pulses, this waiting period is called a $1\uts{ms}$
``delay''. The effect on spin $\sysfnt{2}$ in its rotating frame is to
precess counterclockwise if spin $\sysfnt{1}$ is up, and clockwise
for the same angle if spin $\sysfnt{1}$ is down.  Now, apply a pulse
that rotates spin $\sysfnt{1}$ by $180^\circ$ around the $x$-axis.
This is called an ``inversion'', or in the current context, a
``refocusing'' pulse.  It exchanges the up and down states.  For the
next $1\uts{ms}$, the effect of the coupling on spin $\sysfnt{2}$ is
to undo the earlier rotation. At the end of the second $1\uts{ms}$
delay, one can apply another $180^\circ$ pulse to reverse the
inversion and recover the initial state. The pulse sequence is
depicted in Fig.~\ref{fig:refocussing}.

\pagebreak

\begin{herefig}
\label{fig:refocussing}
\begin{picture}(7,3)(-3.5,-3)
\nputgr{0,0}{t}{width=7in}{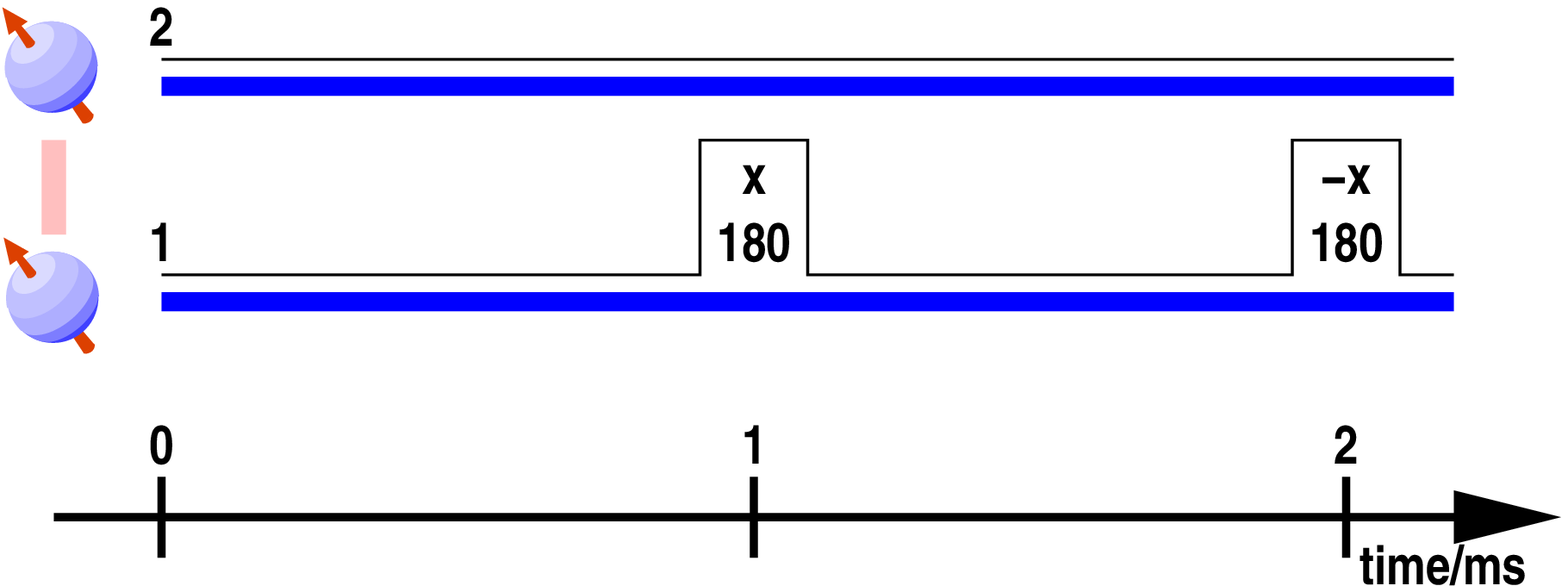}
\end{picture}
\herefigcap{ Pulse sequence for refocusing the coupling.  The sequence
of events is shown with time running from left to right.  The two
spins' lifelines are shown in blue, and the RF power targeted at each
spin is indicated by the black line above. Pulses are applied to spin
$\sysfnt{1}$ only, as indicated by the rectangular rises in RF power
at $1\uts{ms}$ and $2\uts{ms}$. The axis for each pulse is given with the
pulse. The angle is determined by the area under the pulse and is also
given explicitly. Ideally for pulses of this type, the pulse times
(the widths of the rectangles) should be zero. In practice, for hard pulses,
they can be as small as $\approx .01\uts{ms}$.  Any
$\slb{\sigma_z}{1}\slb{\sigma_z}{2}$ coupling's effect is refocused by
the sequence shown, so that the final state of the two spins is the
same as the initial state. The axis for the pair of refocusing
pulses can be changed to any other axis in the plane.}
\end{herefig}

Turning off couplings between more than two nuclear spins can be quite
complicated unless one takes advantage of the fact that non-adjacent
nuclear spins tend to be relatively weakly coupled.  Methods that scale
polynomially with the number of nuclear spins and that can be used to
selectively couple pairs of nuclear spins can be found
in~\cite{leung:qc1999a,jones:qc1999a}.  These techniques can be used
in other physical systems where couplings exist that are difficult to
turn off directly.  An example is qubits represented by the state of
one or more electrons in tightly packed quantum dots.

\subsection{Measurement}

To determine the ``answer'' of a quantum computation it is necessary
to make a measurement. As noted earlier, the technology for making a
projective measurement of individual nuclear spins does not yet
exist. In liquid-state NMR, instead of using just one molecule to
define a single quantum register, we use a large ensemble of molecules
in a test tube.  Ideally, their nuclear spins are all placed in the
same initial state, and the subsequent RF pulses affect each molecule
in the same way. As a result, weak magnetic signals from (say) the
proton spins in TCE add to form a detectable magnetic field called the
``bulk magnetization''.  The signal that is measured in high-field NMR
is the magnetization in the $xy$-plane, which can be picked up by
coils whose axes are placed transversely to the external field.
Because the interaction of any given nuclear spin with the coil is
very weak, the effect of the coil on the quantum state of the spins is
negligible in most NMR experiments.  As a result, it is a good
approximation to think of the generated magnetic fields and their
detection classically. In this approximation, each nuclear spin
behaves like a tiny bar magnet and contributes to the bulk
magnetization.  As the nuclear spins precess, so does the
magnetization. As a result, an oscillating current is induced in the
coil, provided it is electronically configured to be ``tuned'' to the
precession frequency.  By observing the amplitude and phase of this
current over time, we can keep track of the absolute magnetization in
the plane and its phase with respect to the rotating frame. This
process yields information about the qubit states represented by the
state of the nuclear spins.

To see how one can use the bulk magnetization to learn about the qubit
states, consider the TCE molecule with three spin-${1\over 2}$ nuclei
used for information processing. The bulk magnetizations generated by
the protons and the carbons precess at $500\uts{Mhz}$ and
$125\uts{Mhz}$, respectively. The proton and carbon contributions to
the magnetization are detected separately with two coils tuned to
$500\uts{Mhz}$ (proton magnetization) and $125\uts{Mhz}$ (carbon
magnetization).  For simplicity, we restrict our
attention to the two carbons and assume that the protons are not
interacting with the carbons. (It is possible to actively remove such
interactions by using a technique called ``decoupling''.)

At the end of a computation, the qubit state of the two nuclear spins
is given by a density matrix $\rho_q$. We can assume that this state
is the same for each molecule of TCE in the sample.  As we mentioned
earlier, the density matrix is relative to logical frames for each
nuclear spin.  The current phases for the two logical frames with
respect to a rotating reference frame at the precession frequency of
the first carbon are known. If we learn something about the state in
the reference frame, that information can be converted to the desired
logical frame by a rotation around the $z$-axis.  Let $\rho(0)$ be the
state of the two nuclear spins in the reference frame. In this frame,
the state evolves in time as $\rho(t)$ according to a Hamiltonian $H$
that consists of a chemical shift term for the difference in the
precession frequency of the second carbon and of a coupling term. To
a good approximation,
\begin{equation}
H = \pi900\uts{Hz}\slb{\sigma_z}{2} + 
      \pi50\uts{Hz}\slb{\sigma_z}{1}\slb{\sigma_z}{2}.
\end{equation}
The magnetization detected in the reference $x$-direction at time $t$ is given
by
\begin{equation}
M_x(t)= m\;\trace\left(\rho(t)(\slb{\sigma_x}{1}+\slb{\sigma_x}{2})\right),
\label{eq:xmag}
\end{equation}
where $\trace (\sigma)$ denotes the trace, that is, the sum of the
diagonal elements of the matrix $\sigma$.  Eq.~\ref{eq:xmag} links the
magnetization to the Bloch sphere representation.  The constant of
proportionality $m$ depends on the size of the ensemble and the
magnetic moments of the nuclei. From the point of view of NMR, $m$
determines a scale whose absolute size is not relevant. What matters
is how strong this signal is compared to the noise in the system.  For
the purpose of the following discussion, we set $m=1$.

We can also detect the magnetization $M_y(t)$ in the $y$-direction and
combine it with $M_x(t)$ to form a complex number representing
the planar magnetization.
\begin{eqnarray}
M(t)&=&M_x(t)+iM_y(t)\\
    &=&\trace\left(\rho(t)(\slb{\sigma_+}{1}+\slb{\sigma_+}{2})\right),
\end{eqnarray}
where we defined $\sigma_+=\sigma_x+i\sigma_y = \qaop{0}{2}{0}{0}$.
What can we infer about $\rho(0)$ from observing $M(t)$ over time?
For the moment, we neglect the coupling Hamiltonian. Under
the chemical shift Hamiltonian 
$H_{CS}=\pi900\uts{Hz}\slb{\sigma_z}{2}$, $M(t)$ evolves as
\begin{equation}
\begin{array}[b]{rcll}
M(t) &=& \trace\left( e^{-iH_{CS}t}\rho(0) e^{i H_{CS}\;t}(\slb{\sigma_+}{1}+\slb{\sigma_+}{2})\right)\\
     &=& \trace\left( \rho(0) e^{i H_{CS}\;t}(\slb{\sigma_+}{1}+\slb{\sigma_+}{2})e^{-iH_{CS}\;t}\right) &\mbox{using $\trace(AB)=\trace(BA)$,}\\
     &=& \trace\left( \rho(0) (\slb{\sigma_+}{1}+e^{iH_{CS}\;t}\slb{\sigma_+}{2}e^{-iH_{CS}\;t})\right)&\mbox{because $H_{CS}$ acts only on spin $\sysfnt{2}$,}\\
     &=& \trace\left( \rho(0) (\slb{\sigma_+}{1}+e^{i2\pi900\uts{Hz}\; t}\slb{\sigma_+}{2})\right)&\mbox{by multiplying the matrices,}\\
     &=& \trace\left( \rho(0) \slb{\sigma_+}{1}\right)+
         \trace\left( \rho(0) e^{i2\pi900\uts{Hz}\;t}\slb{\sigma_+}{2}\right) &\mbox{because the trace is linear.}
\end{array}
\label{eq:obsrewrite}
\end{equation}
Thus the signal is a combination of a constant signal given by
the first spin's contribution to the magnetization in the plane, and a
signal oscillating with a frequency of $900\uts{Hz}$ with amplitude
given by the second spin's contribution to the planar magnetization.
The two contributions can be separated by Fourier transforming $M(t)$,
which results in two distinct peaks, one at $0\uts{Hz}$ and a second
at $900\uts{Hz}$. See Fig.~\ref{fig:c13peaks}.

\pagebreak

\begin{herefig}
\begin{picture}(7,5)(-3.5,-5)
\nputgr{0,-2.35}{br}{width=3.5in}{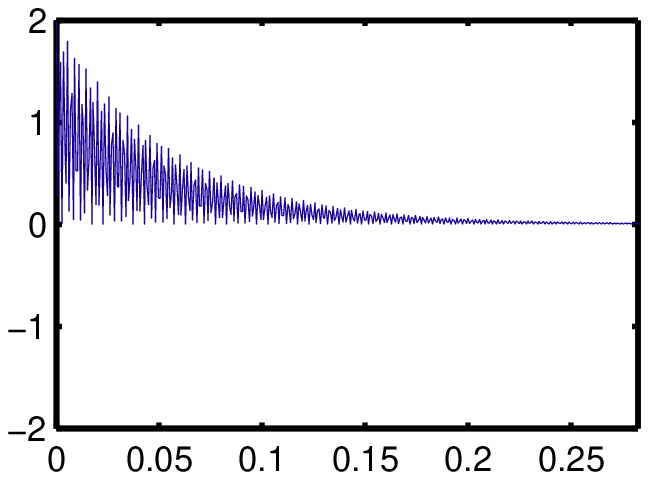}
\nputgr{0,-2.35}{bl}{width=3.5in}{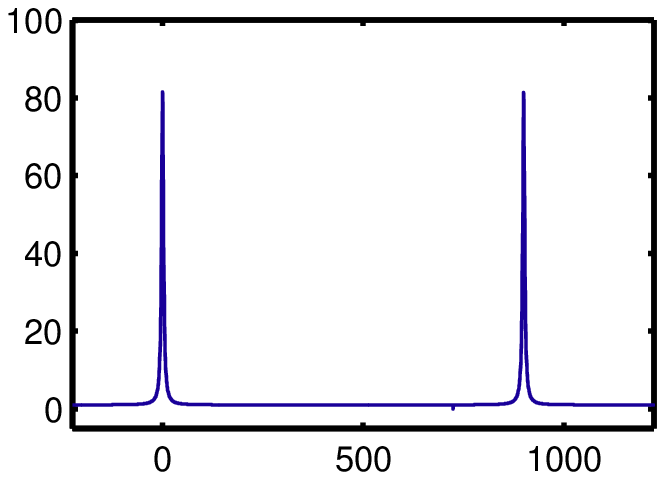}
\nputgr{0,-2.5}{tr}{width=3.5in}{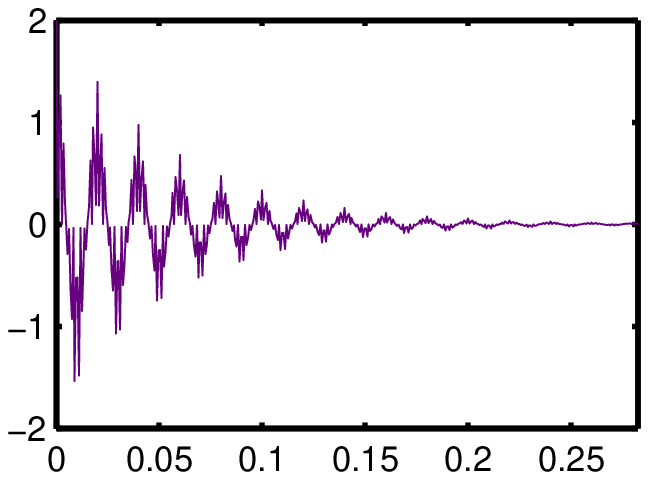}
\nputgr{0,-2.5}{tl}{width=3.5in}{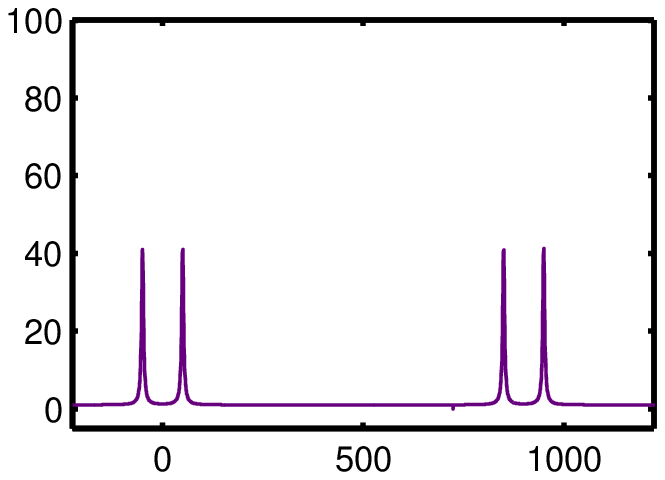}
\nputbox{-.6,-.55}{br}{\large (a)}
\nputbox{2.9,-.55}{br}{\large (b)}
\nputbox{-.6,-3.05}{br}{\large (c)}
\nputbox{2.9,-3.05}{br}{\large (d)}
\end{picture}
\label{fig:c13peaks}
\herefigcap{Simulated magnetization signals (left) and spectra (right).
(a) The $x$-magnetization signal is shown as a function of time for a
pair of uncoupled spins with a relative chemical shift of
$900\uts{Hz}$. The initial spin directions are along the $x$-axis.
The signal (called the ``free induction decay'') decays with a
halftime of $0.0385s$ because of simulated relaxation
processes. Typically, the halftimes are much longer. A short one was
chosen to broaden the peaks for visual effect. (b) This shows the
spectrum for the signal in (a), that is, the Fourier transform of the
combined $x$- and $y$-magnetization.  The spectrum has peaks at
frequencies of $0\uts{Hz}$ (spin $\sysfnt{1}$'s peak) and
$900\uts{Hz}$ (spin $\sysfnt{2}$'s peak) because of the independently
precessing pair of spins.  (c) This is the $x$-magnetization signal
when the two spins are coupled as described in the text.  (d) This
shows the spectrum for the signal in (c) obtained from the combined
$x$- and $y$-magnetization.  Each spin's peak from the
previous spectrum ``splits'' into two. The left and right peaks of
each pair are associated with the other spin being in the state
$\ket{\bitone}$ and $\ket{\bitzero}$, respectively.  The vertical axis
units are relative intensity with the same constant of proportionality
for the two spectra.}
\end{herefig}

\pagebreak

To see how the coupling affects the observed magnetization, we rewrite the
expression for $M(t)$ to take advantage of the fact that the up/down states
are invariant under the full Hamiltonian.
\begin{eqnarray}
M(t) &=& \trace \left(\rho(t)\slb{\sigma_+}{1}\right) + \trace\left(\rho(t)\slb{\sigma_+}{2}\right) \nonumber\\
 &=& \trace \left(\rho(t)\slb{\sigma_+}{1}\slb{\one}{2}\right) + \trace\left(\rho(t)\slb{\one}{1}\slb{\sigma_+}{2}\right) \nonumber\\
     &=& \trace \left(\rho(t)\slb{\sigma_+}{1}(\slb{e_\uparrow}{2} + 
                                          \slb{e_\downarrow}{2})\right) + 
         \trace \left(\rho(t)(\slb{e_\uparrow}{1} + \slb{e_\downarrow}{1})
                        \slb{\sigma_+}{2} \right)
\label{eq:fullmag}
\end{eqnarray}
where $e_\uparrow=\qaop{1}{0}{0}{0}$ and $e_\downarrow=\qaop{0}{0}{0}{1}$.
Using a similar calculation to the one leading to Eq.~\ref{eq:obsrewrite},
the first term can be written as
\begin{eqnarray}
M_1(t) &=& \trace\left( e^{-iH\;t}\rho(0)e^{iHt}
           \slb{\sigma_+}{1}(\slb{e_\uparrow}{2} +
           \slb{e_\downarrow}{2})\right)\\ &=&
           e^{i2\pi50\uts{Hz}\;t}\trace\left(\rho(0)
           (\slb{\sigma_+}{1}\slb{e_\uparrow}{2}\right) +
           e^{-i2\pi50\uts{Hz}\;t}\trace\left(\rho(0)\slb{\sigma_+}{1}\slb{e_\downarrow}{2})\right),
\end{eqnarray}
and similarly for the second term, but with an offset frequency of
$900\uts{Hz}$ because of the chemical shift.  It can be seen that the
zero-frequency signal splits into two signals with frequencies of
$-50\uts{Hz}$ and $50\uts{Hz}$, respectively. The difference between the
two frequencies is the coupling constant. The
amplitudes of the different frequency signals can be used to infer the
expectations of operators such as
$\slb{\sigma_+}{1}\slb{e_\uparrow}{2}$, given by
$\trace\left(\rho(0)\slb{\sigma_+}{1}\slb{e_\uparrow}{2}\right)$.  For $n$
spin-${1\over 2}$ nuclei, the spectral peak
of a nucleus splits into a group of $2^{n-1}$ peaks, each
associated with operators like $\slb{\sigma_+}{a}\slb{e_\uparrow}{b}
\slb{e_\downarrow}{c}\slb{e_\downarrow}{d}\ldots$.
Fig.~\ref{fig:peakgroup} shows a simulated peak group for a nuclear
spin coupled to three other spins.  Expectations of the single spin
operators $\slb{\sigma_x}{a}$ and $\slb{\sigma_y}{a}$ can be obtained
from the real and imaginary parts of the total signal in a peak group
for a nucleus. The positions of the $2^{n-1}$ peaks depend on the
couplings.  If the peaks are all well separated, we can infer 
expectations of product operators with only one $\sigma_x$ or
$\sigma_y$, such as
$\slb{\sigma_x}{a}\slb{\sigma_z}{b}\slb{\one}{c}\slb{\sigma_z}{d}$ by
taking linear combinations with appropriate coefficients of the peak
amplitudes in a peak group.

In addition to the unitary evolution due to the internal Hamiltonian,
relaxation processes tend to decay $\rho(t)$ toward the equilibrium
state. In liquid state, the equilibrium state $\rho_{\mbox{\tiny
thermal}}$ is close to $\one/N$ where $N$ is the total dimension of
the state space.  The difference between $\rho_{\mbox{\tiny thermal}}$
and $\one/N$ is the equilibrium ``deviation'' density matrix and has
magnetization only along the $z$-axis (see Sect.~\ref{sec:initial}).
Because the only observed magnetization is planar, the observed signal
decays to zero as the state relaxes to equilibrium.  To a good
approximation we can write
\begin{equation}
\rho(t) = {1\over N}\one + e^{-\lambda t} \rho'(t) + (\mbox{not observed}),
\end{equation}
where $\rho'(t)$ has trace zero and evolves unitarily under the
Hamiltonian. The effect of the relaxation process is that $M(t)$ has
an exponentially decaying envelope, explaining the conventional name
for $M(t)$, namely, the ``free induction decay'' (FID). Typical half-times for
the decay are $.1\uts{s}$ to $2\uts{s}$ for nuclear spins used for QIP.  A
normal NMR observation consists of measuring $M(t)$ at discrete time
intervals until the signal is too small. The acquired FID is then
Fourier transformed to visualize the amplitudes of the different
frequency contributions.  The shape of the peaks in
Fig.~\ref{fig:c13peaks} reflects the decay envelope. The width
of the peaks is proportional to the decay rate $\lambda$.

For QIP, we wish to measure the probability $p$ that a given qubit,
say the first, labeled $\sysfnt{1}$, is in the state
$\kets{\bitone}{1}$.  We have $1-2p=\trace (\rho \slb{\sigma_z}{1})$,
which is the expectation of $\slb{\sigma_z}{1}$.  One can measure this
expectation by first applying a $90^\circ$ $y$-pulse to qubit
$\sysfnt{1}$, thus changing the state to $\rho'$. This pulse has the
effect of rotating initial, unobservable $z$-magnetization to
observable $x$-magnetization.  From $M(t)$ one can then infer
$\trace(\rho'\slb{\sigma_x}{1})$, which is the desired number. For the
coupled pair of carbons, $\trace(\rho'\slb{\sigma_x}{1})$ is given by
the sum of the real components of the amplitudes of the $50\uts{Hz}$
and the $-50\uts{Hz}$ contributions to $M(t)$.  However, the problem
is that these amplitudes are determined only up to a scale.  A second
problem is that the available states $\rho$ are highly mixed
(close to $\one/N$). The next section discusses how to compensate for
both problems.

As a final comment on NMR measurement, note that the ``back reaction''
on the nuclear spins due to the emission of electromagnetic energy is
weak. This is what enables us to measure the bulk magnetization over
some time. The ensemble nature of the system gives us direct, if
noisy, access to expectations of observables such as $\sigma_z$,
rather than a single answer---$\bitzero$ or $\bitone$. For algorithms
that provide a definite answer, having access only to expectations is
not a problem, because it is easy to distinguish the answer from the
noise. However, using expectations can increase the need for quantum
resources.  For example, Shor's factoring algorithm includes a
significant amount of classical post-processing based on highly random
answers from projective measurements. In order to implement the
algorithm in an ensemble setting, the post-processing must be
performed reversibly and integrated into the quantum computation to
guarantee a definite answer. This post-processing can be done with
polynomial additional quantum resources.

\subsection{The Initial State}
\label{sec:initial}

Because the energy difference between the nuclear spins' up and down
states is so small compared to room temperature, the equilibrium
distribution of states is nearly random.  In the liquid samples used,
equilibrium is established after $10\uts{s}$--$40\uts{s}$ if no RF
fields are being applied.  As a result, all computations
start with the sample in equilibrium. One way to think of this initial
state is that every nuclear spin in each molecule begins in the highly
mixed state $(1-\epsilon)\one/2 +
\epsilon\ketbra{\bitzero}{\bitzero}$, where $\epsilon$ is a small
number (of the order of $10^{-5}$). This is a nearly random state with
a small excess of the state $\ket{\bitzero}$.  The expression for the
initial state derives from the fact that the equilibrium state
$\rho_{\mbox{\tiny thermal}}$ is proportional to $e^{-H/kT}$, where
$H$ is the internal Hamiltonian of the nuclear spins in a molecule (in
energy units), $T$ is the temperature and $k$ is the Boltzman
constant. In our case, $H/kT$ is very small and the coupling terms are
negligible. Therefore
\begin{eqnarray}
e^{-H/kT} &\approx& e^{-\epsilon_1\slb{\sigma_z}{1}/kT}e^{-\epsilon_2\slb{\sigma_z}{2}/kT}\ldots\\
e^{-\epsilon_1\slb{\sigma_z}{1}/kT} &\approx& \one - \epsilon_1\slb{\sigma_z}{1}/kT\\
e^{-H/kT} &\approx& \one - \epsilon_1\slb{\sigma_z}{1}/kT - \epsilon_2\slb{\sigma_z}{2}/kT - \ldots 
\end{eqnarray}
where $\epsilon_l$ is half of the energy difference between
the up and down states of the $l$'th nuclear spin.

Clearly the available initial state is very far from what is needed
for standard QIP. However, it can still be used to perform interesting
computations. The main technique is to use available NMR tools to
change the initial state to a ``pseudopure'' state, which for all
practical purposes behaves like the initial state required by QIP.
The technique is based on three key observations.  First, only the
trace-less part of the density matrix contributes to the
magnetization.  Suppose that we are using $n$ spin-$1\over 2$ nuclei
in a molecule and the density matrix is $\rho$. Then the current
magnetization is proportional to $\trace(\rho \hat m)$, where $\hat m$
is a traceless operator (see Eq.~\ref{eq:fullmag}).  Therefore the
magnetization does not depend on the part of $\rho$ proportional to
the identity matrix. A ``deviation density matrix'' for $\rho$ is any
matrix $\delta$ such that $\delta-\rho=\lambda\one$ for some
$\lambda$. For example, $\epsilon\ketbra{\bitzero}{\bitzero}$ is a
deviation for the equilibrium state of one nuclear spin.
We have
\begin{eqnarray}
\trace(\delta\hat m) 
  &=& \trace((\rho+\lambda\one)\hat m)\nonumber\\
  &=& \trace(\rho\hat m) +\trace(\hat m)\nonumber\\
  &=& \trace(\rho\hat m).
\end{eqnarray}

The second observation
is that all the unitary operations used, as well as the non-unitary
ones to be discussed below, preserve the completely mixed state
$\one/2^n$.\footnote{The intrinsic relaxation process does not 
preserve the completely mixed state.
But its contribution is either negligible over the time scale of
typical experiments or can be removed with the help of subtractive phase
cycling.} Therefore, all future observations of magnetization
depend only on the initial deviation.

The third observation is that all the scales are relative.  In
particular, as will be explained, the probability that the final
answer of a quantum computation is $\bitone$ can be expressed as the
ratio of two magnetizations. It follows that one can arbitrarily
rescale a deviation density matrix.  For measurement, the absolute
size of the magnetizations is not important; the most important issue
is that the magnetizations are strong enough to be observable over the
noise.

To explain the relativity of the scales and introduce ``pseudopure''
states for QIP, we begin with one spin-${1\over 2}$ qubit. Its
equilibrium state has as a deviation
$\delta=\epsilon\ketbra{\bitzero}{\bitzero}$.  If $U$ is the total
unitary operator associated with a computation, then $\delta$ is
transformed to $\delta'=\epsilon
U\ketbra{\bitzero}{\bitzero}U^\dagger$. For QIP purposes, the goal is
to determine what the final probability $p_{\bitone}$ of measuring
$\ket{\bitone}$ is, given that $\ket{\bitzero}$ is the initial state.
This probability can be computed as follows:
\begin{eqnarray}
p_{\bitone} &=&
  \bra{\bitone}U\ketbra{\bitzero}{\bitzero}U^\dagger\ket{\bitone}
  \nonumber\\ &=&
  \trace\left(U\ketbra{\bitzero}{\bitzero}U^\dagger\ketbra{\bitone}{\bitone}\right)\nonumber\\
  &=& \trace\left(U\ketbra{\bitzero}{\bitzero}U^\dagger(\one-\sigma_z)\right)/2
   \nonumber\\
  &=& \left(\trace(U\ketbra{\bitzero}{\bitzero}U^\dagger) -
  \trace(U\ketbra{\bitzero}{\bitzero}U^\dagger\sigma_z)\right)/2
    \nonumber\\ &=&
  \left(1-\trace(U\ketbra{\bitzero}{\bitzero}U^\dagger\sigma_z)\right)/2.
\end{eqnarray}
Thus, the probability
can be determined by measuring the expectations of $\sigma_z$ for
the initial and final states (in different experiments), which yields
the quantities $a =
\trace(\delta\sigma_z)=\epsilon$ and $a' = \trace(\delta'\sigma_z) = \epsilon\,
\trace\left( U\ketbra{\bitzero}{\bitzero}U^\dagger\sigma_z\right)$, 
respectively. The desired answer is $p_{\bitone}=(1-(a/a'))/2$ and
does not depend on the scale $\epsilon$.

The method presented in the previous paragraph for determining the
probability that the answer of a quantum computation is $\bitone$
generalizes to many qubits. The goal is to determine the probability
$p_{\bitone}$ of measuring $\kets{\bitone}{1}$ in a measurement of the
first qubit after a computation with initial state
$\ket{\bitzero\ldots\bitzero}$. Suppose we can prepare the spins in an
initial state with deviation
$\delta=\epsilon\ketbra{\bitzero\ldots\bitzero}{\bitzero\ldots\bitzero}$.
A measurement of the expectations $a$ and $a'$ of $\slb{\sigma_z}{1}$ for the
initial and final states then yields $p_{\bitone}$ as before, by the
formula $p_{\bitone}=(1-(a/a'))/2$.

A state with deviation $\epsilon\ketbra{\psi}{\psi}$ is called a
``pseudopure'' state, because this deviation is proportional to the
deviation of the pure state $\ketbra{\psi}{\psi}$.  With respect to
scale-independent NMR observations and unitary evolution, a
pseudopure state is equivalent to the corresponding pure state.
Because NMR QIP methods are scale independent, we now
generalize the definition of deviation density matrix: $\delta$ is a
deviation of the density matrix $\rho$ if $\epsilon\delta =
\rho+\lambda\one$ for some $\lambda$ and $\epsilon$.

Among the most important enabling techniques in NMR QIP are the
methods that can be used to transform the initial thermal equilibrium
state to a standard pseudopure state with deviation
$\ketbra{\bitzero\ldots\bitzero}{\bitzero\ldots\bitzero}$.  An example
of how that can be done will be given as the second algorithm in
Sect.~\ref{sec:examples}.  The basic principle for each method is to
create, directly or indirectly by summing over multiple experiments, a
new initial state as a sum $\rho_0=\sum_iU_i\rho_{\mbox{\tiny
thermal}}U_i^\dagger$, where the $U_i$ are carefully and sometimes
randomly
chosen~\cite{cory:qc1997a,chuang:qc1997a,knill:qc1997b,sharf:qc2000a}
to ensure that $\rho_0$ has a standard pseudopure deviation.  Among
the most useful tools for realizing such sums are pulsed gradient
fields.

\subsection{Gradient Fields}

Modern NMR spectrometers are equipped with the capability of applying
a magnetic field gradient in any direction for a chosen, brief amount
of time.  If the direction is along the sample's $z$-axis, then while
the gradient is on, the field varies as $B(z)=B_0+\gamma z B_1$, where
$B_0$ is the strong, external field and $B_1$ is the gradient power.
As a result of this gradient, the precession frequency of nuclear
spins depends on their positions' $z$-coordinates.  One of the most
important applications of gradients is NMR imaging because gradients make
it possible to distinguish different parts of the sample.

The effect of applying a $z$-gradient can be visualized for the
situation in which there is only one observable nuclear spin per molecule.
Suppose that the initial deviation density matrix
of each nuclear spin is $\sigma_x$ in the rotating frame. 
After a gradient pulse of duration $t$, the deviation of a
nuclear spin at position $z$ is given by $e^{-i\sigma_z\nu z
t/2}\sigma_x e^{i\sigma_z\nu z t/2} = \cos(\nu zt)\sigma_x+\sin(\nu
zt)\sigma_y$, where the constant $\nu$ depends linearly on the
strength of the gradient and the magnetic moment of the nucleus. See
Fig.~\ref{figbgradient}. The effect of the gradient is a $z$-dependent
change in phase.  The coil used to measure planar magnetization
integrates the contribution to the magnetization of all the nuclei in
the neighborhood of the coil.  Assuming a coil equally sensitive
over the interval between $-a$ and $a$ along the sample's $z$-axis,
the observed total $x$-magnetization is:
\begin{eqnarray}
M_x &=& \int_{-a}^{a}dz\,\trace\left(\sigma_x(\cos(\nu zt)\sigma_x+\sin(\nu zt)\sigma_y)\right) \nonumber\\
&=& \int_{-a}^{a}dz\,\trace\left(\cos(\nu zt)\sigma_x^2+\sin(\nu zt)\sigma_x\sigma_y\right) \nonumber \\
&=& \int_{-a}^{a}dz\,\trace\left(\cos(\nu zt)+i\sin(\nu zt)\sigma_z\right) \nonumber \\
    &=& 2\int_{-a}^a dz\cos(\nu zt).
\end{eqnarray}
For large values of $\nu t$, $M_x\simeq 0$.  In general, a
sufficiently powerful gradient pulse eliminates the planar
magnetization.

\vspace*{.2in}
\begin{herefig}
\begin{tabular}{c@{}c@{}c}
\raisebox{-1.5in}{\begin{picture}(3,3)(-1.5,-3)
\nputgr{0,0}{t}{height=3in}{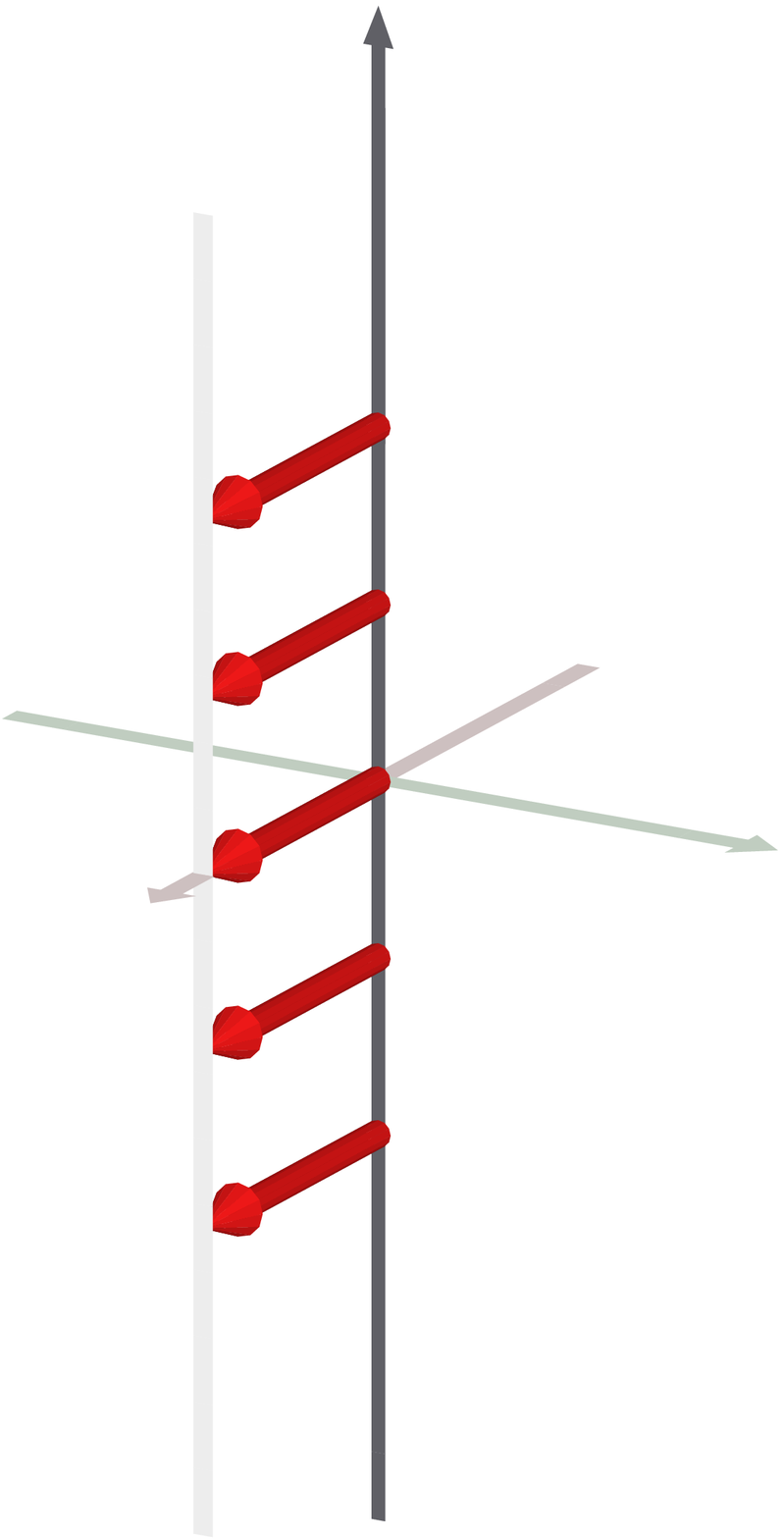}
\nputbox{.1,0}{t}{\Large $z$}
\end{picture}}& 
{\Huge$\stackrel{\textrm{\large gradient}}{\longrightarrow}$}&
\raisebox{-1.5in}{\begin{picture}(3,3)(-1.5,-3)
\nputgr{0,0}{t}{height=3in}{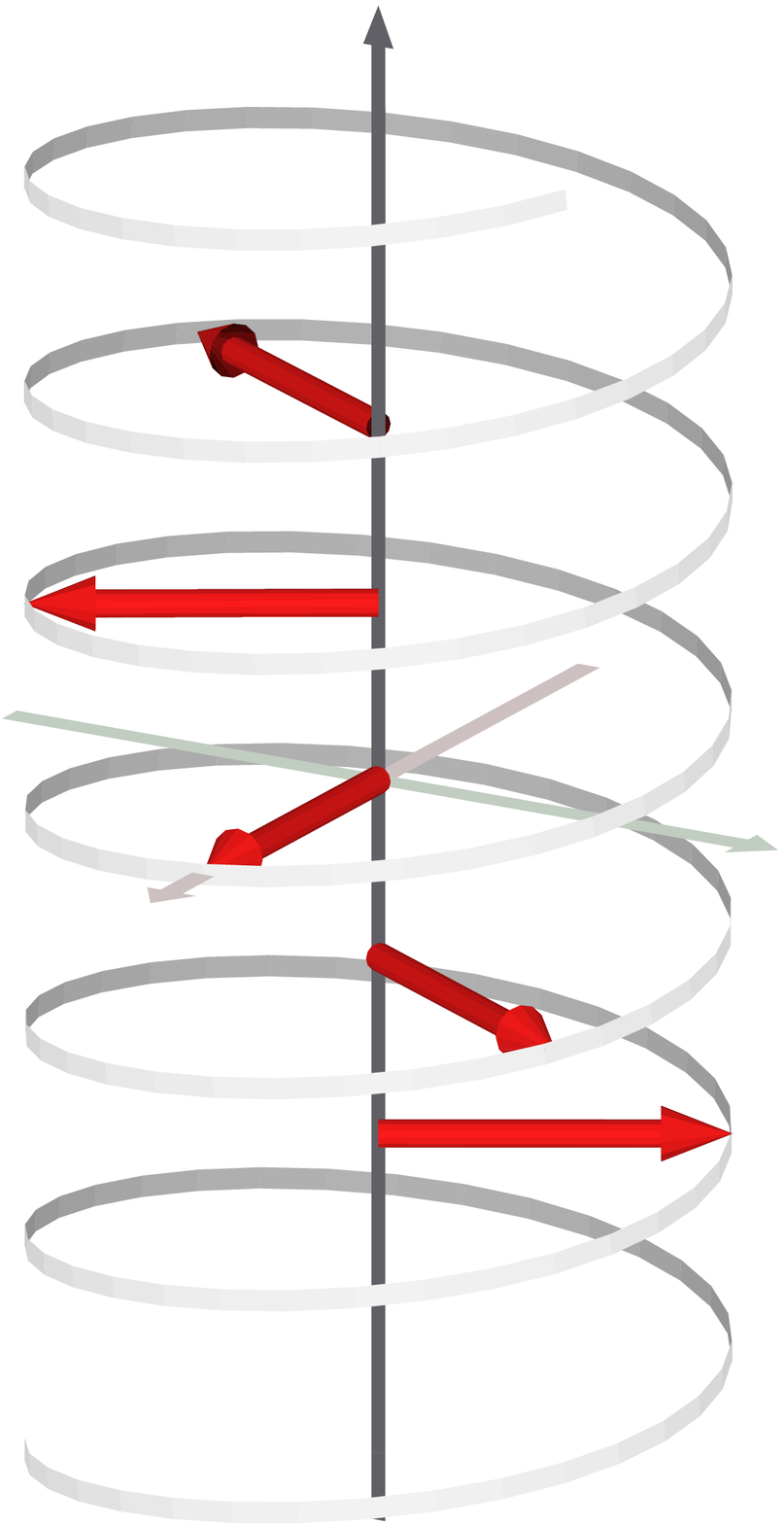}
\nputbox{.1,0}{t}{\Large $z$}
\end{picture}}
\end{tabular}
\label{figbgradient}
\vspace*{.2in}
\herefigcap{
Effect of a pulsed gradient field along the $z$ axis in the rotating
frame.  Initial $x$-magnetization is assumed.  A spin at $z=0$ is not
affected, but the ones above and below are rotated by an amount
proportional to $z$.  As a result, the local planar magnetization
follows a spiral curve. }
\end{herefig}

Interestingly, the effect of a gradient pulse can be reversed if an
opposite gradient pulse is applied for the same amount of time. This
effect is called a ``gradient echo''.  The reversal only works if the
second pulse is applied sufficiently soon. Otherwise, diffusion
randomizes the molecules' positions along the gradient's direction
before the second pulse. If the positions are randomized, then the
phase change from the second pulse is no longer correlated with that
from the first for any given molecule.  The loss of memory of the
phase change from a gradient pulse can be fine-tuned by variations in the
delay between the two pulses in a gradient echo sequence.  This 
method can be
used for applying a controllable amount of phase noise,
which is useful for investigating the effects of noise and the ability to
correct for noise in QIP.

If the gradient pulse is not reversed and the memory of the phase
changes is lost, then the pulse's effect can be described as an
irreversible operation on the state of the nuclear spin. If the
initial state of the nuclear spin in each molecule is $\rho$, then
after the gradient pulse, the spin state of a molecule at position $z$
is given by $\rho(z)= e^{-i\sigma_z\nu z t/2}\rho e^{i\sigma_z\nu z
t/2}$.  Suppose that the positions of the molecules are randomized
over the region that the coil is sensitive to.  Now it is no longer
possible to tell where a given molecule was when the gradient pulse
was applied.  As a result, as far as our observations are concerned,
the state of a molecule is given by $\rho(z)$, where $z$ is random.
In other words, the state is indistinguishable from
\begin{equation}
\rho'={1\over 2a}\int_{-a}^{a}dz \rho(z)
     ={1\over 2a}\int_{-a}^{a}dz
         e^{-i\sigma_z\nu z t/2}\rho e^{i\sigma_z\nu z t/2}.
\end{equation}
Thus the effect of the gradient pulse is equivalent to the operation
$\rho\rightarrow\rho'$ as defined by the above equation. This is an
operation of the type mentioned at the end of the previous section and
can be used for making states such as pseudopure states.  Note that
after the gradients have been turned off, nuclei at different
positions cannot be distinguished by the measurement coil.  It is
therefore not necessary to wait for the molecules' positions to be
randomized.

So far we have described the effects of gradient pulses on isolated
nuclear spins in a molecule. In order to restrict the effect to a
single nuclear spin in a molecule, one can invert the other spins
between a pair of identical gradient pulses in the same direction.
This technique refocuses the gradient for the inverted spins.  An
example of how effects involving multiple nuclear spins can be
exploited is the algorithm for pseudopure state preparation described
in Sect.~\ref{sec:pseudo-pure}.

\section{Examples of Quantum Algorithms for NMR}
\label{sec:examples}

We give three examples of algorithms for NMR QIP.  The first
example is an NMR implementation of the controlled-not gate. The
second consists of a procedure for preparing a type of pseudopure
state. And the last shows how NMR can be used to investigate the
behavior of simple error-correction procedures.  The first two
examples are fundamental to QIP with NMR.  Realizations of the
controlled-not are needed to translate standard quantum algorithms
into the language of NMR, and procedures for making pseudopure states
have to precede the implementation of many quantum algorithms.

\subsection{The Controlled-not}

One of the standard gates used in quantum algorithms is the
controlled-not. The controlled-not gate ($\mb{cnot}$) acts on two
qubits. The action of $\mb{cnot}$ can be described by ``if the first
qubit is $\ket{\bitone}$, then flip the second qubit.''
Consequently, the effect of $\mb{cnot}$ on the logical states is given
by the mapping
\begin{equation}
\begin{array}{rcl}
\mb{cnot}\ket{\bitzero\bitzero} &=&\ket{\bitzero\bitzero}\\
\mb{cnot}\ket{\bitzero\bitone} &=&\ket{\bitzero\bitone}\\
\mb{cnot}\ket{\bitone\bitzero} &=&\ket{\bitone\bitone}\\
\mb{cnot}\ket{\bitone\bitone} &=&\ket{\bitone\bitzero}.
\end{array}
\end{equation}
As an operator, the controlled-not is given by
\begin{equation}
\mb{cnot}=\ketbras{\bitzero}{\bitzero}{1} + 
   \ketbras{\bitone}{\bitone}{1}\slb{\sigma_x}{2} = 
\left((\one+\slb{\sigma_z}{1}) + 
      (\one-\slb{\sigma_z}{1})\slb{\sigma_x}{2}\right)/2.
\end{equation}
The goal is to derive a sequence of NMR operations that realize the
controlled-not.  As discussed in Sect.~\ref{sec:princs}, the unitary
operations that are implementable by simple NMR techniques are
rotations $e^{-i\slb{\sigma_u}{a}\theta/2}$ by $\theta$ around the
$u$-axis, where $u$ is any direction in the plane (RF pulses), and the
two-qubit operations $e^{-i\slb{\sigma_z}{b}\slb{\sigma_z}{c}\phi/2}$
(the $J$-coupling). We call
$e^{-i\slb{\sigma_z}{b}\slb{\sigma_z}{c}\phi/2}$ a rotation by $\phi$
around $\slb{\sigma_z}{b}\slb{\sigma_z}{c}$. This terminology 
reflects the fact that such rotations and their effects on deviation
density matrices can be understood by a generalization of the Bloch
sphere picture called the ``product operator formalism'' introduced by
O.~S\"orensen \emph{et al.}~\cite{sorensen:qc1983a}.

To implement the controlled-not using NMR techniques one can decompose
the gate into a sequence of $90^\circ$ rotations around the main axes
on each of the two qubits, and a $90^\circ$ rotation around
$\slb{\sigma_z}{1}\slb{\sigma_z}{2}$.  One way to find a decomposition
is to first realize that the two-qubit $90^\circ$ rotation
$e^{-i\slb{\sigma_z}{1}\slb{\sigma_z}{2}\pi/4}$ is equivalent to a
combination of two gates, each conditional on the logical state of
qubit $\sysfnt{1}$.  The first gate applies a $90^\circ$ rotation
around the $z$-axis ($e^{-i\slb{\sigma_z}{2}\pi/4}$) to qubit
$\sysfnt{2}$ conditional on qubit $\sysfnt{1}$'s state being
$\kets{\bitzero}{1}$. The second applies the $-90^\circ$ rotation
$e^{i\slb{\sigma_z}{2}\pi/4}$ to qubit $\sysfnt{2}$ conditional on
qubit $\sysfnt{1}$'s state being $\kets{\bitone}{1}$.  By following
the two-qubit rotation with a $-90^\circ$ rotation around $z$-axis
($e^{i\slb{\sigma_z}{2}\pi/4}$) on qubit $\sysfnt{2}$, the total
effect is to cancel the rotation if qubit $\sysfnt{1}$ is in state
$\kets{\bitzero}{1}$; if qubit $\sysfnt{1}$ is in state
$\kets{\bitone}{1}$, the rotations add to a $-180^\circ$ rotation
$e^{i\slb{\sigma_z}{2}\pi/2}=i\slb{\sigma_z}{2}$ on qubit
$\sysfnt{2}$.  If we precede this sequence with
$e^{-i\slb{\sigma_y}{2}\pi/4}$ and follow it by
$e^{i\slb{\sigma_y}{2}\pi/4}$ (this operation is called
``conjugating'' by a $-90^\circ$ $y$-rotation), then the overall
effect is a conditional $-i\slb{\sigma_x}{2}$ operation. Note how the
conjugation rotated the operation's axis according to the Bloch sphere
rules.  The controlled-not is obtained by eliminating the $-i$ with a
$90^\circ$ $z$-rotation on qubit $\sysfnt{1}$. That is, the effect of
the complete sequence is $e^{-i\pi/4}\ketbras{\bitzero}{\bitzero}{1}+
e^{-i\pi/4}\ketbras{\bitone}{\bitone}{2}\slb{\sigma_x}{2}$, which is
the controlled-not up to a global phase.  The decomposition thus
obtained can be represented as a quantum network with rotation gates
as shown in Fig.~\ref{fig:cnot=nmr}. The corresponding NMR pulse
sequence implementation is shown in Fig.~\ref{fig:cnotpulse}.

\begin{herefig}
\label{fig:cnot=nmr}
\begin{picture}(7,2.7)(-3.5,-2.6)
\nputgr{0,0}{t}{}{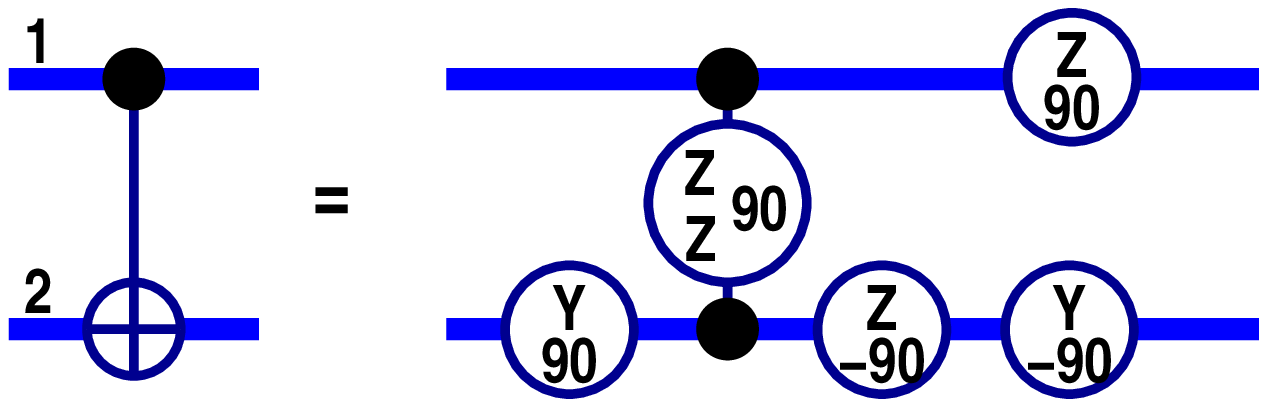}
\end{picture}
\herefigcap{ Quantum network for implementing the controlled-not using
operations available in NMR. The conventions for depicting gates are
as explained in~\cite{knill:qc2001c}. The two one-qubit $z$-rotations
can be implemented by a change in the reference phase of the rotating
frame without applying any RF pulses.  }
\end{herefig}

\begin{herefig}
\label{figcnotcircuit}
\label{fig:cnotpulse}
\begin{picture}(7,3)(-3.5,-3)
\nputgr{0,0}{t}{width=7in}{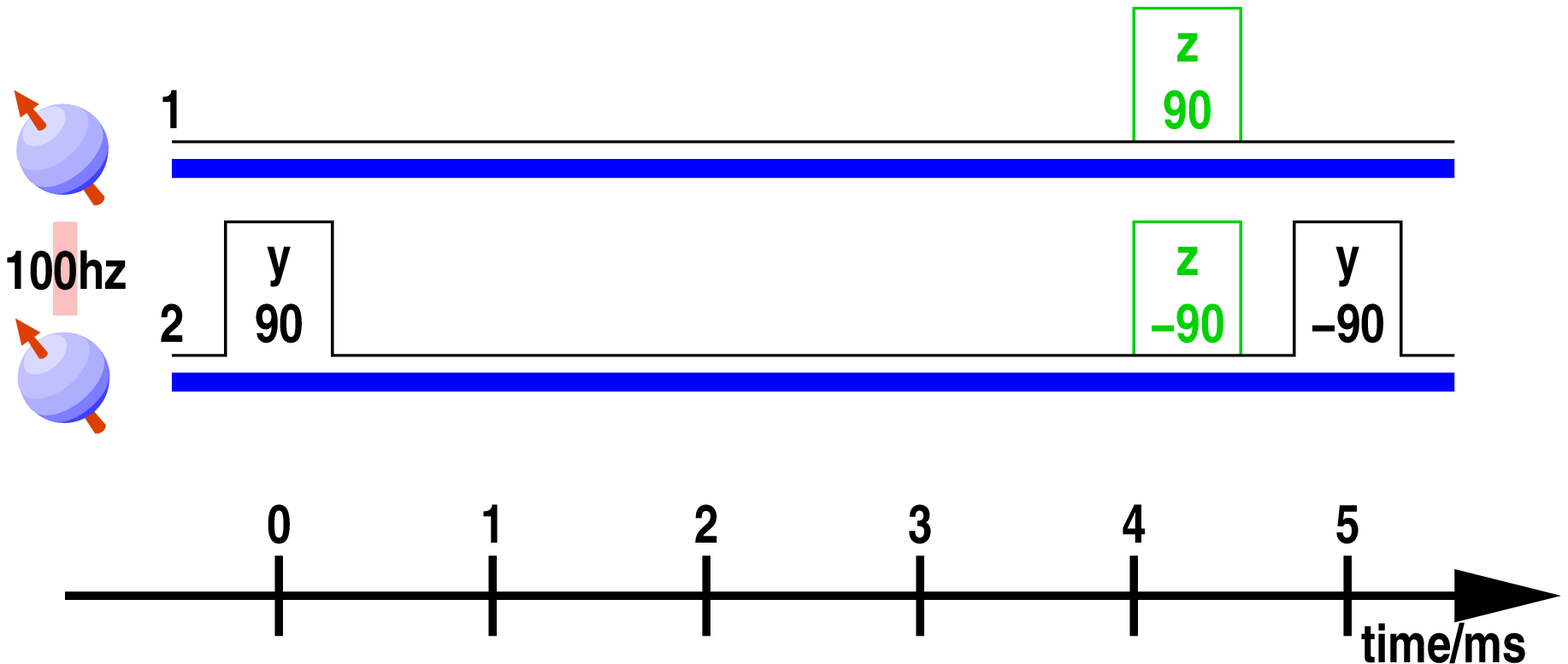}
\end{picture}
\herefigcap{Pulse sequence for realizing the controlled-not.  The
control bit is spin $\sysfnt{1}$ and the target is spin
$\sysfnt{2}$. The pulses are shown using the representation introduced
in Fig.~\ref{fig:refocussing}.  The $z$-pulses (shown in green) are
``virtual'', requiring only a change of reference frame. The placement
of the $z$-pulses between the RF pulses is immaterial, because they
commute with the coupling that evolves in between. The delay between
the two RF pulses is $1/(2J)$ ($5\uts{ms}$ if $J=100\uts{Hz}$), which
realizes the desired two-qubit rotation by internal evolution.  The
$-90^\circ$ $y$-rotation is actually implemented with a $90^\circ$
pulse with axis $-y$. The resulting rotation has the desired effect
up to a global phase.  The pulse widths are exaggerated and should be
as short as possible to avoid errors due to coupling evolution during
the RF pulses. Alternatively, techniques can be used that compensate
for some of these errors~\cite{knill:qc1999a}.}
\end{herefig}

The effect of the NMR pulse sequence that implements the
controlled-not can be visualized for logical initial states with the
help of the Bloch-sphere representation of the states. Such a
visualization is shown for two initial states in
Fig. \ref{figcnotbloch}.

\pagebreak

\begin{herefig}
\label{figcnotbloch}
\begin{tabular}{|r|c|c|}
\hline
&
$\kets{\bitzero}{1}\kets{\bitzero}{2}\rightarrow
 \kets{\bitzero}{1}\kets{\bitzero}{2}$
&
$\kets{\bitone}{1}\kets{\bitzero}{2}\rightarrow
 \kets{\bitone}{1}\kets{\bitone}{2}$\\[3pt]
\hline
(1)&
\raisebox{-.75in}{\begin{picture}(2.5,1.4)(-1.25,-1.4)
\nputgr{0,0}{t}{height=1.4in}{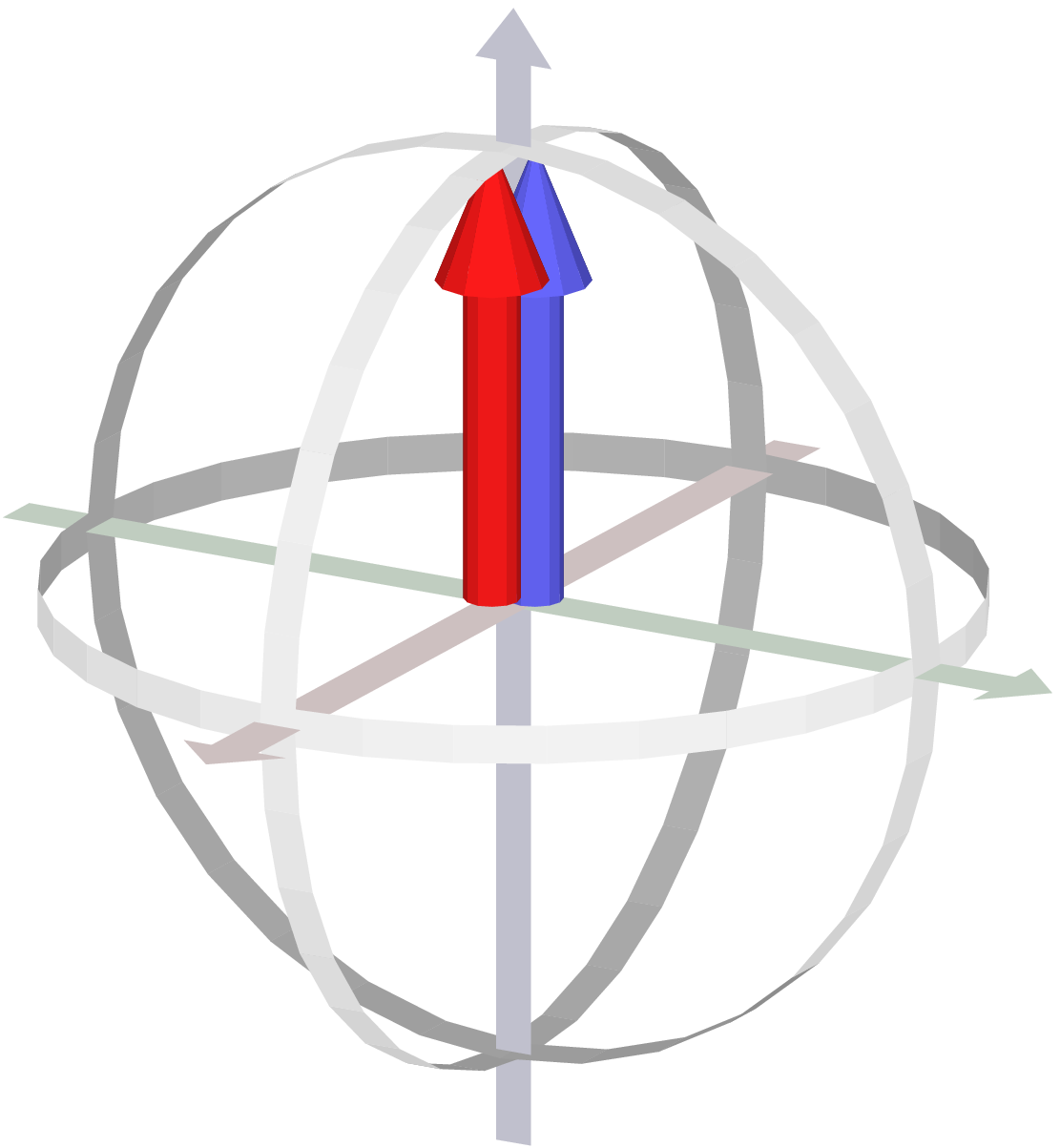}
\end{picture}}&
\raisebox{-.75in}{\begin{picture}(2.5,1.4)(-1.25,-1.4)
\nputgr{0,0}{t}{height=1.4in}{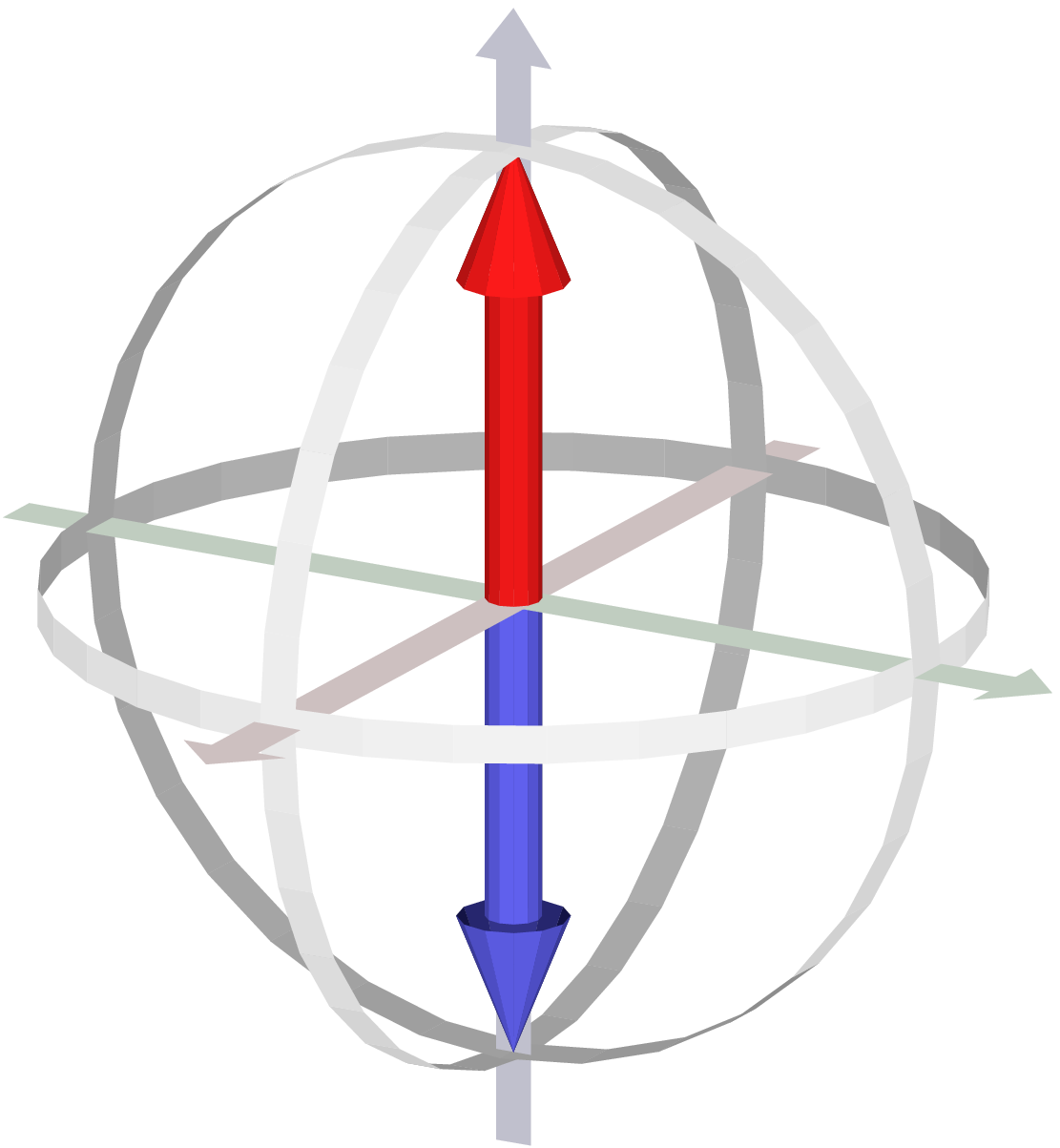}
\end{picture}}\\
\hline
(2)&
\raisebox{-.75in}{\begin{picture}(2.5,1.4)(-1.25,-1.4)
\nputgr{0,0}{t}{height=1.4in}{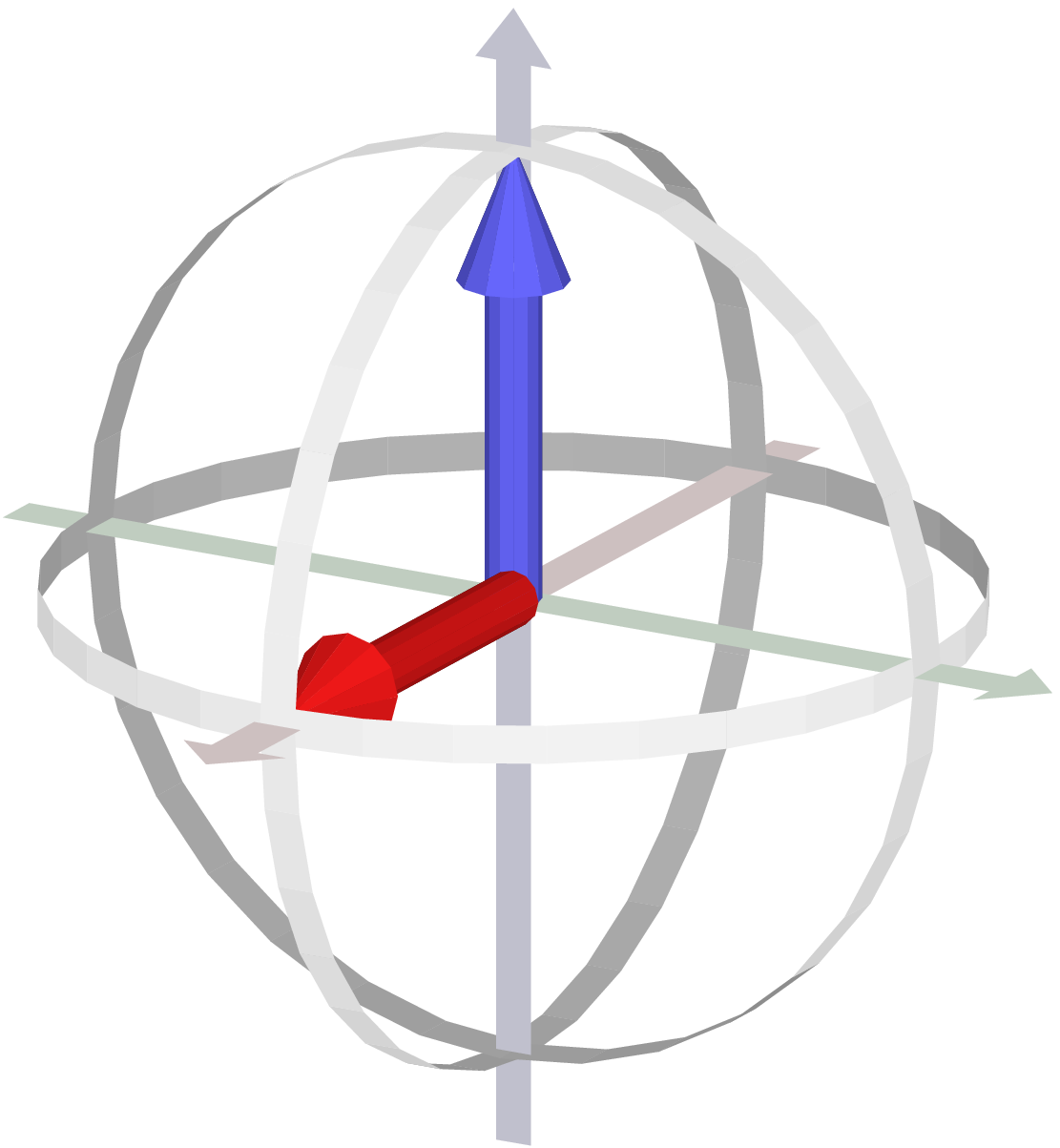}
\end{picture}}&
\raisebox{-.75in}{\begin{picture}(2.5,1.4)(-1.25,-1.4)
\nputgr{0,0}{t}{height=1.4in}{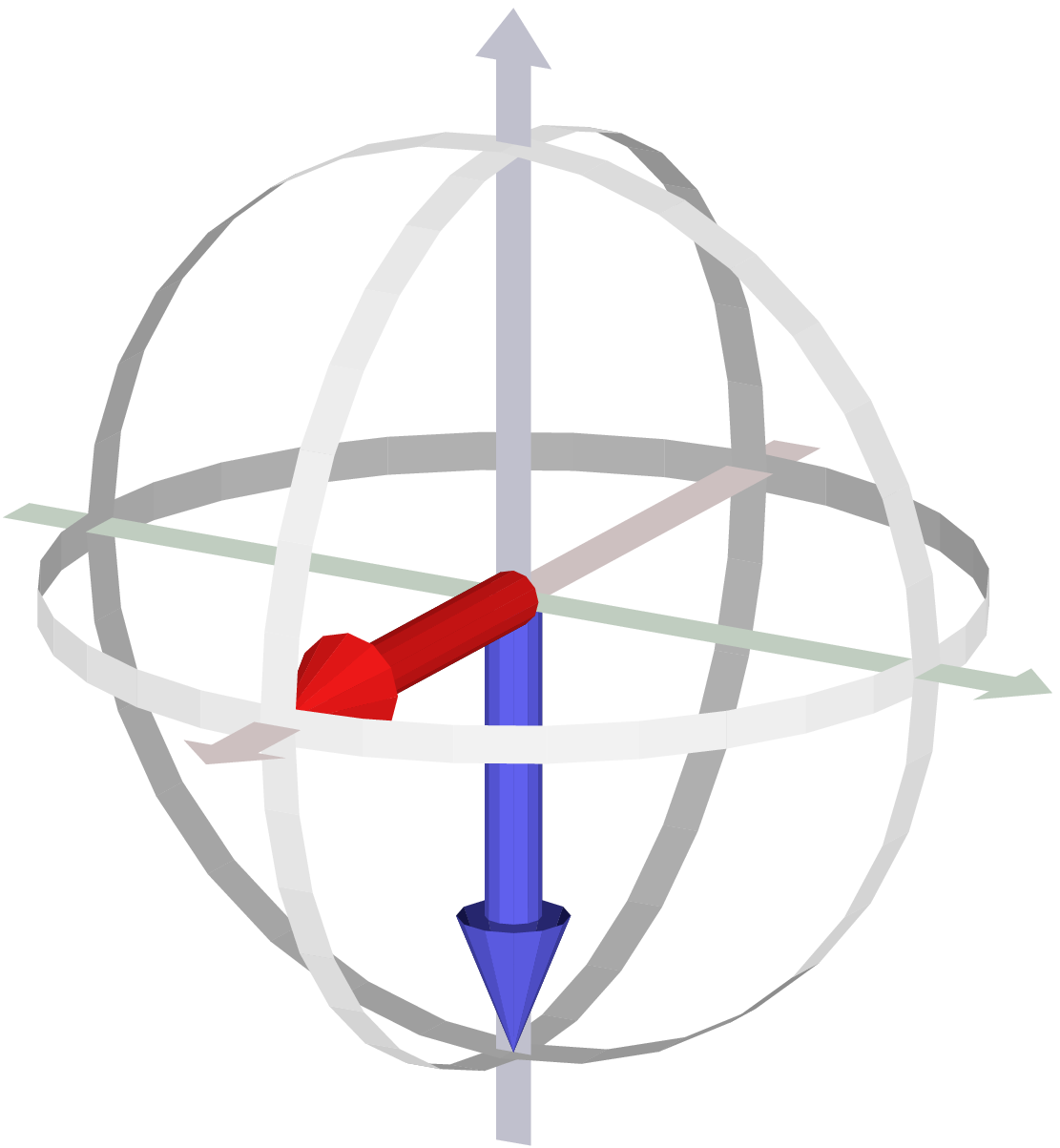}
\end{picture}}\\
\hline
(3)&
\raisebox{-.75in}{\begin{picture}(2.5,1.4)(-1.25,-1.4)
\nputgr{0,0}{t}{height=1.4in}{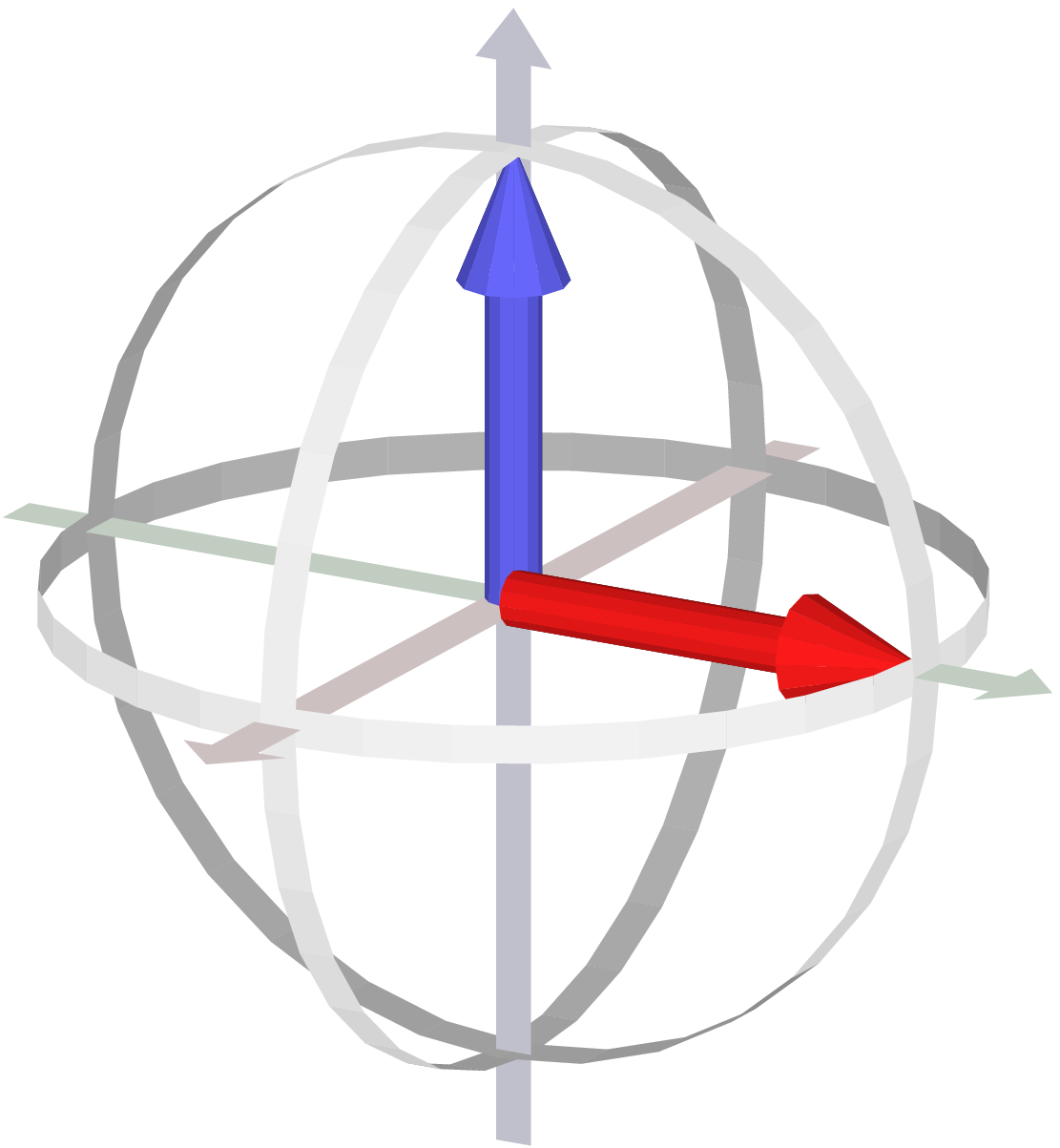}
\end{picture}}&
\raisebox{-.75in}{\begin{picture}(2.5,1.4)(-1.25,-1.4)
\nputgr{0,0}{t}{height=1.4in}{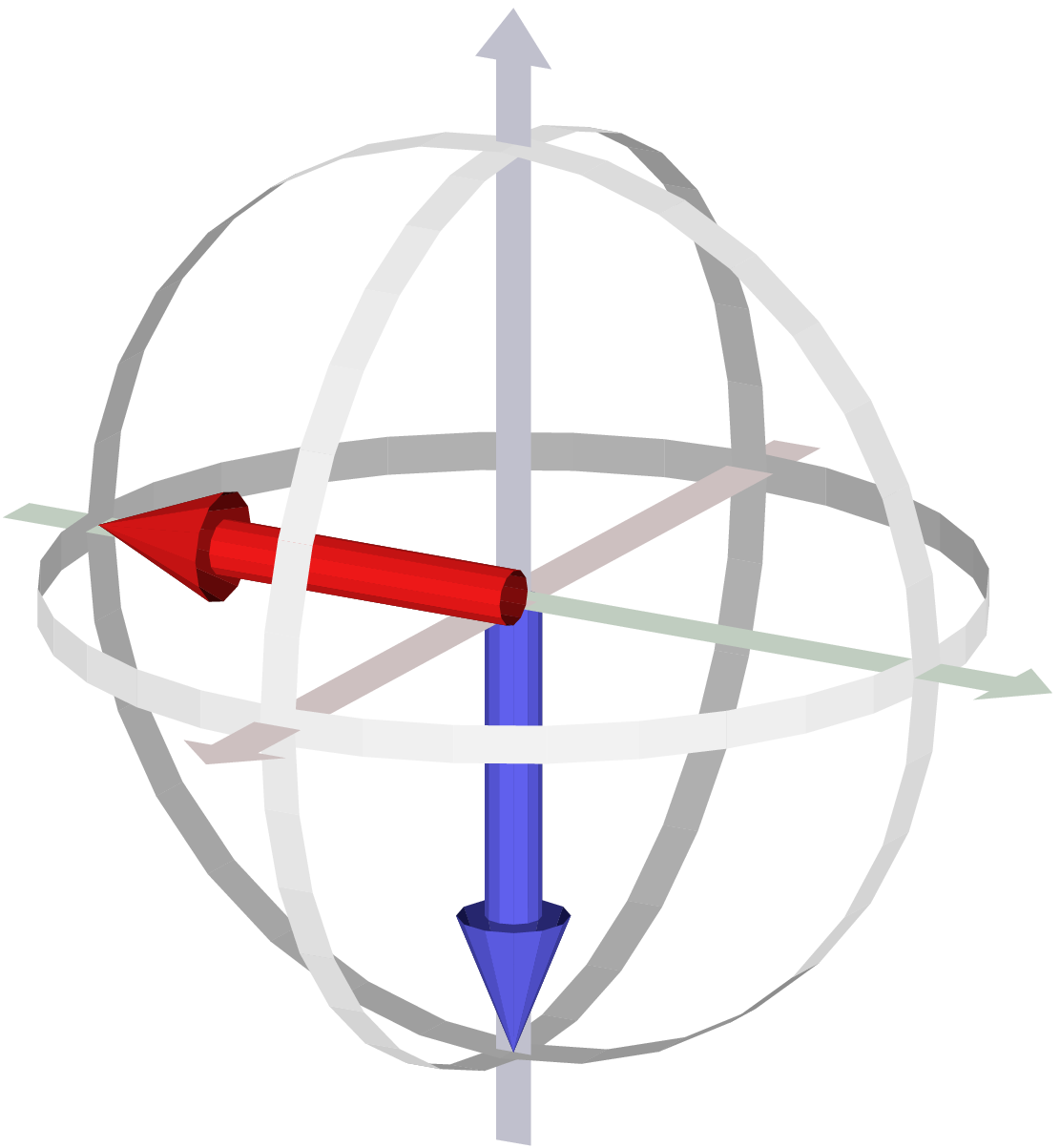}
\end{picture}}\\
\hline
(4)&
\raisebox{-.75in}{\begin{picture}(2.5,1.4)(-1.25,-1.4)
\nputgr{0,0}{t}{height=1.4in}{bloch4a1}
\end{picture}}&
\raisebox{-.75in}{\begin{picture}(2.5,1.4)(-1.25,-1.4)
\nputgr{0,0}{t}{height=1.4in}{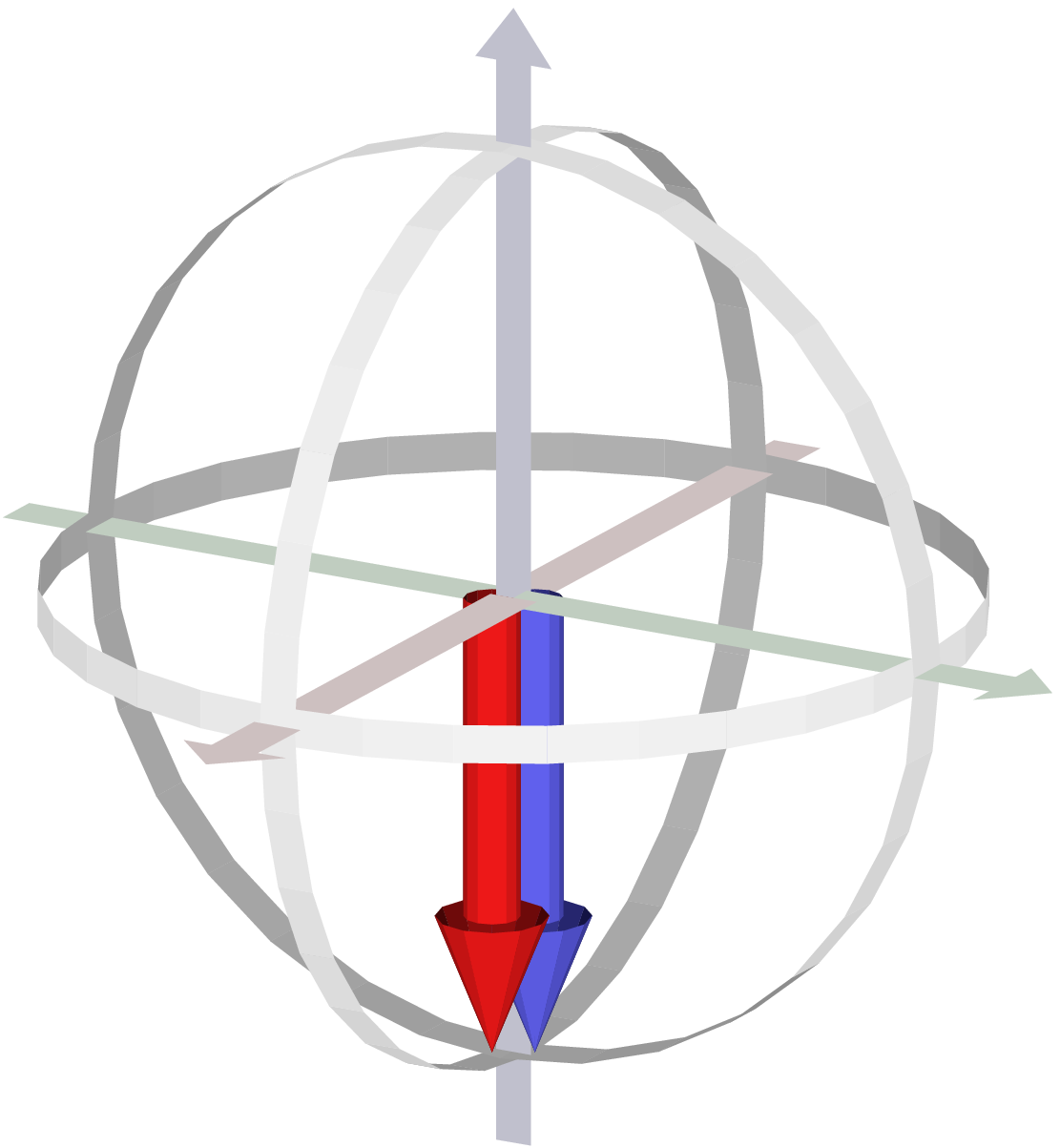}
\end{picture}}\\
\hline
\end{tabular}
\vspace*{.15in}
\herefigcap{
Sequences of states for the controlled-not pulse sequence.  The first
column has both spins initially in the logical $\ket{\bitzero}$ state,
represented by two arrows pointing up.  The blue and red arrows
represent spin $\sysfnt{1}$ and $\sysfnt{2}$, respectively.  The
second column has the first spin initially in the $\ket{\bitone}$
state, indicated by its arrow (blue) pointing down.  The
configurations are shown (1) at the beginning of
the sequence, (2) after the $90^\circ$ $y$-rotation, (3) after the
$J$-coupling (but before the $z$- and $y$-pulses), and (4) at the end
of the sequence. The conditional effect is realized by the second
spin's pointing down at the end of the second column. The effect of
the $J$-coupling causing the evolution from (2)~to (3)~is best
understood as a conditional rotation around the $z$-axis (forward by
$90^\circ$ if the first spin is up; backward if it is down). }
\end{herefig}

The effects of the pulse sequence for the controlled-not can be shown
with the Bloch sphere as in Fig.~\ref{figcnotbloch} only if the
intermediate states are products of states on each qubit.  Things are
no longer so simple if the initial state of the spins is
${1\over\sqrt{2}}\left(\ket{\bitzero}+\ket{\bitone}\right)\ket{\bitzero}
=
{1\over\sqrt{2}}\left(\ket{\bitzero\bitzero}+\ket{\bitone\bitzero}\right)$,
for example. This is representable as spin $\sysfnt{1}$'s arrow
pointing along the $x$-axis, but the $J$-coupling leads to a
superposition of states (a maximally entangled state) no longer
representable by a simple combination of arrows in the Bloch sphere.

\subsection{Creating a Labeled Pseudopure State}
\label{sec:pseudo-pure}

One way to realize the standard pseudopure state starting from the
equilibrium density matrix $\rho_{\mbox{\tiny thermal}}$ is to
eliminate the observable contributions due to terms of
$\rho_{\mbox{\tiny thermal}}$ different from
$\ketbra{\bitzero\ldots\bitzero}{\bitzero\ldots\bitzero}$.  There are
several different methods of accomplishing this.  For example, one can
perform multiple experiments with different pre-processing of the
equilibrium state so that signals from unwanted terms average to zero
(temporal averaging). Or one can use gradients to remove the unwanted
terms in one experiment (spatial averaging).

In this section, we show how to use spatial averaging to prepare a
so-called ``labeled'' pseudopure state on two nuclear spins. In
general, instead of preparing the standard pseudopure state with
deviation $\ketbra{\bitzero\ldots}{\bitzero\ldots}$ on $n$
spin-$1\over 2$ nuclei, one can prepare a ``labeled'' pseudopure
state with deviation
$\slb{\sigma_x}{1}\ketbra{\bitzero\ldots}{\bitzero\ldots}$ on $n+1$
spins.  This state is easily recognizable with an NMR observation
of the first spin: Assuming that all the peaks arising from couplings
to other spins are resolved, the first spin's peak group has $2^n$
peaks corresponding to which logical states the other spins are in. If
the current state is the above labeled pseudopure state, then all the
other spins are in the logical state $\ket{\bitzero}$, which implies
that in the spectrum, only one of the peaks of the first spin's peak
group is visible. See Fig.~\ref{fig:peakgroup}.

\begin{herefig}
\begin{picture}(7,5.5)(-3.5,-2.25)
\nputgr{0,0}{b}{width=6in}{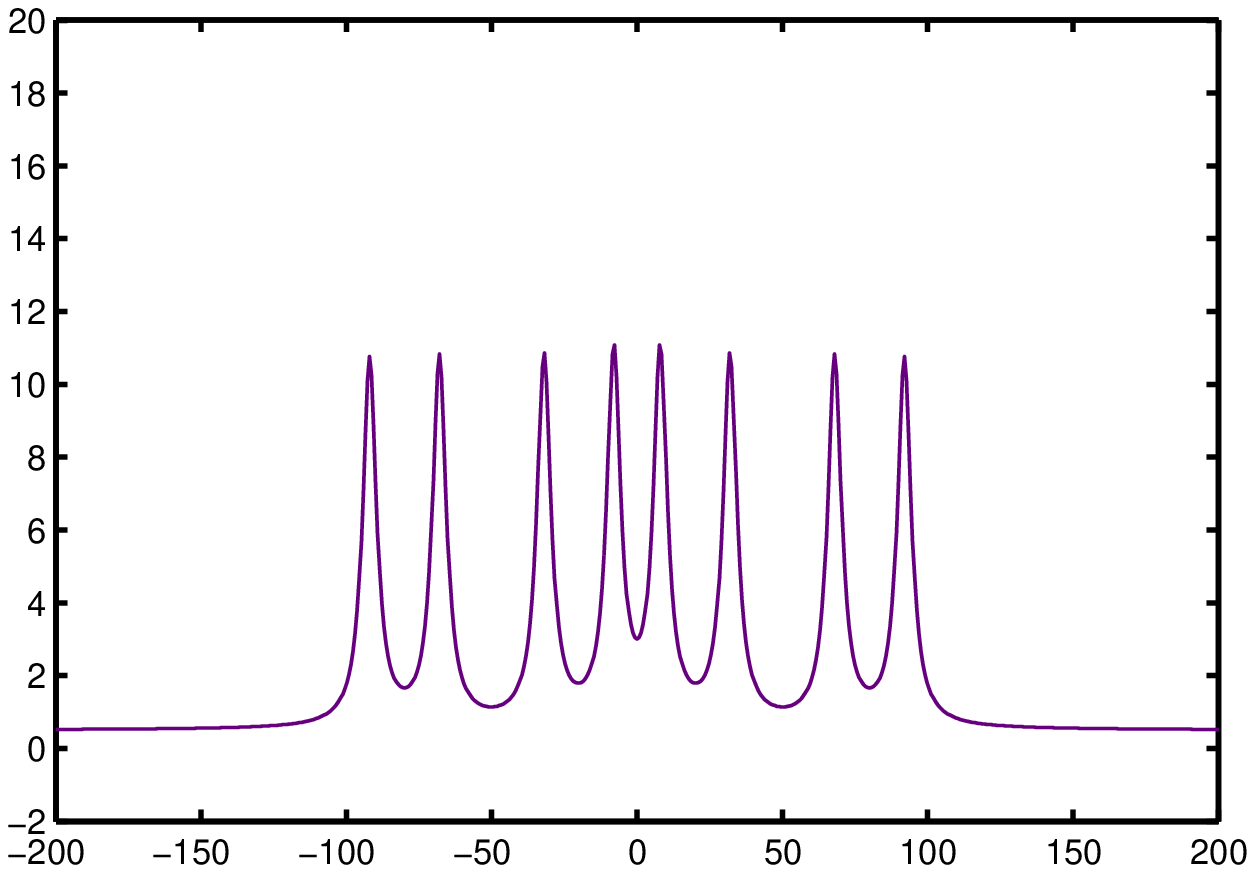}
\nputgr{0,0}{t}{width=6in}{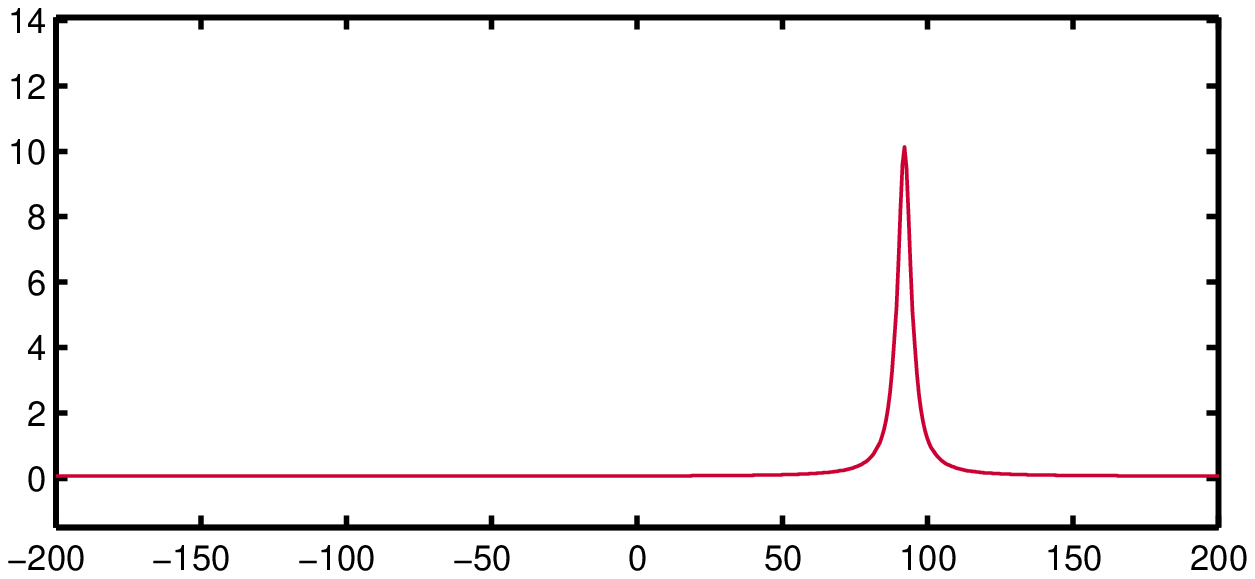}
\nputbox{1.13,2.45}{bl}{\rotatebox{60}{$\sigma_x e_\uparrow e_\uparrow e_\uparrow $}}
\nputbox{.8,2.45}{bl}{\rotatebox{60}{$\sigma_x e_\uparrow e_\uparrow e_\downarrow $}}
\nputbox{.4,2.45}{bl}{\rotatebox{60}{$\sigma_x e_\uparrow e_\downarrow e_\uparrow $}}
\nputbox{.15,2.45}{bl}{\rotatebox{60}{$\sigma_x e_\uparrow e_\downarrow e_\downarrow $}}
\nputbox{-.06,2.45}{bl}{\rotatebox{60}{$\sigma_x e_\downarrow e_\uparrow e_\uparrow $}}
\nputbox{-.35,2.45}{bl}{\rotatebox{60}{$\sigma_x e_\downarrow e_\uparrow e_\downarrow $}}
\nputbox{-.75,2.45}{bl}{\rotatebox{60}{$\sigma_x e_\downarrow e_\downarrow e_\uparrow $}}
\nputbox{-1.03,2.45}{bl}{\rotatebox{60}{$\sigma_x e_\downarrow e_\downarrow e_\downarrow$}}
\end{picture}
\label{fig:peakgroup}
\herefigcap{Relationship of a labeled pseudopure state spectrum 
to a peak group.
The top spectrum shows the
peak group of a simulated nuclear spin coupled to three other spins
with coupling constants of $100\uts{Hz}$, $60\uts{Hz}$, and
$24\uts{Hz}$. The simulation parameters are the same as in
Fig.~\ref{fig:c13peaks}. Given above each peak is the part of the
initial deviation that contributes to the peak.  The spin labels have
been omitted. Each contributing deviation consists of $\sigma_x$ on
the observed nucleus followed by one of the logical (up or down)
states (density matrices) for each of the other spins. The notation is
as defined after Eq.~\ref{eq:fullmag}. The bottom spectrum shows what
is observed if the initial deviation is the standard labeled pseudopure
state.  This state contributes only to the right-most peak, as this
peak is associated with the logical $\ket{\bitzero}$ states on the spins
not observed.}%
\end{herefig}

The labeled pseudopure state can be used as a standard pseudopure
state on $n$ qubits. Observation of the final answer of a computation
is possible by observing spin $\sysfnt{1}$, provided that the coupling
to the answer-containing spin is sufficiently strong for the peaks
corresponding to its two logical states to be well separated.  For
this purpose, the couplings to the other spins need not be resolved in
the peak group.  Specifically, to determine the answer of a
computation, the peaks of the peak group of spin $\sysfnt{1}$ are
separated into two subgroups, the first (second) containing the peaks
associated with the answer-containing spin being in state
$\ket{\bitzero}$ ($\ket{\bitone}$), respectively.  
Comparing the total signal in each of the two peak subgroups gives
the relative probabilities of the two answers ($\bitzero$ or $\bitone$).

The labeled pseudopure state can also be used to investigate the
effect of a process that manipulates the state of one qubit and requires
$n$ additional initialized qubits.  Examples include experimental
verification of one-qubit error-correcting codes as explained in
Sect.~\ref{sec:ec_exp}.

For preparing the two-qubit labeled pseudopure state, consider the
two carbon nuclei in labeled TCE with the proton spin decoupled so that its
effect can be ignored.  A ``transition'' in the density matrix for
this system is an element of the density matrix of the form
$\ketbra{ab}{cd}$, where $a,b,c$, and $d$ are $\bitzero$ or $\bitone$.  Let
$\Delta(ab,cd) = (a-c)+(b-d)$, where in the expression on the right,
$a,b,c$, and $d$ are interpreted as the numbers $0$ or $1$ as appropriate.
Applying a pulsed gradient along the $z$-axis evolves the transitions
according to: $\ketbra{ab}{cd}
\rightarrow e^{i\Delta(ab,cd)\nu z}\ketbra{ab}{cd}$, where $\nu$ is
proportional to the product of the gradient power and pulse time, and
$z$ is the molecule's position along the $z$-coordinate. For example,
$\ketbra{\bitzero\bitone}{\bitone\bitzero}$ has $\Delta=0$ and is not
affected, whereas $\ketbra{\bitzero\bitzero}{\bitone\bitone}$ acquires a
phase of $e^{-i2\nu z}$.  There are only two transitions,
$\ketbra{\bitzero\bitzero}{\bitone\bitone}$ and
$\ketbra{\bitone\bitone}{\bitzero\bitzero}$,
whose acquired phase has a rate of $\Delta=\pm 2$
along the $z$ axis. These transitions are called ``two-coherences''.
The idea is to first
recognize that these transitions can be used to define a labeled
pseudopure ``cat'' state (see below), then to exploit the
two-coherences' unique behavior under the gradient in order to extract the
pseudopure cat state, and finally to ``decode'' to a standard labeled
pseudopure state.  Note that the property that two-coherences' phases
evolve at twice the basic rate is a uniquely quantum phenomenon for
two spins. No such effect is observed for a pair of classical spins.

The standard two-qubit labeled pseudopure state's deviation can be
written as $\rho_{\textrm{\footnotesize
std}_x}=\slb{\sigma_x}{1}{1\over
2}\left(\one+\slb{\sigma_z}{2}\right)$.  We can consider other
deviations of this form where the two Pauli operators are replaced by
a pair of different, commuting products of Pauli operators.  An
example is
\begin{equation}
\rho_{\textrm{\footnotesize cat}_x} =
\left(\slb{\sigma_x}{1}\slb{\sigma_x}{2}\right){1\over 2}\left(\one+\slb{\sigma_z}{1}\slb{\sigma_z}{2}\right),
\end{equation}
where we replaced $\slb{\sigma_x}{1}$ by
$\slb{\sigma_x}{1}\slb{\sigma_x}{2}$ and $\slb{\sigma_z}{2}$ by
$\slb{\sigma_z}{1}\slb{\sigma_z}{2}$, and as announced, the two Pauli
products commute.  We will show that there is a simple sequence of
$90^\circ$ rotations whose effect is to ``decode'' the deviations
$\slb{\sigma_x}{1}\slb{\sigma_x}{2}\rightarrow
\slb{\sigma_x}{1}$ and $\slb{\sigma_z}{1}\slb{\sigma_z}{2}\rightarrow
\slb{\sigma_z}{2}$, thus converting the state $\rho_{\textrm{\footnotesize cat}_x}$ to
$\rho_{\textrm{\footnotesize std}_x}$.
The state $\rho_{\textrm{\footnotesize cat}_x}$ can be expressed
in terms of the transitions as follows:
\begin{equation}
\rho_{\textrm{\footnotesize cat}_x} =
\ketbra{\bitzero\bitzero}{\bitone\bitone}+
\ketbra{\bitone\bitone}{\bitzero\bitzero}.
\end{equation}
It can be seen that $\rho_{\textrm{\footnotesize cat}_x}$
consists only of two-coherences.  Another such state is
\begin{eqnarray}
\rho_{\textrm{\footnotesize cat}_y} &=&
    \left(\slb{\sigma_x}{1}\slb{\sigma_y}{2}\right){1\over 2}\left(\one+\slb{\sigma_z}{1}\slb{\sigma_z}{2}\right)\\
  &=&
    -i\ketbra{\bitzero\bitzero}{\bitone\bitone} 
    +i\ketbra{\bitone\bitone}{\bitzero\bitzero}.
\end{eqnarray}

Suppose that one can create a state that has a deviation of the form
$\rho=\alpha\rho_{\textrm{\footnotesize cat}_x}+\beta\rho_{\textrm{\footnotesize rest}}$ such that
$\rho_{\textrm{\footnotesize rest}}$ contains no two-coherences or zero-coherences.
After a gradient pulse is applied, the state becomes
\begin{equation}
\alpha\left(\cos(2\nu z)\rho_{\textrm{\footnotesize cat}_x}+\sin(2\nu z)\rho_{\textrm{\footnotesize cat}_y}\right)
      +\beta\rho_{\textrm{\footnotesize rest}}(z),
\end{equation}
where $\rho_{\textrm{\footnotesize rest}}(z)$ 
depends periodically on $z$ with spatial frequencies of $\pm\nu$,
not $\pm 2\nu$ or $0$.
We can then decode this state to 
\begin{eqnarray}
\varrho(z) &=&
  \alpha\left(\cos(2\nu z)\rho_{\textrm{\footnotesize std}_x}+\sin(2\nu z)\rho_{\textrm{\footnotesize std}_y}\right) +\beta\rho'_{\textrm{\footnotesize rest}}(z)
  \\
  &=&
  \alpha\left(\cos(2\nu z)\slb{\sigma_x}{1}+\sin(2\nu z)\slb{\sigma_y}{1}\right){1\over 2}\left(\one+\slb{\sigma_z}{1}\right)+\beta\rho'_{\textrm{\footnotesize rest}}(z).
\end{eqnarray}

If one now applies a gradient pulse of twice the total strength and
opposite orientation, the first term is restored to
$\alpha\rho_{\textrm{\footnotesize std}_x}$, but the second term retains non-zero
periodicities along $z$. Thus, if  we no longer use any operations to
distinguish among different molecules along the $z$-axis, or if we let
diffusion erase the memory of the position along $z$, then the second term
is eliminated from observability by being averaged to zero.  The desired
labeled pseudopure state is obtained.  Zero-coherences during the
initial gradient pulse are acceptable provided that the decoding
transfers them to coherences different from zero or two during the
final pulse in order to ensure that they also average to zero.  A pulse
sequence that realizes a version of the above procedure is shown in
Fig.~\ref{ppstate}.

\pagebreak

\begin{herefig}
\begin{picture}(7,5)(-0.25,-4)
\nputgr{0,0}{tl}{}{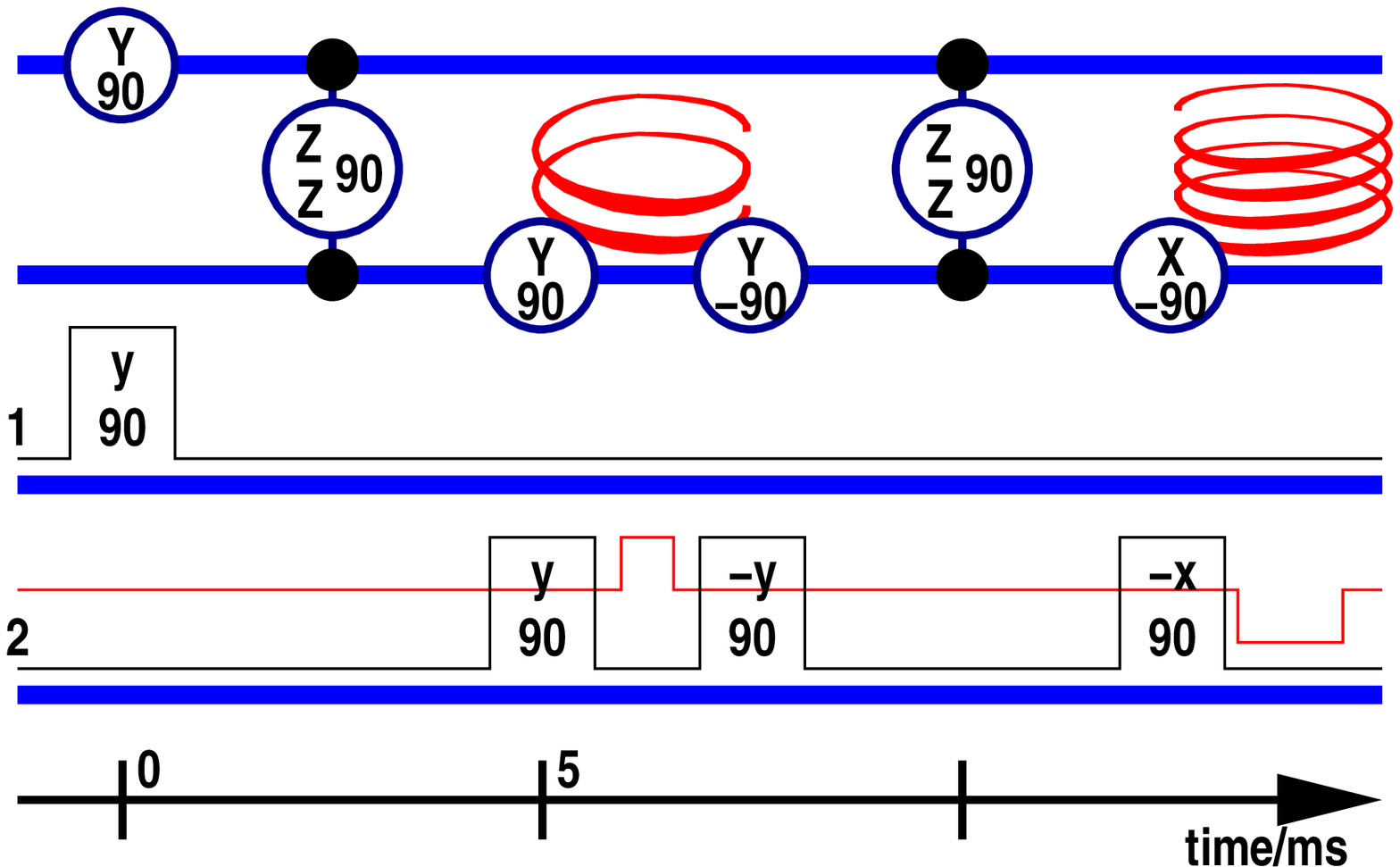}
\nputbox{.1,-.1}{bl}{$(1)$}
\nputbox{.8,-.1}{bl}{$(2)$}
\nputbox{1.75,-.1}{bl}{$(3)$}
\nputbox{2.65,-.1}{bl}{$(4)$}
\nputbox{3.35,-.1}{bl}{$(5)$}
\nputbox{3.75,-.1}{bl}{$(6)$}
\nputbox{4.75,-.1}{bl}{$(7)$}
\nputbox{5.65,-.1}{bl}{$(8)$}
\nputbox{6.35,-.1}{bl}{$(9)$}
\end{picture}
\label{ppstate}
\herefigcap{ Quantum network and pulse sequence to realize a two-qubit
labeled pseudopure state. The network is shown above the pulse
sequence realizing it. A coupling constant of $100\uts{Hz}$ is
assumed. Gradients are indicated by spirals in the network. The
gradient strength is given as the red line in the pulse sequence. The
doubling of the integrated gradient strength required to achieve the
desired ``echo'' is indicated by a doubling of the gradient pulse
time. The numbers above the quantum network are checkpoints used in
the discussion below. The input state's deviation is assumed to be
$\slb{\sigma_z}{1}$. This deviation can be obtained from the equilibrium state
by applying a $90^\circ$ rotation to spin $\sysfnt{2}$ followed by a
gradient pulse along another axis to remove
$\slb{\sigma_z}{2}$. Instead of using a gradient pulse, one can use
phase cycling, which involves performing two experiments, the second
having the sign of the phase in the first $y$ pulse changed, and then
subtracting the measured signals. }
\end{herefig}

\pagebreak

We can follow what happens to an initial deviation density matrix
of $\slb{\sigma_z}{1}$ as the network of Fig.~\ref{ppstate} is
executed. We use product operators with the abbreviations
$I=\one,X=\sigma_x,Y=\sigma_y,Z=\sigma_z$, and, for example
$XY=\slb{\sigma_x}{1}\slb{\sigma_y}{2}$. At the checkpoints indicated
in the figure the deviations are the following
\begin{equation}
\begin{array}{@{}llcl}
(1) & ZI\\
(2) & XI\\
(3) & YZ\\
(4) & YX \propto& \\ 
   & YX+XY &+& YX-XY\\
(5) & \cos(2\nu z)(YX+XY) + \sin(2\nu z)(YY-XX) &+& YX-XY\\ 
(6) & \cos(2\nu z)(YZ+XY)+ \sin(2\nu z)(YY-XZ) &+&  YZ-XY\\ 
(7) & \cos(2\nu z)(-XI+XY) + \sin(2\nu z)(YY-YI) &+& -XI-XY\\ 
(8) & \cos(2\nu z)(-XI-XZ) +\sin(2\nu z)(-YZ-YI) &+& -XI+XZ\\ 
(9) &  -X(I+Z) &+& -(\cos(-2\nu z)X+\sin(-2\nu z)Y)(I-Z).
\end{array}
\end{equation}
Except for a sign, the desired state is obtained. The
right-most term is eliminated after integrating over the sample,
or after diffusion erases memory of $z$.

This method for making a two-qubit labeled pseudopure state can be
extended to arbitrarily many ($n$) qubits with the help of the two
$n$-coherences, which are the transitions with $\Delta=\pm n$.
An experiment implementing this method
can be used to determine how good the available quantum control is.
The quality of the control is determined by a comparison of two spectral
signals: $I_p$, the intensity of the single peak that shows up in the
peak group for spin $\sysfnt{1}$ when observing the labeled
pseudopure state; and $I_0$, the intensity of the same peak in an
observation of the initial deviation after applying a $90^\circ$ pulse
to rotate $\slb{\sigma_z}{1}$ into the plane.  We performed this
experiment on a seven-spin system and determined that $I_p/I_0=
.73\pm.02$. This result implies a total error of $27\pm 2\%$. Because the
implementation has $12$ two-qubit gates, an error
rate of about $2\%$ per two-qubit gate is achievable for nuclear spins
in this setting~\cite{knill:qc1999a}.

\subsection{Quantum Error Correction for Phase Errors}
\label{sec:ec_exp}

Currently envisaged scalable quantum computers require the use of
quantum error correction to enable relatively error-free computation
on a platform of physical systems that are inherently error-prone. For
this reason, some of the most commonly used ``subroutines'' in quantum
computers will be associated with maintaining information in encoded
forms. This observation motivates experimental realizations of quantum
error-correction to determine whether adequate control can be achieved
in order to implement these subroutines and to see in a practical setting that
error-correction has the desired effects. Experiments to date have
included realizations of a version of the three-qubit repetition
code~\cite{cory:qc1998a} and of the five-qubit one-error-correcting
code (the shortest possible such code)~\cite{knill:qc2001a}. In this
section, we discuss the experimental implementation of the former.

In NMR, one of the primary sources of error is phase decoherence of
the nuclear spins due to both systematic and random fluctuations in
the field along the $z$-axis. At the same time, using gradient pulses
and diffusion, phase decoherence is readily induced artificially and
in a controlled way.  The three-bit quantum repetition code
(see~\cite{knill:qc2001d}) can be adapted to protect against phase
errors to first order. Define
$\ket{+}={1\over\sqrt{2}}(\ket{\bitzero}+\ket{\bitone})$ and
$\ket{-}={1\over\sqrt{2}}(\ket{\bitzero}-\ket{\bitone})$.  The code we
want is defined by the logical states
\begin{equation}
\kets{\bitzero}{L} = \ket{+}\ket{+}\ket{+},\hspace*{.5in}
\kets{\bitone}{L} = \ket{-}\ket{-}\ket{-}.
\end{equation}
It is readily seen that the three one-qubit phase errors,
$\slb{\sigma_z}{1},\slb{\sigma_z}{2},\slb{\sigma_z}{3}$ and ``no
error'' ($\one$) unitarily map the code to orthogonal subspaces.  It
follows that this set of errors is correctable.  See the introduction
to quantum error-correction~\cite{knill:qc2001d}.  The simplest way to
use this code is to encode one qubit's state into it, wait for some
errors to happen, and then decode to an output qubit. Success is
indicated by the output qubit's state being significantly closer to
the input qubit's state after error correction.
Without errors between encoding and decoding,
the output state should be the same as the input state, provided that
the encoding and decoding procedures are implemented perfectly.
Therefore, in this case, the experimentally determined difference
between input and output gives a measurement of how well the
procedures were implemented.

To obtain the phase-correcting repetition code from
the standard repetition code, Hadamard transforms or
$90^\circ$ $y$-rotations are applied to each qubit. The quantum
network shown in Fig.~\ref{qeccircuit} was obtained in
this fashion from the network given in~\cite{knill:qc2001d}.

\begin{herefig}
\begin{picture}(7,3)(-3.5,-2.5)
\nputgr{0,.5}{t}{width=6in}{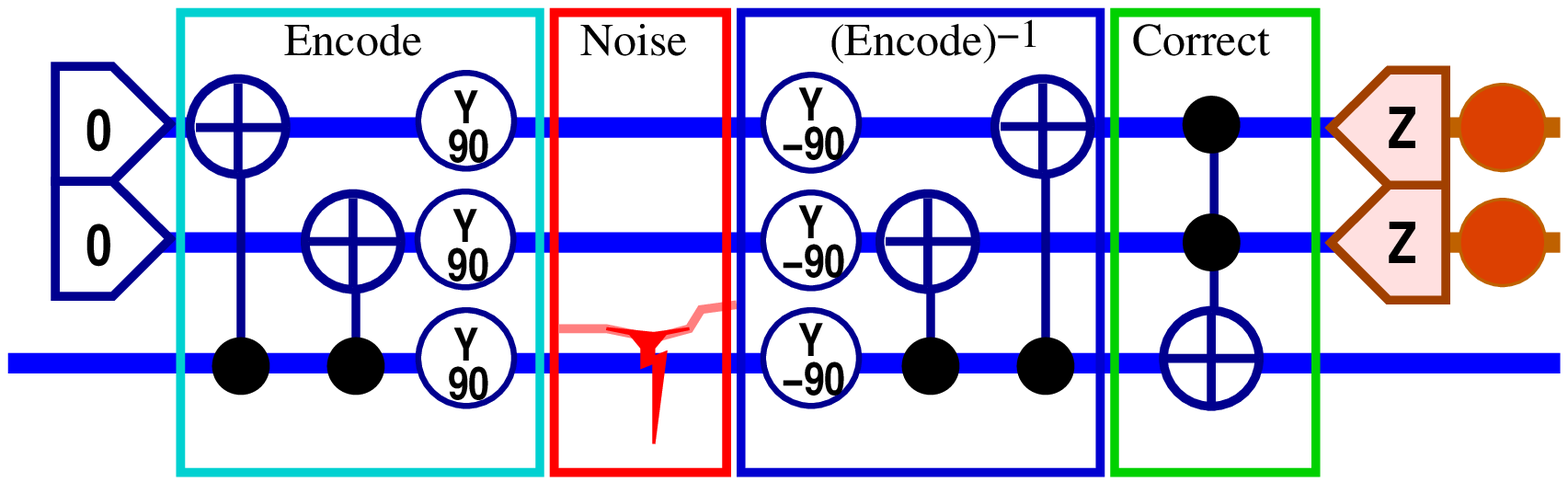}

\nputbox{-3.1,-.88}{r}{\scalebox{1.3}{$\ket{\psi}$}}

\nputbox{-2.3,-1.3}{t}{$1.$}
\nputbox{-2.5,-1.5}{t}{$
\alpha\ket{\bitzero}+\beta\ket{\bitone}\rightarrow \left\{
\overbrace{\begin{array}{l}
  \alpha\ket{{\bitzero\bitzero}}\ket{\bitzero}\\
  +\\
  \beta\ket{{\bitzero\bitzero}}\ket{\bitone}
\end{array}}\right.$}

\nputbox{-.9,-1.3}{t}{$2.$}
\nputbox{-1.3,-1.5}{tl}{$
\overbrace{\begin{array}{l}
  \alpha\ket{+{+}+}\\
  +\\
  \beta\ket{-{-}-}
\end{array}}$}

\nputbox{-.2,-1.3}{t}{$3.$}
\nputbox{+.4,-1.5}{tr}{$
\overbrace{\begin{array}{l}
  \alpha\ket{+{+}{\color{red}-}}\\
  +\\
  \beta\ket{-{-}{\color{red}+}}
\end{array}}$}

\nputbox{+1.2,-1.3}{t}{$4.$}
\nputbox{+1.45,-1.5}{tr}{$
\overbrace{\begin{array}{l}
  \alpha\ket{{\bitone{\bitone}}}\ket{\bitone}\\
  +\\
  \beta\ket{{\bitone{\bitone}}}\ket{\bitzero}\end{array}}$}

\nputbox{+2,-1.3}{t}{$5.$}
\nputbox{+1.5,-1.5}{tl}{$
\left.\overbrace{\begin{array}{l}
  \alpha\ket{{\bitone{\bitone}}}\ket{\bitzero}\\
  +\\
  \beta\ket{{\bitone{\bitone}}}\ket{\bitone}
\end{array}}\right\}\rightarrow\alpha\ket{\bitzero}+\beta\ket{\bitone}
$}
\nputbox{3.1,-.88}{l}{\scalebox{1.3}{$\ket{\psi}$}}

\end{picture}
\label{qeccircuit}
\herefigcap{Quantum network for the three-qubit phase-error-correcting
repetition code. The bottom qubit is encoded with two controlled-nots
and three $y$-rotations. In the experiment, either physical or
controlled noise is allowed to act. The encoded information is then
decoded.  For the present purposes, it is convenient to separate the
decoding procedures into two steps: The first is the inverse of the
encoding procedure, the second consists of a Toffoli gate that uses
the error information in the syndrome qubits (the top two) to restore
the encoded information.  The Toffoli gate in the last step flips the
output qubit conditionally on the syndrome qubits' state being
$\ket{\bitone\bitone}$. This gate can be realized with NMR-pulses and
delays by using more sophisticated versions of the implementation of
the controlled-not.  The syndrome qubits can be ``dumped'' at the end
of the procedure. The behavior of the network is shown for a generic
state in which the bottom qubit experiences a $\sigma_z$
error. See also~\cite{knill:qc2001d}.}
\end{herefig}

To determine the behavior and the quality of the implementation for
various $\sigma_z$-error models in an actual NMR realization, one can
use as initial states labeled pseudopure states with deviations
$\sigma_u\ketbra{\bitzero\bitzero}{\bitzero\bitzero}$ for
$u=x,y,z$. Without error, the total output signal on spin $\sysfnt{1}$
along $\sigma_u$ for each $u$ should be the same as the input signal.
Some of the data reported in~\cite{cory:qc1998a} is shown in
Fig.~\ref{qecfidelity}.

\begin{herefig}
\begin{picture}(6,3)(-3,0)
\nputgr{0,-4}{b}{width=6in}{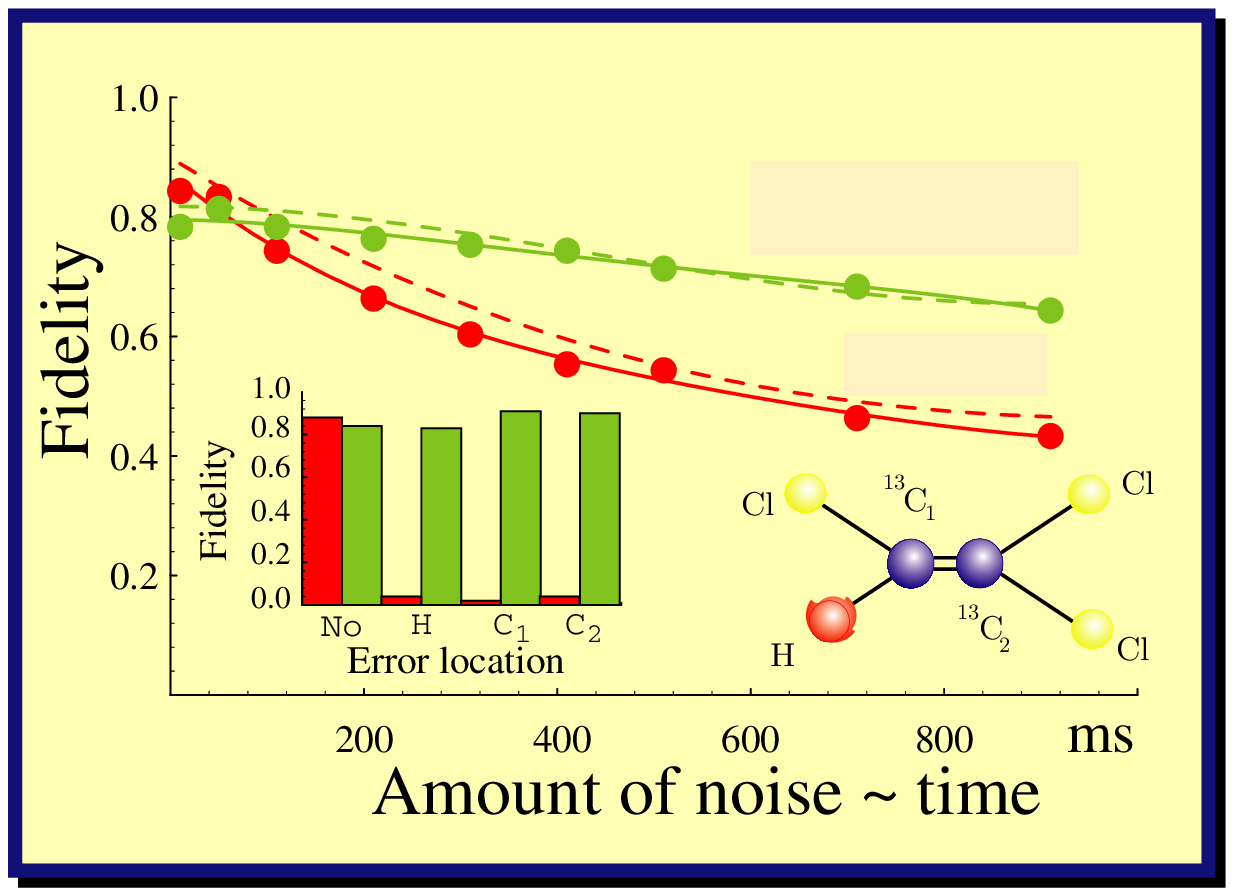}
\end{picture}
\label{qecfidelity}
\herefigcap{ Experimentally obtained fidelities for the
error-correction experiment.  The inset bar graph shows fidelities for
explicitly applied errors. The fidelities $f$ (technically, the
``entanglement'' fidelities) are an average of the signed ratios $f_u$
of the input to the output signals for the initial deviations
$\sigma_u\ketbra{\bitzero\bitzero}{\bitzero\bitzero}$ with $u=x,y,z$.
Specifically, $f={1\over 4}(1+f_x+f_y+f_z)$.  The reduction from $1$
of the green bars (showing fidelity for the full procedure) is due to
errors in our implementation of the pulses and from relaxation
processes.  The red bars are the fidelity for the output before the
last error-correction step, and they contain the effects of the
errors. The main graph shows the fidelities for the physical
relaxation process. Here, the evolution consisted of a delay varying
up to $1000\uts{ms}$.  The red curve is the fidelity of the output
qubit before the final Toffoli gate that corrects the errors based on
the syndrome. The green curve is the fidelity of the output after the
Toffoli gate.  The effect of error-correction can be seen by a
significant flattening of the curve because correction of first-order
(that is, single) phase errors implies that residual, uncorrected
(that is, double or triple) phase errors increase quadratically in
time.  The green curve starts lower than the red one because of
additional errors incurred by the implementation of the the Toffoli
gate.  The dashed curves are obtained by simulation using estimated
phase relaxation rates with half times of $2\uts{s}$ (proton),
$0.76\uts{s}$ (first carbon) and $0.42\uts{s}$ (second carbon).
Errors in the data points are approximately $0.05$. The molecule used
was TCE. For a more thorough implementation and analysis of a
three-qubit phase-error correcting code, see~\cite{sharf:qc1999a}.}
\end{herefig}

Work on benchmarking error-control methods using liquid-state NMR is
continuing. Other experiments include the implementation of a
two-qubit code with an application to
phase-errors~\cite{leung:qc1999b} and the verification of the shortest
non-trivial noiseless subsystem on three qubits~\cite{viola:qc2001a}.
The latter demonstrates that for some physically realistic noise
models, it is possible to store quantum information in such a way that
it is completely unaffected by the noise.

\section{Discussion}

\subsection{Overview of Contributions to QIP}

Important issues in current experimental efforts toward realizing QIP
are to find ways of achieving the necessary quantum control and to
determine whether sufficiently low error-rates are
possible. Liquid-state NMR is the only extant system (as of 2002) with
the ability to realize relatively universal manipulations on more than
two qubits (restricted control has been demonstrated in four
ions~\cite{sackett:qc2000a}). For this reason, NMR serves as a useful
platform for developing and experimentally verifying techniques for
QIP and for establishing simple procedures for benchmarking
information processing tasks.  The ``cat-state'' and the various
error-correction benchmarks~\cite{knill:qc1999a,knill:qc2001a} consist
of a set of quantum control steps and measurement procedures that can
be used with any general-purpose QIP system to determine, in a device
independent way, the degree of control achieved.  The demonstration of
error rates in the few percent per non-trivial operation is
encouraging. For existing and proposed experimental systems other than
NMR, achieving such error rates is still a great challenge.

Prior research in NMR, independent of quantum information, has proved
to be a rich source of basic quantum control techniques useful for
physically realizing quantum information in other settings.  We
mention four examples. The first is the development of sophisticated
shaped-pulse techniques that can selectively control transitions or
spins while being robust against typical errors.  These techniques are
finding applications to quantum control involving laser
pulses~\cite{warren:qc1993a} and are likely to be very useful when
using coherent light to accurately control transitions in atoms or
quantum dots, for example. The second is the recognition that there
are simple ways in which imperfect pulses can be combined to eliminate
systematic errors such as those associated with miscalibration of
power or side-effects on off-resonant nuclear spins.  Although many of
these techniques were originally developed for such problems as
accurate inversion of spins, they are readily generalized to other
quantum gates~\cite{levitt:qc1982a,cummins:qc1999a}. The third example
is decoupling used to reduce unwanted external interactions.  For
example, a common problem in NMR is to eliminate the interactions
between proton and labeled carbon nuclear spins to observe
``decoupled'' carbon spins. In this case, the protons constitute an
external system with an unwanted interaction.  To eliminate the
interaction, it is sufficient to invert the protons frequently.
Sophisticated techniques for ensuring that the interactions are
effectively turned off independent of pulse errors have been developed
(See, for example,~\cite{ernst:qc1994a}). These techniques have been
greatly generalized and shown to be useful for actively creating
protected qubit subsystems in any situation in which the interaction
has relatively long correlation
times~\cite{viola:qc1998a,viola:qc1999a}.  Refocusing to undo unwanted
internal interactions is our fourth example.  The technique for
``turning off'' the coupling between spins that is so important for
realizing QIP in liquid-state NMR is a special case of much more
general methods of turning off or refocusing Hamiltonians. For
example, a famous technique in solid state NMR is to reverse the
dipolar coupling Hamiltonian using a clever sequence of $180^\circ$
pulses at different phases (see, for example, \cite{ernst:qc1994a},
page 48). Many other proposed QIP systems suffer from such internal
interactions while having similar control opportunities.

The contributions of NMR QIP research extend beyond those directly
applicable to experimental QIP systems.  It is due to NMR that the
idea of ensemble quantum computation with weak measurement was
introduced and recognized as being, for true pure initial states, as
powerful for solving algorithmic problems as the standard model of
quantum computation. (It cannot be used in settings involving quantum
communication.)  One implication is that to a large extent, the usual
assumption of projective measurement can be replaced by any
measurement that can statistically distinguish between the two states
of a qubit.  Scalability still requires the ability to ``reset''
qubits during the computation, which is not possible in liquid-state
NMR. Another interesting concept emerging from NMR QIP is that of
``computational cooling''~\cite{schulman:qc1998a}, which can be used
to efficiently extract initialized qubits from a large number of noisy
qubits in initial states that are only partially biased toward
$\ket{\bitzero}$. This is a very useful tool for better exploiting
otherwise noisy physical systems.

The last example of interesting ideas arising from NMR studies is the
``one-qubit'' model of quantum computation~\cite{knill:qc1998c}. This
is a useful abstraction of the capabilities of liquid-state NMR. In
this model, it is assumed that initially, one qubit is in the state
$\ket{\bitzero}$ and all the others are in random states. Standard
unitary quantum gates can be applied and the final measurement is
destructive. Without loss of generality, one can assume that all
qubits are re-initialized after the measurement. This model can
perform interesting physics simulations with no known efficient
classical algorithms.  On the other hand, with respect to oracles, it
is strictly weaker than quantum computation. It is also known that it
cannot ``faithfully'' simulate quantum computers~\cite{ambainis:qc2000a}.

\subsection{Capabilities of Liquid-State NMR}

One of the main issues in liquid-state NMR QIP is the highly mixed
initial state. The methods for extracting pseudopure states are not
practical for more than $10$ (or so) nuclear spins. The problem is
that for these methods, the pseudopure state signal decreases
exponentially with the number of qubits prepared while the noise level
is constant.  This exponential loss limits the ability to explore and
benchmark standard quantum algorithms even in the absence of
noise. There are in fact ways in which liquid-state NMR can be
usefully applied to many more qubits.  The first and less practical is
to use computational cooling for a (unrealistically) large number of
spins to obtain less mixed initial states. Versions of this technique
have been studied and used in NMR to increase signal to
noise~\cite{glaser:qc1998a}.  The second is to use the one-qubit model
of quantum computation instead of trying to realize pseudopure
states. For this purpose, liquid-state NMR is limited only by
relaxation noise and pulse control errors, not by the number of
qubits.  Noise still limits the number of useful operations, but
non-trivial physics simulations are believed to be possible with less
than 100 qubits~\cite{lloyd:qc1996a}. Remarkably, a one-qubit quantum
computer can efficiently obtain a significant amount of information
about the spectrum of a Hamiltonian that can be emulated on a quantum
computer~\cite{knill:qc1998c,somma:qc2001a,miquel:qc2001a}.
Consequently, although QIP with molecules in liquid state cannot
realistically be used to implement standard quantum algorithms
involving more than about $10$ qubits, its capabilities have the
potential of exceeding the resource limitations of available classical
computers for some applications.

\subsection{Prospects for NMR QIP}

There are many more algorithms and benchmarks that can be usefully
explored using the liquid-state NMR platform.  We hope to soon have a
molecule with ten or more useful spins and good properties for QIP.
Initially this molecule can be used to extend and verify the behavior of
existing scalable benchmarks. Later, experiments testing basic ideas
in physics simulation or more sophisticated noise-control methods are
likely.

Liquid-state NMR QIP is one of many ways in which NMR can be used for
quantum information. One of the promising proposals for quantum
computation is based on phosphorus embedded in
silicon~\cite{kane:qc1997a} and involves controlling phosphorus
nuclear spins using NMR methods. In this proposal, couplings and
frequencies are controlled with locally applied voltages. RF pulses
can be used to implement universal control.  It is also possible to
scale up NMR QIP without leaving the basic paradigms of liquid-state
NMR while adding such features as high polarization, the ability to
dynamically reset qubits (required for scalability) and much faster
two-qubit gates.  One proposal for achieving this goal is to use
dilute molecules in a solid state matrix instead of molecules in
liquid~\cite{cory:qc2000a}. This approach may lead to pure-state
quantum computation for significantly more than ten qubits.

NMR QIP has been a useful tool for furthering our understanding of the
experimental challenges of quantum computation.  We believe that NMR
QIP will continue to shed light on important issues in physically
realizing quantum information.

\pagebreak

\noindent{\bf Acknowledgements}: We thank Nikki Cooper and Ileana Buican
for their extensive encouragement and editorial help.

\vspace*{\baselineskip}

\makeaddress

\bibliographystyle{unsrt}
\bibliography{journalDefs,qc}

\section{Glossary}

\begin{description}
\setlength{\itemsep}{0pt}\setlength{\parskip}{0pt}\setlength{\parsep}{0pt}

\item[\textbf{Bloch sphere}.]
A representation of the state space of a qubit using the unit
sphere in three dimensions. See Fig.~\ref{figbloch}.

\item[\textbf{Crosstalk}.] 
In using physical control to implement a gate, crosstalk refers to
unintended effects on qubits not involved in the gate.

\item[\textbf{Decoupling}.]
A method for ``turning off'' the interactions between two sets of
spins. In NMR, this task can be achieved if one applies a rapid
sequence of refocusing pulses to one set of spins. The other set of
spins can then be controlled and observed as if independent of the
first set.

\item[\textbf{Deviation of a state}.]
If $\rho$ is a density matrix for a state and $\rho=\alpha\one +
\beta\sigma$, then $\sigma$ is a deviation of $\rho$.

\item[\textbf{Ensemble computation}.]
Computation with a large ensemble of identical and independent
computers. Each step of the computation is applied identically to the
computers. At the end of the computation, the answer is determined
from a noisy measurement of the fraction $p_{\bitone}$ of the
computers whose answer is ``$\bitone$''.  The amount of noise is
important for resource accounting: To reduce the noise to
below $\epsilon$ requires increasing the resources used by a factor of
the order of $1/\epsilon^2$.

\item[\textbf{Equilibrium state}.]
The state of a quantum system in equilibrium with its environment.  In
the present context, the environment behaves like a heat bath at
temperature $T$ and the equilibrium state can be written as $\rho =
e^{-H/kT}/Z$, where $H$ is the effective internal Hamiltonian of the
system and $Z$ is determined by the identity $\trace\rho = 1$.

\item[\textbf{FID}.]  Free induction decay. To obtain a spectrum
on an NMR spectrometer after having applied pulses to a sample, one
measures the decaying planar magnetization induced by the nuclear
spins as they precess. The $x$- and $y$-components $M_x(t)$ and
$M_y(t)$ of the magnetization as a function of time are combined to
form a complex signal $M(t)=M_x(t)+iM_y(t)$. The record of $M(t)$ over
time is called the FID, which is Fourier-transformed to yield the
spectrum.

\item[\textbf{Inversion}.] A pulse that flips the 
component of the spin along the $z$-axis.  Note that any $180^\circ$
rotation around an axis in the $xy$-plane has this effect.

\item[\textbf{$J$-coupling}.]
The type of coupling present between two nuclear spins in
a molecule in the liquid state. 

\item[\textbf{Labeled molecule}.]
A molecule in which some of the nuclei are substituted by less
common isotopes. A common labeling for NMR QIP involves
replacing the naturally abundant carbon isotope $^{12}$C,
with the spin-${1\over 2}$ isotope $^{13}$C.

\item[\textbf{Larmor frequency}.]
The precession frequency of a nuclear spin in a magnetic field.
It depends linearly on the spin's magnetic moment and the
strength of the field.

\item[\textbf{Logical frame}.]
The current frame with respect to which the state of a qubit carried
by a spin is defined.  There is an absolute (laboratory) frame
associated with the spin observables $\sigma_x,\sigma_y,$ and $\sigma_z$.  The
observables are spatially meaningful. For example, the magnetization
induced along the $x$-axis is proportional to
$\trace(\sigma_x\ketbra{\psi}{\psi})$, where $\ket{\psi}$ is the
physical state of the spin. Suppose that the logical frame is obtained
from the physical frame with a rotation by an angle of $\theta$ around
the $z$-axis.  The observables for the qubit are then given by
$\slb{\sigma_x}{L}=\cos(\theta)\sigma_x+\sin(\theta)\sigma_y$,
$\slb{\sigma_y}{L}=\cos(\theta)\sigma_y-\sin(\theta)\sigma_z$, and
$\slb{\sigma_z}{L}=\sigma_z$.  As a result, the change to the logical
frame transforms the physical state to a logical state according to
$\kets{\phi}{L} = e^{i\sigma_z\theta/2}\ket{\psi}$.  That is, the
logical state is obtained from the physical state by a $-\theta$
rotation around the $z$-axis.  A resonant logical frame is used in NMR
to compensate for the precession induced by the strong external field.

\item[\textbf{Magnetization}.] The magnetic field induced
by an ensemble of magnetic spins. The magnitude of the magnetization
depends on the number of spins, the extent of alignment and
the magnetic moments.

\item[\textbf{Nuclear magnetic moment}.]
The magnetic moment of a nucleus determines the strength of the
interaction between its nuclear spin and a magnetic field.  The
precession frequency $\omega$ of a spin ${1\over 2}$ nucleus is given
by $\mu B$, where $\mu$ is the nuclear magnetic moment and $B$ the
magnetic field strength. For example, for a proton,
$\mu=42.7\uts{Mhz}/\uts{T}$.

\item[\textbf{NMR spectrometer}.]
The equipment used to apply RF pulses to and observe precessing
magnetization from nuclear spins. Typical spectrometers consist of a
strong, cylindrical magnet with a central bore in which there is a
``probe'' that contains coils and a sample holder. The probe is
connected to electronic equipment for applying RF currents to the
coils and for detecting weak oscillating currents induced by the
nuclear magnetization.

\item[\textbf{Nuclear spin}.]
The quantum spin degree of freedom of a nucleus. It is characterized
by its total spin quantum number, which is a multiple of ${1\over 2}$.
Nuclear spins with spin ${1\over 2}$ are  two-state quantum systems
and can therefore  be used as qubits immediately.

\item[\textbf{Nutation}.] The motion of a spin in a strong
$z$-axis field caused by a resonant pulse.

\item[\textbf{Nutation frequency}.]
The angular rate at which a resonant pulse causes nutation of a precessing
spin around an axis in the plane.

\item[\textbf{One-qubit quantum computing}.]
The model of computation in which one can initialize any number of qubits
in the state where qubit $\sysfnt{1}$ is in the state
$\kets{\bitzero}{1}$ and all the other qubits are in a random
state. One can then apply one- and two-qubit unitary quantum gates and make
one final measurement of the state of qubit $\sysfnt{1}$ after which
the system is reinitialized. The model can be used to determine
properties of the spectral density function of a Hamiltonian which can
be emulated by a quantum computer~\cite{knill:qc1998c}.

\item[\textbf{Peak group}.]
The spectrum of an isolated nuclear spin consists of one peak at its
precession frequency. If the nuclear spin is coupled to others,
this peak ``splits'' and multiple peaks are observed near
the precession frequency. The nuclear spin's peak group
consists of these peaks.

\item[\textbf{Precession}.]
An isolated nuclear spin's state can be associated with a spatial
direction using the Bloch sphere representation.  If the direction
rotates around the $z$-axis at a constant rate, we say that it
precesses around the $z$-axis.  The motion corresponds to that of a
classical top experiencing a torque perpendicular to both the $z$-axis
and the spin axis. For a nuclear spin, the torque can be caused
by a magnetic field along the $z$-axis.

\item[\textbf{Projective measurement}.]
A measurement of a quantum system determined by a complete set of
orthogonal projections whose effect is to apply one of the projections
to the system (``wave function collapse'') with a probability determined
by the amplitude squared of the projected state. Which projection
occurred is known after the measurement.  The simplest example is that
of measuring qubit $\sysfnt{q}$ in the logical basis.  In this case,
there are two projections, namely,
$P_{\bitzero}=\ketbras{\bitzero}{\bitzero}{q}$ and
$P_{\bitone}=\ketbras{\bitone}{\bitone}{q}$.  If the initial state of
all the qubits is $\ket{\psi}$, then the probabilities of the two
measurement outcomes $\bitzero$ and $\bitone$ are
$p_{\bitzero}=\bra{\psi}P_{\bitzero}\ket{\psi}$ and
$p_{\bitone}=\bra{\psi}P_{\bitone}\ket{\psi}$, respectively.  The
state after the measurement is $P_\bitzero=\ket{\psi}/\sqrt{p_{\bitzero}}$
for outcome $\bitzero$ and $P_\bitone=\ket{\psi}/\sqrt{p_{\bitone}}$ for
outcome $\bitone$.

\item[\textbf{Pseudopure state}.]
A state with deviation given by a pure state $\ketbra{\psi}{\psi}$.

\item[\textbf{Pulse}.]
A transient field applied to a quantum system.  In the case of NMR
QIP, pulses are rotating magnetic fields (RF pulses) whose effects are
designed to cause specific rotations of the qubit states carried by
the nuclear spins.

\item[\textbf{Refocusing pulse}.]
A pulse that causes a $180^\circ$ rotation around an axis in the
plane. A typical example of such a rotation is $e^{-i\sigma_x\pi/2} =
-i\sigma_x$, which is a $180^\circ$ $x$-rotation.

\item[\textbf{Resonant RF pulse}.]
A pulse whose field oscillates at the same frequency as the precession
frequency of a target nuclear spin. Ideally, the field is in the
plane, rotating at the same frequency and in the same direction as the
precession. However, as long as the pulse field is weak compared to
the precession frequency (that is, by comparison, its nutation
frequency is small), the nuclear spin is affected only by the
co-rotating component of the field. As a result, other planar
components can be neglected, and a field oscillating in a constant
direction in the plane has the same effect as an ideal resonant field.

\item[\textbf{RF pulse}.]
A pulse resonant at radio frequencies. Typical frequencies used in NMR
are in this range.

\item[\textbf{Rotating frame}.]
A frame rotating at the same frequency as the precession frequency
of a spin.

\item[\textbf{Rotation}.]
In the context of spins and qubits, a rotation around $\sigma_u$ by an
angle $\theta$ is an operation of the form $e^{-i\sigma_u\theta/2}$.
The operator $\sigma_u$ may be any unit combination of Pauli
matrices. This defines an axis in three-space, and in the Bloch sphere
representation, the operation has the effect suggested by the
terminology.

\item[\textbf{Spectrum}.]
In the context of NMR, the Fourier transform of an FID.

\item[\textbf{Weak measurement}.]
A measurement involving only a weak interaction with the measured
quantum system. Typically, the measurement is ineffective unless an
ensemble of these quantum systems is available so that the effects of
the interaction add up to a signal detectable above the noise.  The
measurement of nuclear magnetization used in NMR is weak in this
sense.

\end{description}

\end{document}